\documentclass[prd,preprint,longbibliography,
  lengthcheck,onecolumn,notitlepage,
  preprintnumbers,superscriptaddress]{revtex4-1}

\usepackage{graphicx}
\usepackage[usenames,dvipsnames]{xcolor}
\graphicspath{{figures/}}

\usepackage{hyperref}
\hypersetup{colorlinks=true,citecolor=blue,linkcolor=blue,urlcolor=blue}

\usepackage{diagbox}
\usepackage{booktabs}

\usepackage{amsmath,amssymb,amsfonts}
\usepackage{mathrsfs}
\usepackage{slashed}
\usepackage{bm}
\usepackage{bbm}

\usepackage[draft]{changes} 

\usepackage[utf8]{inputenc}
\usepackage{newtxtext}
\usepackage[mathcal]{euscript}

\renewcommand*{\d}{\mathop{}\!\mathrm{d}}
\newcommand{\nn}{\nonumber}

\definecolor{darkgreen}{RGB}{0,120,0}

\begin{document}

\title{
  Kernel methods for evolution of generalized parton distributions 
}
\author{A. Freese}
\email{afreese@jlab.org}
\affiliation{Theory Center, Jefferson Lab, Newport News, Virginia 23606, USA}

\author{D. Adamiak}
\affiliation{Theory Center, Jefferson Lab, Newport News, Virginia 23606, USA}

\author{I. Clo\"{e}t}
\affiliation{Physics Division, Argonne National Laboratory, Argonne, Illinois 60439, USA}

\author{W. Melnitchouk}
\affiliation{Theory Center, Jefferson Lab, Newport News, Virginia 23606, USA}

\author{J.-W. Qiu}
\affiliation{Theory Center, Jefferson Lab, Newport News, Virginia 23606, USA}
\affiliation{\mbox{Department of Physics, William \& Mary, Williamsburg, Virginia 23187, USA}}

\author{N. Sato}
\affiliation{Theory Center, Jefferson Lab, Newport News, Virginia 23606, USA}

\author{M. Zaccheddu}
\affiliation{Theory Center, Jefferson Lab, Newport News, Virginia 23606, USA}

\begin{abstract}
Generalized parton distributions (GPDs) characterize the 3-dimensional structure of hadrons, combining information about their internal quark and gluon longitudinal momentum distributions and transverse position within the hadron.
The dependence of GPDs on the factorization scale $Q^2$ allows one to connect hard exclusive processes involving GPDs at disparate energy and momentum scales, which is needed in global analyses of experimental data.
In this work we explore how finite element methods can be used to construct fast and differentiable $Q^2$ evolution codes for GPDs in momentum space, which can be used in a machine learning framework.
We show numerical benchmarks of the methods' accuracy, including a comparison to an existing evolution code from PARTONS/APFEL++, and provide a repository where the code can be accessed.
\end{abstract}

\preprint{JLAB-THY-24-4249}

\maketitle

\section{Introduction}
\label{sec:intro}

Protons and neutrons, known collectively as nucleons, are the fundamental building blocks of all atomic nuclei, making up the bulk of the visible mass in the universe.
More than 50 years of study have revealed that nucleons are not static but have a complex and dynamic internal structure in terms of elementary quarks and gluons (collectively, partons), whose interactions are governed by quantum chromodynamics (QCD) and its strong color force~\cite{Ellis:1996mzs, Collins:2011zzd}.
Because of color confinement, quarks and gluons cannot be seen directly in particle detectors, and reliably quantifying the partonic structure of hadrons is a unique and unprecedented intellectual challenge in all of science.
With the help of asymptotic freedom~\cite{Gross:1973id, Politzer:1973fx}, at short distances the color interaction becomes weaker, however, allowing approximate calculations to be made within perturbation theory.
A rigorous theoretical framework to connect measurements of hadrons to information about their partonic constituents is provided through QCD factorization, which enables us to systematically study the partonic structure indirectly in terms of well defined and universal quantum correlation functions (QCFs)~\cite{Collins:1989gx, Collins:1998be, Collins:2011zzd}.

With advances over the last decades in both theory and experiment, high precision experiments in which highly energetic leptons are scattered diffractively from the proton (with the proton remaining intact) can now be routinely performed, ushering a new era of nuclear tomography in which the internal landscape and images of quarks and gluons inside the proton can be quantitatively explored.
With the intense 12~GeV beam of electrons at Jefferson Lab and the future higher energy Electron-Ion Collider (EIC), we anticipate reconstructing a variety of physical features of confined systems of partons inside nucleons and nuclei, which are systematically encoded into the various partons' QCFs.
In particular, generalized parton distributions (GPDs)~\cite{Ji:1996nm, Radyushkin:1997ki, Ji:1998pc, Radyushkin:2000uy, Diehl:2003ny} are one important class of QCFs that provide access to the spatial distribution of quarks and gluons, their emergent mechanical properties encoded in the energy-momentum tensor~\cite{Polyakov:2002yz, Polyakov:2018zvc, Polyakov:2018rew, Lorce:2018egm}, as well as the spin~\cite{Jaffe:1989jz, Ji:1996ek, Leader:2013jra} and mass decompositions~\cite{Lorce:2017xzd, Lorce:2018uyy, Metz:2020vxd, Lorce:2021xku} of the hadronic system.

Experiments measuring scattering cross sections of leptons and hadrons provide indirect access to QCFs.
When scattering occurs with large momentum transfer, QCD factorization allows us to express the measured cross sections in terms of convolutions between short-distance processes, calculable perturbatively in QCD, and the various partonic QCFs, which represent the long-distance and universal structure information of nucleons and nuclei.
(Corrections to this convolution are suppressed by inverse powers of the large four-momentum transfer squared.)
The separation of the experimentally measured cross section into a long-distance QCF and a short-distance scattering process is not unique; however, the experimentally measured cross sections do not depend on how the factorization is implemented.
This invariance allows the derivation of renormalization group equations (RGEs), which describe how the QCFs depend on the momentum scales at which they are extracted.
The RGE evolution of the QCFs is key to the predictive power of QCD, and should be calculated precisely in any comparison between theory and experiment, or any phenomenological extraction of QCFs from data using QCD factorization.

In the convolution formula of the factorization formalism, the momenta of active partons participating in the hard, short-distance process are integrated out, which makes the extraction of these universal QCFs from experimental data a challenging inverse problem.
Solving this inverse problem involves developing a simulation-based workflow to infer the partonic QCFs.
There are different levels at which such simulation-based analyses can be realized, including the traditional approach of partial simulation-based analysis, where the observational data are transformed into a space, such as cross sections or asymmetries, that can be directly compared with theory. 
However, such transformations are limited by irreducible systematic uncertainties arising from the removal of effects introduced by the experimental simulations. 
Alternatively, one can implement the full end-to-end simulation pipeline to reconstruct QCFs directly from the observational data, which are phase space samples from particle reactions (or ``events''). 
Although such an inference strategy is ideal, it comes at a high computational cost that requires the use of large-scale computing facilities, and the existing software toolkits are not currently developed for these purposes. 
In addition, with the rise of deep learning and associated auto-differentiation programming capabilities, the simulation-based framework allows for the utilization of AI/ML methods and tools. 
These advancements also have the potential to mitigate model biases in reconstructing QCFs and enhance optimization algorithms through the use of exact gradients.

In this paper, we address one of the most computationally intensive components in the simulation pipeline for GPD analysis. 
Specifically, we focus on solving the GPD evolution equations, which are numerically demanding, as described in the following sections.
Our solution is developed using methods based on finite-element theory, allowing us to formulate the problem as ordinary tensor manipulations that are suitable for modern GPUs capable of efficiently parallelizing the computations.

This work is organized as follows. 
In Sec.~\ref{sec:theory} we give a basic overview of GPD evolution, and in Sec.~\ref{sec:pixel} present a pedagogical description of the ``kernel'' or matrix method---essentially, a finite element method---for performing evolution in discretized $x$ space.
In Sec.~\ref{sec:main} we describe in detail how specific finite element methods can be used to numerically solve the evolution equations.
For both pedagogy and broadness, we examine two different finite element methods, which we refer to as the ``Simple Method'' and ``Refined Method'', and perform comparative benchmarks of these methods against each other (and against existing codes) in Sec.~\ref{sec:benchmarks}.
Finally, we summarize our findings and conclude in Sec.~\ref{sec:end}.

\section{Renormalization group evolution}
\label{sec:theory}

Analytically, GPD evolution is described by a collection of coupled integro-differential equations:
\begin{align}
  \label{eqn:evolution}
  \frac{\d H_a(x,\xi,Q^2)}{\d \log Q^2}
  =
  \sum_{b}
  \int_{-1}^1 \d y \,
  K_{ab}(x,y,\xi,Q^2)\,
  H_b(y,\xi,Q^2)
  \,,
\end{align}
where $H_a(x,\xi,Q^2)$ is a GPD for a parton with flavor $a \in \{g, u, d, s, c, b \}$, written as a function of the parton light-cone momentum fraction $x$, skewness $\xi$, momentum transfer squared $t$, and the renormalization scale $Q^2$, and where $K_{ab}(x,y,\xi,Q^2)$ is the evolution kernel.
While the GPD also depends on a momentum transfer squared $t$, the evolution kernel does not, so any numerical method for performing evolution is $t$~independent.
In this work, for ease of notation we will suppress the $t$ dependence of the GPDs throughout.
The evolution kernel can be perturbatively expanded as:
\begin{align}
  K_{ab}(x,y,\xi,Q^2)
  =
  \sum_{n=1}^\infty
  \left(
  \frac{\alpha_{s}(Q^2)}{2\pi}
  \right)^n
  K_{ab}^{(n)}(x,y,\xi,Q^2)
  \,,
\end{align}
where $\alpha_s$ is the QCD coupling, and we refer to the $n=1$ contribution as leading order (LO), the $n=2$ as next-to-leading order (NLO), and so on.
The methods we develop here will be applicable to any order in perturbative QCD, however, in this paper (and the first code release) we only deal with LO.
Explicit formulas for the kernels are given in Appendix~\ref{sec:kernels}.

\subsubsection{Partial diagonalization of evolution equations}

As usual for evolution of parton distributions~\cite{Ellis:1996mzs}, we convert from a physical flavor basis to an evolution basis.
First, the plus and minus versions of the quark GPDs are defined as:
\begin{subequations}
\begin{align}
  H_q^+(x,\xi,Q^2)
  & \equiv
  H^q(x,\xi,Q^2)
  -
  H^q(-x,\xi,Q^2)
  =
  H^q(x,\xi,Q^2)
  +
  H^{\bar{q}}(x,\xi,Q^2)\,,
  \label{eqn:Hqplus}
  \\
  H_q^-(x,\xi,Q^2)
  & \equiv
  H^q(x,\xi,Q^2)
  +
  H^q(-x,\xi,Q^2)
  =
  H^q(x,\xi,Q^2)
  -
  H^{\bar{q}}(x,\xi,Q^2)
  \,,
  \label{eqn:Hqminus}
\end{align}
\end{subequations}
where the latter equality in each equation arises from the crossing symmetry relation $H^{\bar{q}}(x,\xi,Q^2) = -H^q(-x,\xi,Q^2)$.
The minus-type GPDs $H_q^-(x,\xi,Q^2)$ are nonsiglets and fully decouple in their evolution.
From the plus-type GPDs, the following nonsinglet combinations can be constructed:
\begin{subequations}
\begin{align}
  H_{T3}
  & \equiv
  H_u^+
  -
  H_d^+\,,
  \\
  H_{T8}
  & \equiv
  H_u^+
  +
  H_d^+
  -
  2 H_s^+\,,
  \\
  H_{T15}
  & \equiv
  H_u^+
  +
  H_d^+
  +
  H_s^+
  -
  3 H_c^+\,,
  \\
  H_{T24}
  & \equiv
  H_u^+
  +
  H_d^+
  +
  H_s^+
  +
  H_c^+
  -
  4 H_b^+\,,
\end{align}
\label{eqn.nonsinglet}
\end{subequations}
which likewise decouple in their evolution.
The quark singlet combination:
\begin{align}
  H^{\rm S}
  & \equiv
  \sum_q H_q^+
  \,,
\label{eqn.singlet}
\end{align}
couples to the gluon GPD in its evolution.
Note that for clarity the dependence on $(x,\xi,Q^2)$ in Eqs.~(\ref{eqn.nonsinglet}) and (\ref{eqn.singlet}) is omitted.
We refer to the nonsinglet and singlet mixtures of quark distributions as the evolution basis.

A common scheme for accounting for quark masses is the zero-mass variable flavor number scheme~\cite{Collins:1986mp}, which we employ in our analysis.
In this scheme, evolution of heavy quark distributions is turned off when $Q^2$ is below the mass squared of that quark flavor.
When $Q^2$ passes the mass threshold, the heavy quark is turned on, but treated as if it has zero mass.
For example, in the evolution basis, when $Q^2 < m_b^2$ the GPDs $H_{T24}$ and $H_{T35}$ are not evolved, but simply set equal to the singlet GPD $H^{\rm S}$.
When $Q^2 > m_b^2$, the GPD $H_{T24}$ is evolved on its own as an independent nonsinglet GPD.

\subsubsection{Distributions in kernels}

The evolution kernels are distributions rather than functions.
In general, they break down into three pieces:
\begin{align}
  \label{eqn:kern_break_down}
  K(x,y,\xi,Q^2)
  =
  K^{(R)}(x,y,\xi,Q^2)
  +
  \big[ K^{(P)}(x,y,\xi,Q^2) \big]_+
  +
  K^{(D)}(Q^2)\,
  \delta(y-x)
  \,,
\end{align}
where $K^{(R)}$ is a regular function, $\big[ K^{(P)}(x,y,\xi,Q^2) \big]_+$ is a ``plus'' distribution, and the function $K^{(D)}$ is associated with the Dirac $\delta$ distribution $\delta(y-x)$.
As distributions, the latter two are defined in terms of integrals along with test functions.
The $\delta$ distribution is defined as:
\begin{align}
  \int_{-1}^1 \d y \,
  f(y)\, \delta(y-x)
  =
  f(x)
  \,,
\end{align}
and the plus distribution is defined via:
\begin{align}
  \label{eqn:plus}
  \int_{-1}^1 \d y \,
  \big[g(x,y)\big]_+
  f(y)
  =
  \int_{-1}^1 \d y \,
  g(x,y)
  \big( f(y) - f(x) \big)
  +
  f(x)
  \int_{-1}^1 \d y \,
  \big( g(x,y) - g(y,x) \big)
  \,.
\end{align}
When applying this formula to the GPD evolution kernels in Appendix~\ref{sec:kernels}, it should be stressed that the step functions all appear inside the plus brackets, which is necessary to obtain finite results from these integrals.

\section{Kernel method}
\label{sec:pixel}

The kernel method is a specific type of finite element method \cite{Karniadakis:2005abc, Strang:2008anl}, in which a parton distribution $H(x)$ is represented by its interpolation from a set of discrete points $\{x_i, H(x_i)\}$.
This method has been used in several prior codes for the evolution of forward parton distribution functions, such as HOPPET~\cite{Salam:2008qg} and APFEL~\cite{Bertone:2013vaa}.

In the kernel formulation, each real variable is replaced by a collection
of discrete values.
For instance, the parton momentum fraction $x \in [-1,1]$ may be replaced by a grid of linearly spaced points
\begin{align}
  x_i
  =
  -1 + \frac{2(i-1)}{n_x-1}
\end{align}
for $i \in \{1,2,\ldots,n_x\}$.
We refer to such a discrete representation as a pixelation.
The linear spacing is not strictly necessary---and, in fact, geometric spacing may be more convenient in some cases---but makes illustration simpler.
Any one-variable function of $x$ is similarly represented by a collection of discrete values; a function $f(x)$ is represented in this formulation through the values:
\begin{align}
  f_i
  =
  f(x_i)
  \,,
\end{align}
where $f(x_i)$ is the evaluation of the original, continuous function at $x=x_i$.
In a problem where multiple variables are present, each variable is discretized, and a function of $n$ variables is represented as a rank-$n$ tensor.
For instance, the three-variable GPD becomes a rank-three tensor:
\begin{align}
  H(x,\xi,Q^2)
  \to
  H_{ijk}
  =
  H(x_i, \xi_j, Q^2_k)
  \,,
\end{align}
where $\xi_j$ and $Q^2_k$ are the discrete values of $\xi$ and $Q^2$ that are tabulated.
For a GPD that depends on additional variables (such as $t$ or a replica index), extra indices can be added and the rank of the tensor increased, and evolution can be parallelized over the added indices.

Derivatives and integrals of pixelated functions are pertinent, especially since the GPD evolution equation is an integro-differential equation.
The simplest manner to implement the derivative is to use finite differences between adjacent discrete values:
\begin{align}
  f'(x_i < x < x_{i+1})
  \approx
  \frac{f_{i+1} - f_{i}}{x_{i+1}-x_{i}}
  \,,
  \label{eqn:fprime}
\end{align}
and a similar implementation of integrals can be made using the trapezoid rule:
\begin{align}
  \int_{x_a}^{x_b} \d x \,
  f(x)
  \approx
  \frac{1}{2}
  \sum_{i=a}^{b-1}
  (f_{i+1} + f_i)( x_{i+1} - x_i )
  \,.
  \label{eqn:ftrapezoid}
\end{align}
The formulas in Eqs.~(\ref{eqn:fprime}) and (\ref{eqn:ftrapezoid}) both amount to assuming that $f(x)$ is given by a piecewise linear interpolation of its pixelation $\{(x_i,f_i)\}$.
This suggests the basic implementation can be improved by using a better interpolation, such as modified cubic Hermite splines or Lagrange interpolation---which motivates the methods developed in Sec.~\ref{sec:main}.
In fact, for the integral in particular, the pixelation can be interpolated to Gaussian weight points, allowing Gaussian quadrature to be used to estimate integrals.
This will be explained in more depth in Sec.~\ref{sec:main}.

A centrally important point is that pixelation turns convolution integrals into tensor contractions.
For instance, given an integral equation of the form:
\begin{align}
  \label{eqn:cff}
  \mathcal{H}(\xi,Q^2)
  =
  \int_{-1}^1 \d x\,
  C(x,\xi)\,
  H(x,\xi,Q^2)
  \,,
\end{align}
which is relevant to the relationship between GPDs and Compton form factors, the kernel formulation of this relation would be:
\begin{align}
  \label{eqn:cff:pixel}
  \mathcal{H}_{jk}
  \approx
  \sum_i
  C_{ij}\,
  H_{ijk}
  \,,
\end{align}
where $C_{ij} \neq C(x_i, \xi_j)$, but is a tensor constructed specifically to make Eq.~(\ref{eqn:cff:pixel}) (approximately) true.
That is, $C_{ij}$ in effect represents for the action of convolution with the function $C(x,\xi)$, or, in other words, stands in for $C \otimes$ rather than $C(x,\xi)$ itself.

We illustrate this using the trapezoidal rule as an integral estimator for ease of demonstration, deferring discussion of more sophisticated methods to Sec.~\ref{sec:main}.
The trapezoid rule applied to Eq.~(\ref{eqn:cff}) gives:
\begin{align}
  \mathcal{H}(\xi_j, Q^2_k)
  \approx
  \frac{1}{2}
  \sum_{i=1}^{n_x-1}
  \Big(
  C(x_i, \xi_j)\,
  H(x_i, \xi_j, Q^2_k)
  +
  C(x_{i+1}, \xi_j,
  H(x_{i+1}, \xi_j, Q^2_k)
  \Big)
  (x_{i+1} - x_i)
  \,.
\end{align}
The sum on the right-hand side can be rearranged to give:
\begin{align}
  \mathcal{H}(\xi_j, Q^2_k)
  \approx
  \frac{ C(x_1, \xi_j) (x_{2} - x_1) }{2}
  H_{1jk}
  +
  \sum_{i=2}^{n_x-1}
  \frac{ C(x_i, \xi_j) (x_{i+1} - x_{i-1}) }{2}
  H_{ijk}
  +
  \frac{ C(x_{n_x}, \xi_j) (x_{n_x} - x_{n_x-1}) }{2}
  H_{n_xjk}
  \,,
\end{align}
which has exactly the form of Eq.~(\ref{eqn:cff:pixel}) if
\begin{align}
  C_{ij}
  =
  \frac{1}{2}
  \left\{
    \begin{array}{lcl}
      C(x_1, \xi_j) (x_2 - x_1) &:& i=1 \\
      C(x_i, \xi_j) (x_{i+1} - x_{i-1}) &:& 1 < i < n_x \\
      C(x_{n_x}, \xi_j) (x_{n_x} - x_{n_x-1}) &:& i=n_x
    \end{array}
    \right.
  \,.
\end{align}
We stress that the use of the trapezoidal rule in this demonstration is simply for ease of explanation and that in practice more refined methods, as will be discussed below, are used.
However, we have demonstrated that by exploiting the linearity of integration, we are able to factorize the convolution into the GPD and the matrix that acts on it.
The convolution matrix, $C_{ij}$, is independent of the GPD $H_j$.
This property will hold for Wilson coefficients and evolution kernels alike---and allows the most numerically expensive parts of code to be executed without knowing the exact GPD, and for the results to be stored as a matrix in memory and reused.
This is a property that we exploit in the evolution algorithm.

\section{Finite element approaches to evolution equations}
\label{sec:main}

After decoupling the singlet and nonsinglet flavor combinations
and pixelizing the $x$ dependence of the GPDs, the evolution equations can be written:
\begin{align}
  \label{eqn:evo:NS}
  \frac{\d H_i^{\mathrm{NS}}(\xi,Q^2)}{\d \log Q^2}
  =
  \sum_{j=1}^{n_x}
  K_{ij}^{\mathrm{NS}}(\xi,Q^2)\,
  H_j^{\mathrm{NS}}(\xi,Q^2)
\end{align}
for the nonsinglet mixtures, where $H^{\rm NS}$ represents any of the combinations in Eqs.~(\ref{eqn:Hqminus}) or (\ref{eqn.nonsinglet}), and
\begin{align}
  \label{eqn:evo:S}
  \frac{\d}{\d \log Q^2}
  \left[
    \begin{array}{c}
      H_i^{\rm S}(\xi,Q^2)
      \\
      H_i^g(\xi,Q^2)
    \end{array}
    \right]
  =
  \sum_{j=1}^{n_x}
  \bigg[
    \begin{array}{cc}
      K_{ij}^{\rm SS}(\xi,Q^2)
      &
      K_{ij}^{{\rm S}g}(\xi,Q^2)
      \\
      K_{ij}^{g{\rm S}}(\xi,Q^2)
      &
      K_{ij}^{gg}(\xi,Q^2)
    \end{array}
    \bigg]
  \bigg[
    \begin{array}{c}
      H_j^{\rm S}(\xi,Q^2)
      \\
      H_j^g(\xi,Q^2)
    \end{array}
    \bigg]
\end{align}
for the coupled singlet and gluon GPDs.
In both equations, the kernel matrices $K_{ij}(\xi,Q^2)$ are components of the matrix necessary to make the discretized evolution equations faithful approximations of the continuum evolution equation---they are \emph{not} the kernels evaluated at $(x,y) = (x_i,y_j)$.

We will consider two specific strategies for constructing the kernel matrices, which we refer to as the Simple Method and the Refined Method.
Before discussing the specifics of each method, we will first give a generic overview of what these strategies have in common.

Both methods go beyond using a trapezoidal rule for evaluating the integral in Eq.~(\ref{eqn:evolution}), employing instead Gaussian quadrature rules.
The use of quadrature rules requires evaluating the integrand at the quadrature points---e.g., roots of the Legendre polynomial $P_n(x)$ in the case of $n$th order Gauss Legendre quadrature.
These roots don't necessarily lie on the grid $\{x_j\}$ at which we have defined our initial scale GPD, so that interpolation from our starting grid to the quadrature grid is required.
To illustrate the workings of this, we can consider the case where the evolution kernel is a regular function (without $\delta$ or plus prescription distributions).
The evolution equation can then be approximated in the form:
\begin{align}
  \label{eqn:evo:interp1}
  \frac{\d H^{A}(x_i,\xi,Q^2)}{\d \log Q^2}
  \approx
  \sum_B
  \sum_{j=1}^{n_x}
  \sum_{g=1}^{n_g}
  w_g
  K^{AB}(x_i, y_g, \xi,Q^2)\,
  L_{gj}(x_i,\xi)\,
  H^B(x_j,\xi,Q^2)
  \,,
\end{align}
where $\{y_g\}$ are the Gaussian evaluation points and $\{w_g\}$ are their associated weights, and the indices $AB$ refer to singlet and gluon ($Sg$), or are trivial for the nonsinglet case.
The tensor $L_{gj}(x_i,\xi)$ is a matrix that interpolates from the grid points $\{x_j\}$ to the Gaussian quadrature points $y_g$, which may in general depend on the target momentum fraction $x_i$ and $\xi$ (as it does in the Refined Method, discussed below).
This can be done provided the interpolation method is linear (in the sense that it distributes over addition)---as it is for modified cubic Hermite splines or Lagrange interpolation, for instance.

The approximate form in Eq.~(\ref{eqn:evo:interp1}) can be written in the form of either Eq.~(\ref{eqn:evo:NS}) or (\ref{eqn:evo:S}) through the identification:
\begin{align}
  K_{ij}^{AB}(\xi,Q^2)
  \equiv
  \sum_{g=1}^{n_g}
  w_g
  K^{AB}(x_i, y_g, \xi,Q^2)\,
  L_{gj}(x_i,\xi)
  \,.
\end{align}
The Simple Method calculates the effective kernel matrices $K_{ij}^{AB}(\xi,Q^2)$ through this method in particular---aided by the further simplification that it uses fixed Gauss-Legendre quadrature, so that the interpolation matrix $L_{gj}$ is independent of $x_i$ and $\xi$.

Another way of viewing the interpolation is as an expression of the initial scale GPD in terms of some basis functions:
\begin{align}
  H(x,\xi,Q^2)
  \approx
  \sum_n
  H_n(\xi,Q^2)\,
  p_n(x)\,,
\end{align}
where $\{p_n(x)\}$ is the set of basis functions.
This is possible precisely because the interpolation method is linear (in the sense of distribution over addition), which is a prerequisite for the interpolation matrix $L_{gj}$ to even be constructed.
In this case, the $\{p_n(x)\}$ are most naturally chosen to be interpolation basis functions.
The convolution in Eq.~(\ref{eqn:evo:interp1}) could instead be expressed as:
\begin{align}
  \label{eqn:evo:interp2}
  \frac{\d H^A(x_i,\xi,Q^2)}{\d \log Q^2}
  =
  \sum_{j=1}^{n_x}
  \sum_n
  \int_{-1}^1 \d y \,
  K^{AB}(x_i,y,\xi,Q^2)\,
  p_n(y)\,
  H^B_n(\xi,Q^2)
  \,.
\end{align}
The two approaches in Eqs.~(\ref{eqn:evo:interp1}) and (\ref{eqn:evo:interp2}) are equivalent.
Indeed, the construction of the interpolation matrix $L_{gj}$ requires one to choose a basis of interpolation: $L_{gj}$ in effect evaluates the $j$th basis function at the Gaussian quadrature point $y_g$, or rather, $L_{gj} = p_j(y_g)$.

Although the expressions in Eqs.~(\ref{eqn:evo:interp1}) and (\ref{eqn:evo:interp2}) are equivalent, their conceptual dissimilarities suggest different numerical strategies.
The Simple Method explored below is designed around (\ref{eqn:evo:interp1}), and aims to cast the evolution as a matrix equation in every step of the calculation.
A motivation behind this design is to construct a code consisting entirely of
tensor operations, so the code can be rendered entirely in PyTorch.
This leads to a choice of interpolation basis that is relatively simple,
so that its matrix form may be computed analytically.
It also uses fixed Gaussian quadrature,
since the target grid of the interpolation matrix has to be specified beforehand.

In contrast, the Refined Method is designed around Eq.~(\ref{eqn:evo:interp2}).
Since the interpolation matrix does not need to be explicitly constructed, more flexibility is afforded in how numerical integration is performed.
This allows more accurate integration to be achieved, in a manner that depends on the target $x_i$ and $\xi$ values.
This requires non-tensor operations to be used in building the kernel matrices, and also makes the matrix construction slower than in the Simple Method.
However, the final result will still be a single evolution matrix that can be applied to any initial scale GPD, and the increased computation time for building the matrix is of no consequence to the time it takes to perform evolution with it.

\subsection{Simple Method}
\label{sec:simple}

As a pedagogical step, we consider the Simple Method first.
This method is based on the Gaussian-Legendre quadrature rule, which allows for the exact calculation of the integral of a polynomial of degree $2n-1$ through the formula:
\begin{equation}
  \label{eqn:gauss_quad}
  \int_{-1}^{1} \d x\,
  f(x)
  \approx
  \sum_{g=1}^{n_g} w_g f(y_g)
  \,,
\end{equation}
where $y_g$ is the $g$th root of the Legendre polynomial $P_n(x)$ of degree $n$, and $w_g$ are the Gaussian weights:
\begin{equation}
  w_g = \frac{2}{(1-y_g^2)[P'_n(y_g)]^2}
  \,.
\end{equation}
Within the kernel methods, the $H$ functions are evaluated at discretized domain points $x_j$ defined through a linearly-spaced grid.
However, to use Eq.~(\ref{eqn:gauss_quad}) to estimate matrices, one needs to know the evaluation of the $H$ function at the Legendre polynomial roots $y_g$.
For instance, if we assume (for now) that the kernel $K(x,y,\xi,Q^2)$ is a regular function, then we would estimate:
\begin{equation}
  \label{eqn:int_gauss_quad}
  \int_{-1}^{1} \d y\,
  K(x,y,\xi,Q^2) H(y,\xi,Q^2)
  \approx
  \sum_{g=1}^{n_g} w_g\,
  K(x_i,y_g,\xi_l,Q^2_q)\,
  H(y_g,\xi_l,Q^2_q)
  \,,
\end{equation}
so that we must interpolate from the $\{x_j\}$ grid to $\{y_g\}$.
To this end, we use modified cubic Hermite splines (see Appendix~\ref{sec:cubic_spline}).
Since this interpolation method is linear, it can be implemented through a matrix multiplication as follows:
\begin{equation}
  H(y_g,\xi,Q^2)
  =
  \sum_{j=1}^{n_x}
  L(y_g,x_j)\,
  H(x_j,\xi,Q^2)
  \,,
\end{equation}
where $L(y_g,x_j)$ is an interpolation matrix (see Appendix~\ref{sec:cubic_spline}).
This implementation then allows us to rewrite the right-hand side
of Eq.~(\ref{eqn:int_gauss_quad}) in the following way:
\begin{align}
  \label{eqn:matrix_integr_reg}
  \sum_{g=1}^{n_g}
  w_g\,
  K(x_i,y_g,\xi_l,Q^2_q)\,
  H(y_g,\xi_l,Q^2_q)
  &\approx
  \sum_{j=1}^{n_x} C_{ijlq} H_{jlq}
  \,,
\end{align}
where
\begin{equation}
    C_{ijlq}
    =
    \sum_{g=1}^{n_g}w_g\, K_{iglq}\, L_{gj}
    \,,
\end{equation}
and is fully independent of the function $H$.
In this particular case, when the kernel is a regular function, this leads to a  discretized form of the integro-differential equation:
\begin{equation}
  \frac{\d H(x_i,\xi_l,Q^2_q)}{\d \log Q^2}
  =
  \sum_{j=1}^{n_x} C_{ijlq}\, H_{jlq}
  \,,
\end{equation}
where the $C$ matrices can be used to obtain the derivative of the GPDs in a matrix form, and will be used within the Runge-Kutta algorithm to solve the differential equation.

In general, however---as stated in Eq.~(\ref{eqn:kern_break_down})---the kernels
that appear in the evolution equations are in general distributions, and in particular can be broken into three pieces.
One of these pieces is a regular function, which has been discussed above.
The result can be written:
\begin{equation}
  \int_{-1}^{1} \d y\,
  K^{(R)}(x,y,\xi,Q^2)\,
  H(y,\xi,Q^2)
  \approx
  \sum_{j=1}^{n_x}
  K^{(R)}_{ijlq}\,
  H_{jlq}
  \,,
\end{equation}
where
\begin{align}
  K^{(R)}_{ijlq}
  =
  \sum_{g=1}^{n_g}
  w_g\,
  K(x_i, y_g, \xi_l, Q^2_q)\,
  L_{gj}
  \,.
\end{align}
For the $\delta$ function part of the kernel $K^{(D)}$, we have:
\begin{equation}
  \int_{-1}^{1} \mathrm{d} y\,
  K^{(D)}(Q^2)\, \delta(y-x)\, H(y,\xi,Q^2)
  =
  \sum_{j=1}^{n_x}
  K^{(D)}(Q^2_q)\,
  \delta_{ij}\, H_{jlq}
  \,,
\end{equation}
where the $\delta$ distribution has in effect been transformed into a Kronecker delta, $\delta_{ij}$.
The last type of kernel that appears inside the evolution equations is $K^{(P)}$, which contains plus prescription distributions.
As shown in Eq.~(\ref{eqn:plus}), an integral with a plus distribution can be split into two pieces:
\begin{eqnarray}
  \int_{-1}^1 \d y \,
  \big[ K^{(P)}(x,y,\xi,Q^2) \big]_+\, H(y,\xi,Q^2)
  &=&
  \int_{-1}^1 \d y \,
  K^{(P)}(x,y,\xi,Q^2) \Big( H(y,\xi,Q^2) - H(x,\xi,Q^2) \Big)
  \nonumber\\
  &+&
  H(x,\xi,Q^2)
  \int_{-1}^1 \d y \,
  \Big( K^{(P)}(x,y,\xi,Q^2) - K^{(P)}(y,x,\xi,Q^2) \Big)
  \,.
\label{eqn:KPH}
\end{eqnarray}
The first integral on the right-hand side of Eq.~(\ref{eqn:KPH}) can be estimated using Gaussian quadrature and modified cubic Hermite splines, just like the regular integral:
\begin{align}
  \int_{-1}^1 \d y \,  K^{(P)}(x,y,\xi,Q^2)\, 
  \Big( H(y,\xi,Q^2) - H(x,\xi,Q^2) \Big)
  &\approx
  \sum_{j=1}^{n_x} K^{(P1)}_{ijlq}\, H_{jlq}
  \,.
\end{align}
where
\begin{align}
  K^{(P1)}_{ijlq}
  =
  \sum_{g=1}^{n_g}
  w_g\,
  K^{(P)}(x_i,y_g,\xi_l,Q^2_q)
  \Big( L(y_g,x_j) - \delta_{ij} \Big)
  \,.
\end{align}
The second integral on the right-hand side of Eq.~(\ref{eqn:KPH}) can be evaluated analytically, with the results given in Eqs.~(\ref{eqn:qq:cst}) and (\ref{eqn:gg:cst}) of Appendix~\ref{sec:kernels}.
For fixed values of $x$ and $\xi$, this results in a constant factor multiplying the GPD:
\begin{equation}
  K^{(P2)}_{il}
  =
  K^{(P2)}(x_i,\xi_l)
  =
  \int_{-1}^1 \d y \,
  \Big( K^{(P)}(x,y,\xi,Q^2) - K^{(P)}(y,x,\xi,Q^2) \Big).
\end{equation}
We can rewrite the full integral for the plus prescription part as:
\begin{equation}
    \int_{-1}^1 \d y \,
    \big[ K^{(P)}(x,y,\xi,Q^2) \big]_+\, H(y,\xi,Q^2)
    \approx
    \sum_{j=1}^{n_x} K^{(P)}_{ijlq}\, H_{jlq}
    \,,
\end{equation}
where the matrix form of the kernel is given by
a sum of the two parts we just obtained:
\begin{equation}
  K^{(P)}_{ijlq}
  =
  \sum_{g=1}^{n_g}
  w_g\,
  K^{(P1)}_{iglq}
  \Big( L_{gj} - \delta_{ij} \Big)
  +
  \delta_{ij} K^{(P2)}_{il}
  \,.
\end{equation}

In summary, the Simple Method discretizes the evolution kernels by breaking down the kernel into the three distribution types in Eq.~(\ref{eqn:kern_break_down}), and estimates convolution integrals as matrix multiplication equations:
\begin{subequations}
\begin{eqnarray}
  \int_{-1}^1 \d y \,
  \big[ K^{(P)}(x,y,\xi,Q^2) \big]_+\, H(y,\xi,Q^2)
  &\approx&
  \sum_{j=1}^{n_x} K^{(P)}_{ijlq}\, H_{jlq}\, ,
  \\
  \int_{-1}^{1} \d y\,
  K^{(R)}(x,y,\xi,Q^2)\, H(y,\xi,Q^2)
  &=&
  \sum_{j=1}^{n_x} K^{(R)}_{ijlq}\, H_{jlq}\, ,
  \\
  \int_{-1}^{1} \d y\,
  K^{(D)}(Q^2)\, \delta(y-x)\, H(y,\xi,Q^2)
  &=&
  \sum_{j=1}^{n_x} K^{(D)}_q  \delta_{ij}\, H_{jlq}
  \,.
\end{eqnarray}
\label{eq:SimpleMethod}
\end{subequations}

\vspace*{-0.5cm}

\subsection{Refined Method}
\label{sec:interpixel}

Going beyond the Simple Method, we next propose a more refined method for estimating kernel matrix elements.
We call this the interpixel method, which stands for ``interpolation pixel.''
The interpixels are effectively interpolation basis functions, and provide a convenient way to calculate elements of the kernel matrix $K_{ij}(\xi,Q^2)$.
In effect, this Refined Method is closely follows the method used by the existing PDF evolution codes HOPPET~\cite{Salam:2008qg} and APFEL~\cite{Bertone:2013vaa}, as well as the recent finite element evolution code used by PARTONS~\cite{Bertone:2023jeh}.

\vspace*{-0.5cm}

\subsubsection{The interpixel method}

A crucial point to stress is that the evolution equations (\ref{eqn:evolution}) are linear in the GPDs.
Thus, the evolution equations are in effect solved once they have been solved for a set of basis functions.
Since a discretization $\{(x_i, H_i(\xi,Q^2)\}$ of the GPDs can be approximated in the continuum by an interpolation, we can utilize an interpolation basis.
For instance, using order $N_L$ Lagrange interpolation, the interpolating polynomial of a discrete set of data points
    $\big\{(x_i, y_i) | 0 \leq i \leq N_L\big\}$
can be written as a sum of basis functions:
\begin{align}
  L\big[
    \{(x_i, y_i)\}
    \big](x)
  =
  \sum_{i=0}^{N_L}
  l_i(x)\,
  H_i(\xi,Q^2)
  \,,
\end{align}
where the basis functions can be written:
\begin{align}
  l_i(x)
  =
  \prod_{\substack{0 \leq j \leq N_L\\j\neq i}}
  \frac{x - x_j}{x_i - x_j}
  \,.
\end{align}
\begin{figure}[t]
  \includegraphics[width=0.49\textwidth]{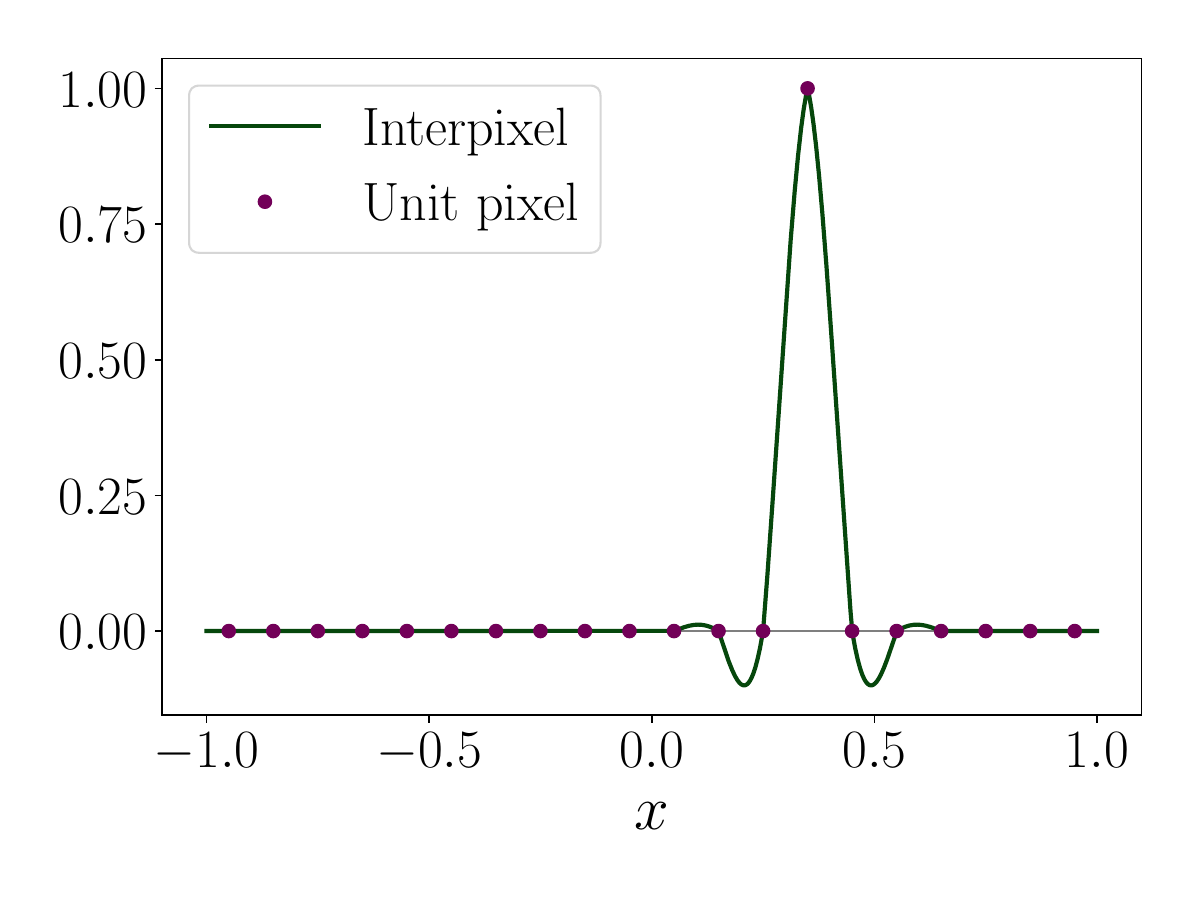}
  \caption{
    A demonstration of what an interpixel function might look like.
    The discrete $x$ values $x_i = (2i-1)/N - 1$ with $N=20$ were used, corresponding to the midpoints of $20$ equally wide intervals spanning the domain $[-1,1]$. The interpixel is constructed by pasting pieces of
    fifth-order Lagrange basis functions together (see text).
    This interpixel is equal to a piecewise polynomial interpolation of a function that is 1 at $x_i = 0.35$ and 0 at all other $\{x_i\}$.
  }
  \label{fig:interpixel}
\end{figure}
Interpolating the entire GPD using a single Lagrange polynomial of order $N_L = N_x-1$ would be problematic, since Lagrange interpolation is subject to the Runge phenomenon.
Instead, we use \emph{piecewise} Lagrange interpolation to represent the continuum limit GPD.
Using $N_L = 5$ as an example, for a given $x$ value we can locate three $x_i$ points to the left and to the right of $x$ (and if there are fewer than three points to either side, we use all the points on that side and six minus that number of points on the other), and use those six points to build the basis functions $l_i(x)$ for this $x$ value.
The $i$th basis function for piecewise polynomial interpolation, referred to as an interpixel, is thus constructed by pasting Lagrange basis functions constructed from different subsets of six $x_i$ values together.
The $i$th interpixel is equal to the piecewise Lagrange interpolation of a data set $\{(x_j, y_j) | 1 \leq j \neq N_x \}$ with $y_j = \delta_{ij}$.
Such an interpixel is illustrated in Fig.~\ref{fig:interpixel} for the case of $N_L=5$ (fifth-order piecewise Lagrange interpolation).

To distinguish from direct Lagrange interpolation, we use $P[\{(x_i,y_i)\}](x)$ to represent piecewise Lagrange interpolation.
We also introduce several other notational shorthands.
Let $\bm{H}(\xi,Q^2)$ represent the discrete collection $\{x_i, H_i(\xi,Q^2)\}$ of GPD points, written as a column matrix, so that $P[\bm{H}(\xi,Q^2)](x)$ represents its piecewise polynomial interpolation.
Additionally, let $\bm{e}_i$ represent a discretized GPD whose $i$th component is 1 and all other components zero: $(\bm{e}_i)_j = \delta_{ij}$.
Thus, $P[\bm{e}_i](x)$ is the $i$th interpixel.

\begin{figure}[t]
  \includegraphics[width=\textwidth]{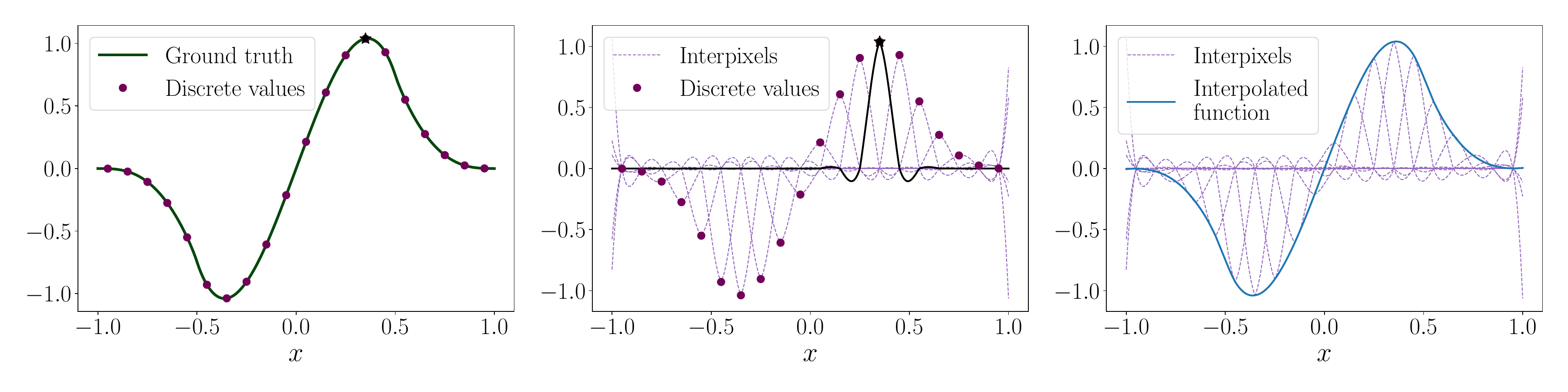}
  \caption{
    Illustration of the workings of the interpixel method.
    A model GPD~\cite{Kroll:2012sm} is used as a proxy for the ``ground truth'' (solid green curve, left panel), which is also represented by a collection of discrete values (purple dots, left and middle panels).
    Each individual discrete value is translated into an interpolation basis function---an interpixel (dashed purple curves, middle and right panels)---through piecewise polynomial interpolation.
    In the middle panel, a single interpixel is shown as a solid black curve for better visibility, and the value it interpolates is shown as a black star.
    The sum of the interpixels gives the piecewise polynomial interpolation of the original discrete values (solid blue curve, right panel), which is a reasonably good approximation of the ground truth.
  }
  \label{fig:interpixeldemo}
\end{figure}

Piecewise polynomial interpolation has the following useful properties:
\begin{enumerate}
  \item
    Linearity:
    \begin{align}
      \label{eqn:linear}
      P\big[
        \bm{H}_1(\xi,Q^2)
        +
        \bm{H}_2(\xi,Q^2)
        \big](x)
      =
      P\big[ \bm{H}_1(\xi,Q^2) \big](x)
      +
      P\big[ \bm{H}_2(\xi,Q^2) \big](x)
      \,.
    \end{align}
  \item
    Projection property:
    \begin{align}
      \label{eqn:projection}
      P\big[ P[\bm{H}](x) \big](x)
      =
      P[\bm{H}](x)
      \,.
    \end{align}
\end{enumerate}
Since $\bm{H}$ is a finite column matrix, the $\{\bm{e}_i\}$ furnish a basis for the possible GPD discretizations (assuming a fixed $\{x_i\}$ grid):
\begin{align}
  \bm{H}(\xi,Q^2)
  =
  \sum_{i=1}^{n_x}
  H_i(\xi,Q^2)\,
  \bm{e}_i
  \,.
\end{align}
The linearity property of polynomial interpolation further entails that:
\begin{align}
  \label{eqn:interpixel}
  P\big[
    \bm{H}(\xi,Q^2)
    \big](x)
  =
  \sum_{i=1}^N
  H(x_i, \xi, Q^2)\,
  P[\bm{e}_i](x)
  \,,
\end{align}
thus confirming that the interpixels are a basis for the interpolated GPDs.
A numerical demonstration of how interpixels can be used as a basis for interpolated GPDs is shown in Fig.~\ref{fig:interpixeldemo}.

\subsubsection{Kernel matrix in the interpixel method}

Proceeding now to obtain the kernel matrix in the interpixel method, we make use of the linearity of the GPD evolution equation (\ref{eqn:evolution}).
Placing Eq.~(\ref{eqn:interpixel}) into the evolution equation gives:
\begin{align}
  \sum_i
  P[\bm{e}_i](x)
  \frac{\d H(x_i,\xi,Q^2)}{\d \log Q^2}
  =
  \sum_i
  H(x_i,\xi,Q^2)
  \int_{-1}^1 \d y \,
  K(x,y,\xi,Q^2)\,
  P[\bm{e}_i](y)
  \,,
  \label{eqn:evolneqn}
\end{align}
Taking the polynomial interpolation of both sides of Eq.~(\ref{eqn:evolneqn})---and utilizing the projection property (\ref{eqn:projection}) of polynomial interpolation---we obtain:
\begin{align}
  \sum_i
  P[\bm{e}_i](x)
  \frac{\d H(x_i,\xi,Q^2)}{\d \log Q^2}
  =
  \sum_j
  H(x_j,\xi,Q^2)
  P\left[
    \int_{-1}^1 \d y \,
    K(x,y,\xi,Q^2)\,
    P[\bm{e}_j](y)
    \right](x)
  \,,
  \label{eqn:evolneqnPe}
\end{align}
where the right-hand side has been reindexed using $j$ instead of $i$.
To find the polynomial interpolation of the integral on the right-hand side of Eq.~(\ref{eqn:evolneqnPe}), we evaluate the integral at the $x$ points $\{ x_1, x_2, \ldots, x_N \}$.
Linearity of polynomial interpolation tells us that:
\begin{align}
  P\left[
    \int_{-1}^1 \d y \,
    K(x,y,\xi,Q^2)
    P[\bm{e}_i](y)
    \right](x)
  =
  \sum_i
  P[\bm{e}_i](x)
  \int_{-1}^1 \d y \,
  K(x_i,y,\xi,Q^2)
  P[\bm{e}_j](y)
  \,.
\end{align}
Therefore the evolution equation can be written as:
\begin{align}
  \sum_i
  P[\bm{e}_i](x)
  \frac{\d H(x_i,\xi,Q^2)}{\d \log Q^2}
  =
  \sum_i
  P[\bm{e}_i](x)
  \sum_j
  \left(
  \int_{-1}^1 \d y \,
  K(x_i,y,\xi,Q^2)
  P[\bm{e}_j](y)
  \right)
  H(x_j,\xi,Q^2)
  \,.
\end{align}
Within the interpixel method, the kernel matrix elements then become:
\begin{align}
  \label{eqn:kernel:interpixel}
  K_{ij}(\xi,Q^2)
  =
  \int_{-1}^1 \d y \,
  K(x_i,y,\xi,Q^2)
  P[\bm{e}_j](y)
  \,.
\end{align}
Crucially, this integral has not yet been discretized, and we are at liberty to employ any numerical method at our disposal to its evaluation, without constraints from the $x_i$ grid.

With this liberty, we have developed a code (implemented primarily in Fortran, but with a user interface in Python) based around using Eq.~(\ref{eqn:kernel:interpixel}) to compute elements of the kernel matrix.
To numerically perform this integral, we first decompose $K(x,y,\xi,Q^2)$ according to Eq.~(\ref{eqn:kern_break_down}).
For the plus prescription piece, Eq.~(\ref{eqn:plus}) is used to rewrite integrals over the plus-type piece, giving:
\begin{eqnarray}
  \label{eqn:kernel:breakdown}
  K_{ij}(\xi,Q^2)
  &=&
  \int_{-1}^1 \d y \,
  K^{(R)}(x_i,y,\xi,Q^2)
  P[\bm{e}_j](y)
  +
  \int_{-1}^1 \d y \,
  K^{(P)}(x_i,y,\xi,Q^2)
  \Big(
  P[\bm{e}_j](y)
  -
  P[\bm{e}_j](x_i)
  \Big)
  \nonumber\\
  &+&
  \left\{
    \int_{-1}^1 \d y \,
    \Big(
    K^{(P)}(x_i,y,\xi,Q^2)
    -
    K^{(P)}(y,x_i,\xi,Q^2)
    \Big)
    +
    K^{(D)}(Q^2)
    \right\}
  P[\bm{e}_j](x_i)
  \,.
\end{eqnarray}
The integral on the second line of Eq.~(\ref{eqn:kernel:breakdown}) is performed analytically, with results given in in Eqs.~(\ref{eqn:qq:cst}) and (\ref{eqn:gg:cst}) of Appendix~\ref{sec:kernels}.
The remaining integrals, on the first line of (\ref{eqn:kernel:breakdown}), are performed numerically.
The principal difficulties in these integrals are potential jump discontinuities in the integrand at $y = \pm \xi$ and $y = \pm |x|$, which can be mitigated by breaking the integration domain into six regions regions with the following seven endpoints:
  \begin{align}
    \nonumber
    a_0
    &=
    -1,
    \\
    \nonumber
    a_1
    &=
    \mathrm{min}(-\xi, -|x|),
    \\
    \nonumber
    a_2
    &=
    \mathrm{max}(-\xi, -|x|),
    \\
    a_3
    &=
    0,
    \\
    \nonumber
    a_4
    &=
    \mathrm{min}(\xi, |x|),
    \\
    \nonumber
    a_5
    &=
    \mathrm{max}(\xi, |x|),
    \\
    \nonumber
    a_6
    &=
    1
    \,,
  \end{align}
so that:
\begin{align}
  \int_{-1}^1 \d y \,
  \to
  \sum_{n=1}^6
  \int_{a_{n-1}}^{a_n} \d y \,
  \,.
\end{align}
Over each of these six regions, the 15-point Gauss-Kronrod rule is used.
This is found to have outstanding accuracy, with negligible improvement from further refinements.
The same numerical integration method is also used to estimate the regular part of the kernel.
As in the Simple Method, the $\delta$ distribution piece is trivial.

\subsection{Solving the evolution equations}

As ordinary differential matrix equations, the evolution equations (\ref{eqn:evo:NS}) and (\ref{eqn:evo:S}) can be solved numerically using standard methods, such as fourth-order Runge-Kutta (RK4) method.
Letting $Q^2_{\text{ini}}$ be an initial renormalization scale, $Q^2_{\text{fin}}$ the final scale, and $Q^2_{\text{mid}} = \sqrt{Q_{\text{fin}}^2 Q_{\text{ini}}^2}$ be their geometric midpoint, the RK4 method can be used to solve for $\bm{H}(\xi, Q_{\text{fin}}^2)$ in terms of $\bm{H}(\xi, Q^2_{\text{ini}})$ via:
\begin{align}
  \label{eqn:solution}
  H_i(\xi, Q^2_{\text{fin}})
  =
  \sum_{j=1}^{n_x}
  M_{ij}(\xi, Q^2_{\text{ini}} \to Q^2_{\text{fin}})\,
  H_j(\xi, Q^2_{\text{ini}}),
\end{align}
where
\begin{subequations}
  \label{eqn:evomatrix}
  \begin{align}
    M_{ij}(\xi, Q^2_{\text{ini}} \rightarrow Q^2_{\text{fin}})
    =
    \delta_{ij}
    +
    \frac{1}{6}
    \log\frac{Q^2_{\text{fin}}}{Q^2_{\text{ini}}}
    \Big(
    M_{ij}^{(1)}(\xi)
    +
    2
    M_{ij}^{(2)}(\xi)
    +
    2
    M_{ij}^{(3)}(\xi)
    +
    M_{ij}^{(4)}(\xi)
    \Big),
  \end{align}
and where
  \begin{align}
    M_{ij}^{(1)}(\xi)
    &=
    K_{ij}(\xi,Q_{\text{ini}}^2),
    \\
    M_{ij}^{(2)}(\xi)
    &=
    \sum_{k=1}^{n_x}
    K_{ik}(\xi,Q^2_{\text{mid}})
    \left(
    \delta_{kj}
    +
    \frac{1}{2} \log\frac{Q_{\text{fin}}^2}{Q_{\text{ini}}^2}
    M_{kj}^{(1)}(\xi)
    \right),
    \\
    M_{ij}^{(3)}(\xi)
    &=
    \sum_{k=1}^{n_x}
    K_{ik}(\xi,Q^2_{\text{mid}})
    \left(
    \delta_{kj}
    +
    \frac{1}{2} \log\frac{Q_{\text{fin}}^2}{Q_{\text{ini}}^2}
    M_{kj}^{(2)}(\xi)
    \right),
    \\
    M_{ij}^{(4)}(\xi)
    &=
    \sum_{k=1}^{n_x}
    K_{ik}(\xi,Q_{\text{fin}}^2)
    \left(
    \delta_{kj}
    +
    \log\frac{Q_{\text{fin}}^2}{Q_{\text{ini}}^2}
    M_{kj}^{(3)}(\xi)
    \right)
    \,.
  \end{align}
\end{subequations}
The elements of the kernel matrix needed here can be found using the methods in Sec.~\ref{sec:simple} (Simple Method) or Sec.~\ref{sec:interpixel} (Refined Method) above.

Summarizing, Eq.~(\ref{eqn:solution}) tells us that GPD evolution from an initial scale $Q^2_{\text{ini}}$ to a target scale $Q^2_{\text{fin}}$ can be carried out as mere matrix multiplication, and Eqs.~(\ref{eqn:evomatrix}) tell us how to compute the evolution matrix.
Most significantly, Eqs.~(\ref{eqn:evomatrix}) suggest that the evolution matrix ${\bm{M}(\xi, Q^2_{\text{ini}} \to Q^2_{\text{fin}})}$ is \emph{independent of the initial GPD}.
The most expensive computations of the GPD evolution can thus be carried out in a way that is agnostic about the form of the GPD, and the results stored in memory as a collection of matrices $\bm{M}(\xi, Q^2_{\text{ini}} \to Q^2_{\text{fin}})$ that can be reused to evolve any initial scale GPD.
This can significantly reduce the computational cost of fitting a model scale GPD to data across a large range of $Q^2$, by eliminating the need to redo costly integration calls with each trial set of parameters.

Further accuracy refinements can be achieved by iterating the RK4 method.
In practice, we consider a grid of $Q^2$ points $Q^2_i \in \{Q^2_0,\, Q^2_1, \ldots, Q^2_N\}$.
The matrix for evolving the GPD from $Q^2_{i-1}$ to $Q^2_i$ is obtained by a single application of the RK4 algorithm.
To evolve between nonadjacent $Q^2$ values, one need only multiply the respective evolution matrices.
Evolving from $Q^2_i$ to $Q^2_{i+2}$, for instance, can be represented as:
\begin{align}
  M_{ij}(\xi, Q^2_i \to Q^2_{i+2})
  =
  \sum_{k=1}^{n_x}
  M_{ik}(\xi, Q^2_{i+1} \to Q^2_{i+2})\,
  M_{kj}(\xi, Q^2_i \to Q^2_{i+1})
  \,,
\end{align}
which illustrates the utility of performing the evolution through matrix multiplication. \\

\section{Numerical benchmarks of evolution codes}
\label{sec:benchmarks}

In this section, we consider several accuracy benchmarks for the computer codes
we have written using the two methods outlined above.

\subsection{Kernel benchmarks}

We first consider numerical accuracy benchmarks for the kernel matrices themselves.
To do this, we must consider how faithfully the discretized evolution formulas (\ref{eqn:evo:NS}) and (\ref{eqn:evo:S}) reproduce the original, continuum evolution equation (\ref{eqn:evolution}).
This requires us to use a specific model GPD must in the benchmark, for which we choose the GK model~\cite{Kroll:2012sm} as a reasonably accurate representation of what a realistic GPD might look like.

The benchmarks are performed by comparing the right-hand side of Eq.~(\ref{eqn:evolution}) to that of (\ref{eqn:evo:NS}) in the nonsinglet sector
or (\ref{eqn:evo:S}) in the singlet sector.
The former, continuum case is used as the ``ground truth.''
In practice, the ``ground truth'' must itself be computed numerically, but by using adaptive Gaussian quadrature, the integral can be determined up to machine precision---which, compared to the numerical error in the kernel method, can be considered arbitrarily accurate for all practical purposes.
The two kernel methods---the Simple Method of Sec.~\ref{sec:simple} and the Refined Method of Sec.~\ref{sec:interpixel}---are then used to compute the kernels appearing in Eqs.~(\ref{eqn:evo:NS}) and (\ref{eqn:evo:S}).
The shifts calculated from the continuum and from the kernel method are compared, with a relative error calculated as:
\begin{align}
  \text{error}_i
  =
  \frac{
    \left|
    \sum_{j=1}^{n_x}
    K_{ij}(\xi,Q^2)\,
    H_j(\xi,Q^2)
    -
    \int_{-1}^1 \d y\, K(x_i, y, \xi,Q^2)\,
    H(y,\xi,Q^2)
    \right|
  }{
    \left|
    \int_{-1}^1 \d y\, K(x_i, y, \xi,Q^2)\,
    H(y,\xi,Q^2)
    \right|
  }
  \,.
\end{align}

The benchmarks for the Simple Method use $n_x=101$, while for the Refined Method we use $n_x=100$; the grids in each method are designed differently, with the Simple Method using $n_x=101$ linearly spaced points from $-1$ to $1$, and the Refined Method using the $100$ midpoints of the intervals delineated by those $101$ points.
For the Simple Method, we find an extremely large number of Gaussian evaluation points are needed to get an accurate result (see Fig.~\ref{fig:bench:gauss} in Appendix~\ref{sec:addbench}, which has a variety of additional benchmarks besides the main ones considered here).
The Refined Method has a much smaller number of evaluation points: $90 = 6 \times 15$, since the domain $-1 \leq y \leq 1$ is broken into six regions, each of which is integrated using the 15-point Gauss-Kronrod rule.

\begin{figure}
  \includegraphics[width=0.33\textwidth]{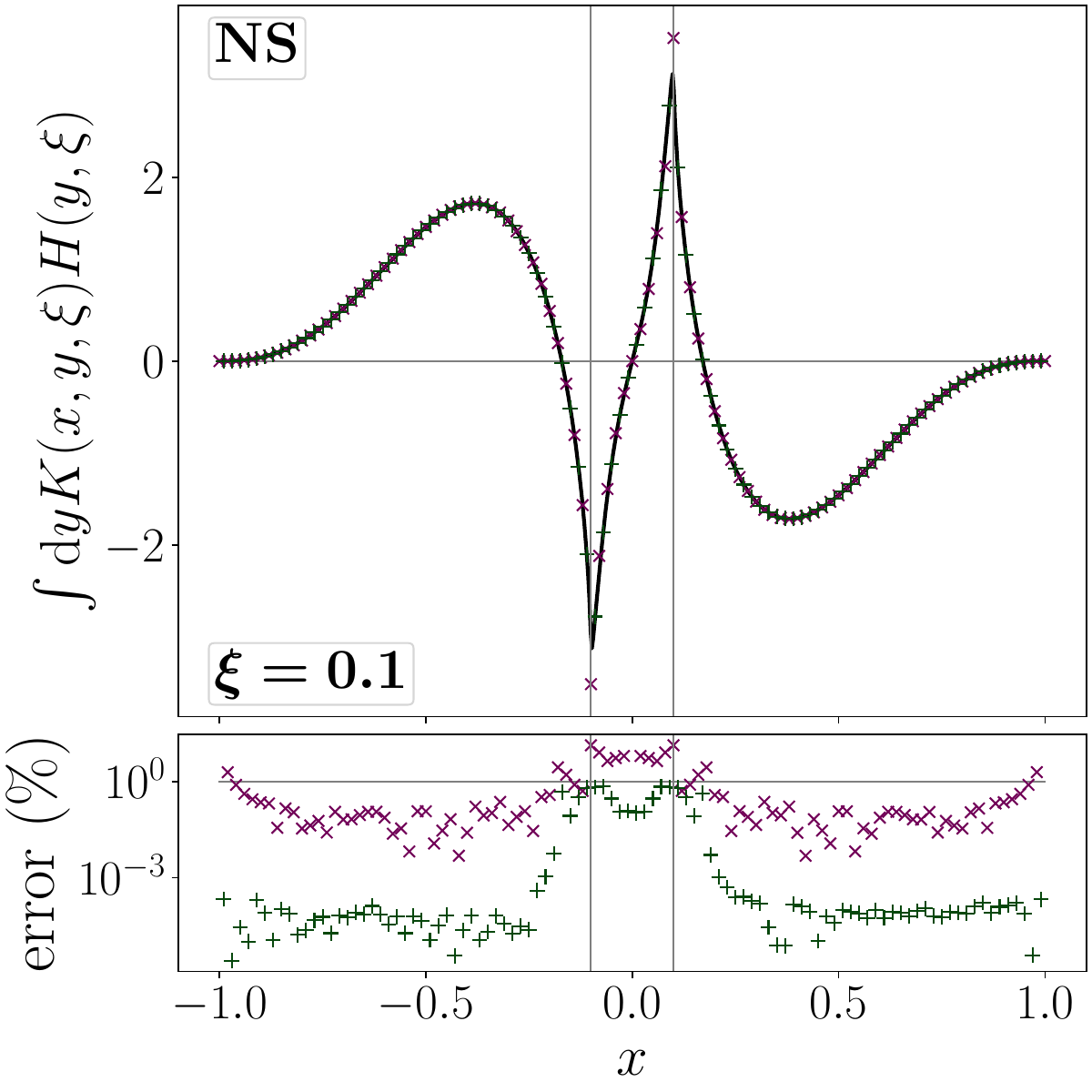}
  \includegraphics[width=0.33\textwidth]{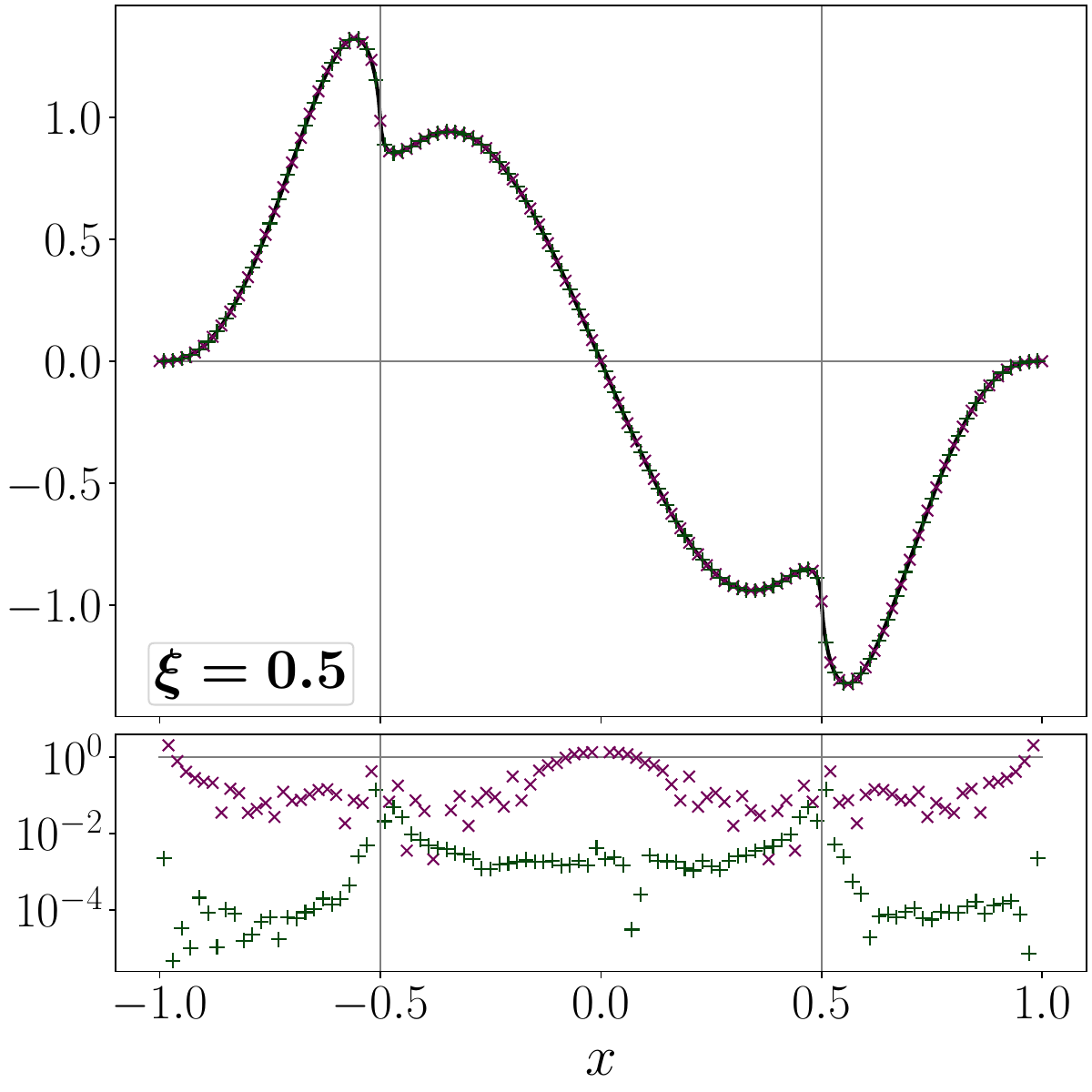}
  \includegraphics[width=0.33\textwidth]{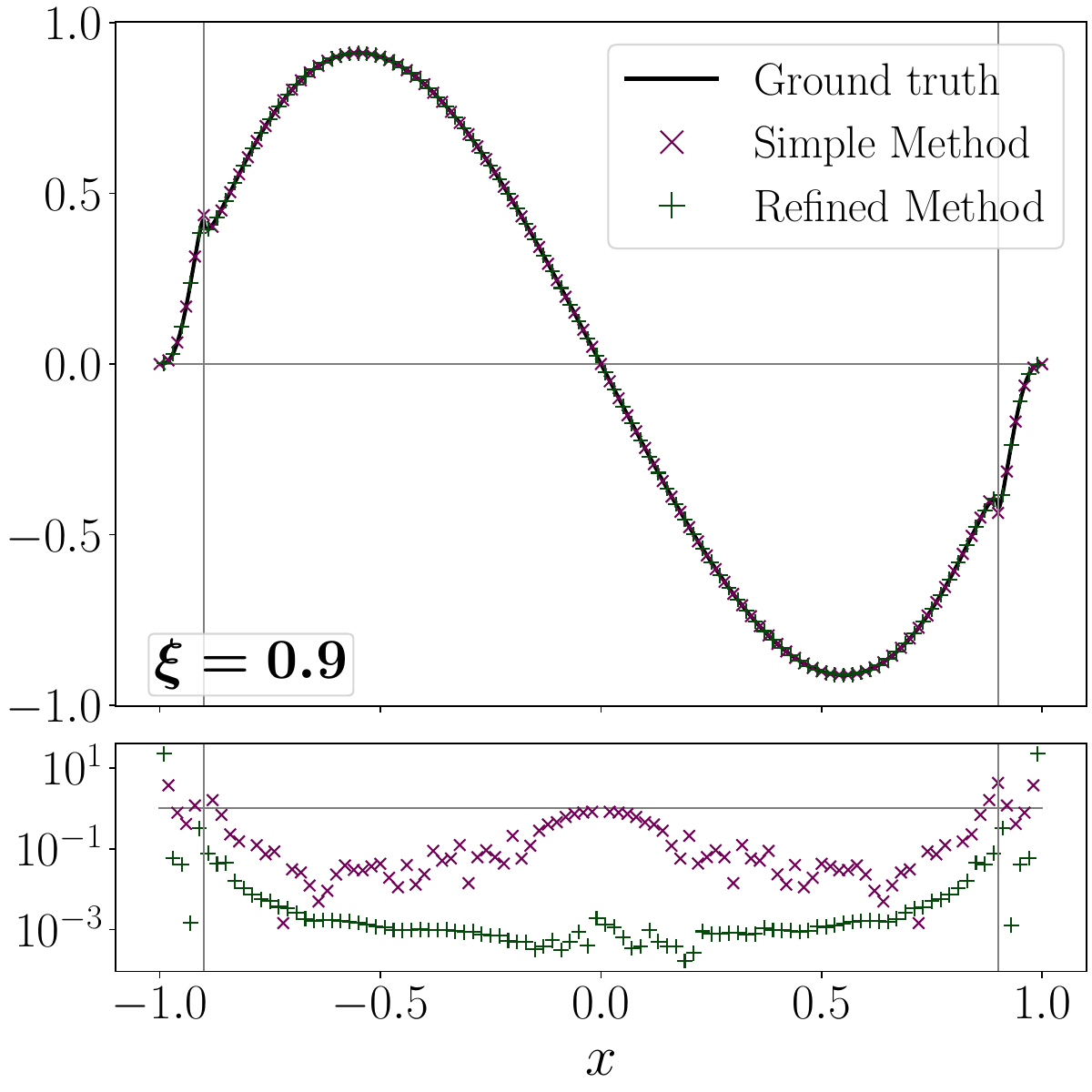}
  \caption{
    Accuracy benchmark for the right-hand side of the evolution equation
    (\ref{eqn:evolution}), at LO in the nonsinglet sector.
    This figure compares both the Simple Method (purple $\times$'s) and the Refined Method (green $+$'s) to a ``ground truth,'' the latter determined with the GK model~\cite{Kroll:2012sm} using adaptive quadrature.
    For the Simple Method, we use $n_g = 5000$ Gaussian weight points and $N_x = 101$; for the Refined Method, we use $N_x = 100$.
    The left panel shows $\xi = 0.1$ to represent the lower limit of the intended domain of applicability for the methods; the middle panel shows $\xi=0.5$ to represent a central $\xi$ value; and the right panel uses $\xi=0.9$ to represent a large $\xi$ value.
  }
  \label{fig:bench:xi}
\end{figure}

In Fig.~\ref{fig:bench:xi}, we show an accuracy benchmark for the nonsinglet kernel at three skewness values.
The left panel uses $\xi=0.1$, which is the lowest $\xi$ value for which results of the Methods should be trusted; it is also the lower limit of the intended domain to which our codes will be applied for phenomenology~\cite{Yu:2024abc}.
The values $\xi=0.5$ and $\xi=0.9$ are also shown in the middle and right panels, to illustrate moderate and large skewnesses.
In all three panels, the Refined Method performs better than the Simple Method in terms of accuracy.

Several trends in Fig.~\ref{fig:bench:xi} can be observed.
First, the codes are more accurate at larger $\xi$ values.
The error in the ERBL region ($|x| < \xi$) increases as $\xi$ decreases, becoming $\sim 1\%$ in the Refined Method, and $ > 1\%$ in the Simple Method.
The codes thus become untrustworthy below $\xi=0.1$.
In Appendix~\ref{sec:addbench}, we show demonstrations of the numerical analysis codes breaking down at smaller skewnesses.

Second, the Refined Method has spikes in its error at $x\sim\pm\xi$.
This occurs because $x=\pm\xi$ are not present in the $x$ grid that the Refined Method uses, and because the GPDs on points off the grid are estimated using polynomial interpolation---which means that the GPD at $x = \pm \xi$ is estimated using a smooth function.
However, the GPD is not actually differentiable at this point~\cite{Collins:1998be}, which means irreducible numerical error will occur at these points.
Since this numerical error is $\sim 1\%$, however, we have chosen to consider this error tolerable and to defer further refinements to the Refined Method until the future, when higher precision becomes necessary.
Similar trends are present in the singlet kernels, which for completeness we show in Appendix~\ref{sec:addbench}.

\subsection{Polynomiality tests}

It is important that the evolution code preserve the analytic properties of GPDs---at least as well as a discretization in $x$ space itself can preserve them.
A crucial property to preserve is polynomiality~\cite{Ji:1998pc}, in which Mellin moments of the GPD are given by polynomials in $\xi$:
\begin{align}
  \int_{-1}^{+1} \d x \,
  x^{s-1} H^q(x,\xi,t,Q^2)
  &=
  \sum_{\substack{n=0\\2|n}}^{s}\,
  \xi^n
  A^q_{s,n}(t,Q^2),
  \\
  \int_{-1}^{+1} \d x \,
  x^{s-1} H^g(x,\xi,t,Q^2)
  &=
  \sum_{\substack{n=0\\2|n}}^{s+1}\,
  \xi^n
  A^g_{s,n}(t,Q^2)
  \,,
\end{align}
where $s$ is an integer, and we have restored the $t$ dependence in these expressions to emphasize their general form.
Polynomiality is a consequence of Lorentz covariance~\cite{Ji:1998pc},
and the polynomial is even in $\xi$ due to time reversal symmetry~\footnote{
  A small number of GPDs, such as $H_4(x,\xi,t,Q^2)$ in spin-one targets, produce odd polynomials in $\xi$ instead, owing to the Lorentz tensor multiplying them in the correlator being time reversal odd~\cite{Berger:2001zb}. The standard vector and axial GPDs for a proton target are all T-even and thus produce even polynomials in $\xi$.}.

When $s$ is even for quarks (odd for gluons), the $\xi^s$ ($\xi^{s+1}$ for gluons) term in this expansion is associated with the Polyakov-Weiss D-term~\cite{Polyakov:1999gs}.
This term has support only in the ERBL region and is not present in several commonly-used models of GPDs, including the GK model which we use here as a benchmark.
Renormalization group evolution is incapable of producing a D-term where none was present, so any nonzero $\xi^s$ ($\xi^{s+1}$ for gluons) term produced by the evolution code is a measure of accumulated numerical error.

In addition, there are several other special constraints that should hold.
Firstly, for quark distributions the lowest moment $A^q_{1,0}(0,Q^2)$ should be constant with respect to evolution, and should equal the net quark number (number of quarks minus number of antiquarks) for that flavor.
Secondly, for the quark singlet and gluon distributions, the quark and gluon momentum fractions are given by:
\begin{align}
  \sum_q \langle x_q \rangle(Q^2)
  =
  \frac{1}{2}
  A^{\rm S}_{2,0}(0,Q^2)
  \,,
  \qquad
  \langle x_g \rangle(Q^2)
  =
  \frac{1}{2}
  A^g_{1,0}(0,Q^2)
  \,,
\end{align}
where the factor $\frac{1}{2}$
counteracts double-counting that arises from
the symmetry relations
    $H^q(-x,\xi,Q^2) = -H^{\bar{q}}(x,\xi,Q^2)$
and
    $H^g(-x,\xi,Q^2) = H^g(x,\xi,Q^2)$.
These should obey the parton momentum sum rule:
\begin{align}
  \label{eqn:sum:mom}
  \sum_q \langle x_q \rangle(Q^2)
  +
  \langle x_g \rangle(Q^2)
  =
  1
  \,,
\end{align}
which is also invariant under evolution.

Using the Refined Method for evolution, we perform several numerical polynomial tests.
The tests are carried out on a discretized GPD grid by using the trapezoidal rule to perform the integration, which gives a function in terms of $\xi$:
\begin{align}
  M_s(\xi,Q^2)
  &=
  \int_{-1}^{+1} \d x \,
  x^{s-1} H(x,\xi,Q^2)
  \approx
  \frac{1}{2}
  \sum_{i=0}^{N_x-1}
  \Big( x_{i+1}^{s-1} H_{i+1}(\xi,Q^2) + x_i^{s-1} H_i(\xi,Q^2) \Big)
  ( x_{i+1} - x_i )
  \,,
\end{align}
where we again suppress the $t$ dependence in the GPD.
A polynomial fit is then performed to the moment $M_s(\xi,Q^2)$, with a polynomial of order $s$ used for the fit function, and the coefficients as the fit parameters.
Since small $\xi$ and large $\xi$ are more likely to deviate from the expected behavior, we restrict the fit to the range $0.2 \leq \xi \leq 0.8$.
Deviation of odd-$n$ coefficients from zero, as well as the highest-order coefficient from zero, are measures of polynomiality breaking.
Additionally, deviations of the discretized GPD from the polynomial fit (fit residuals) may indicate polynomiality breaking; these will most clearly be seen at $\xi$ values not used to fit the polynomial.

\begin{figure}
  \includegraphics[width=0.33\textwidth]{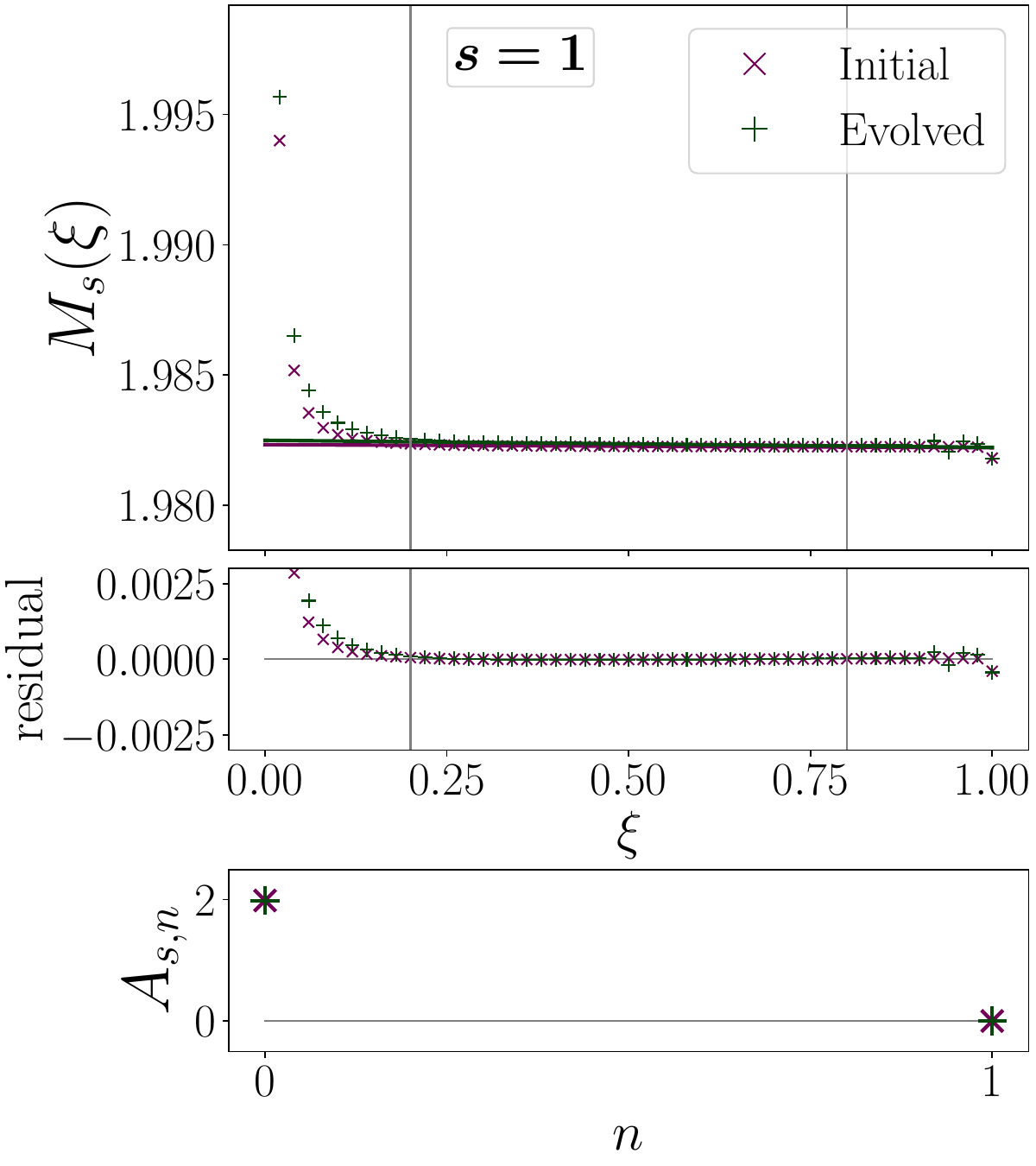}
  \includegraphics[width=0.33\textwidth]{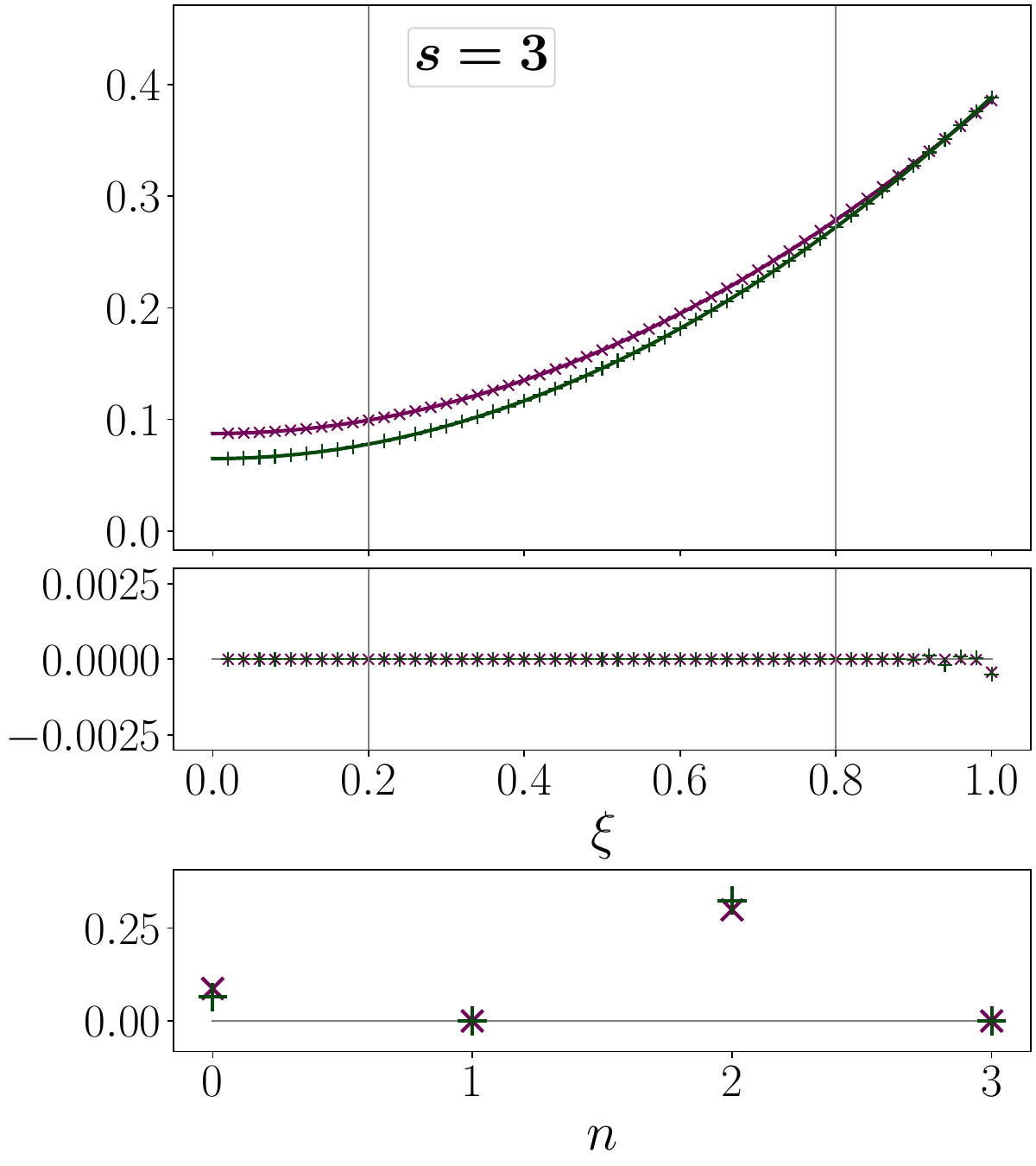}
  \includegraphics[width=0.33\textwidth]{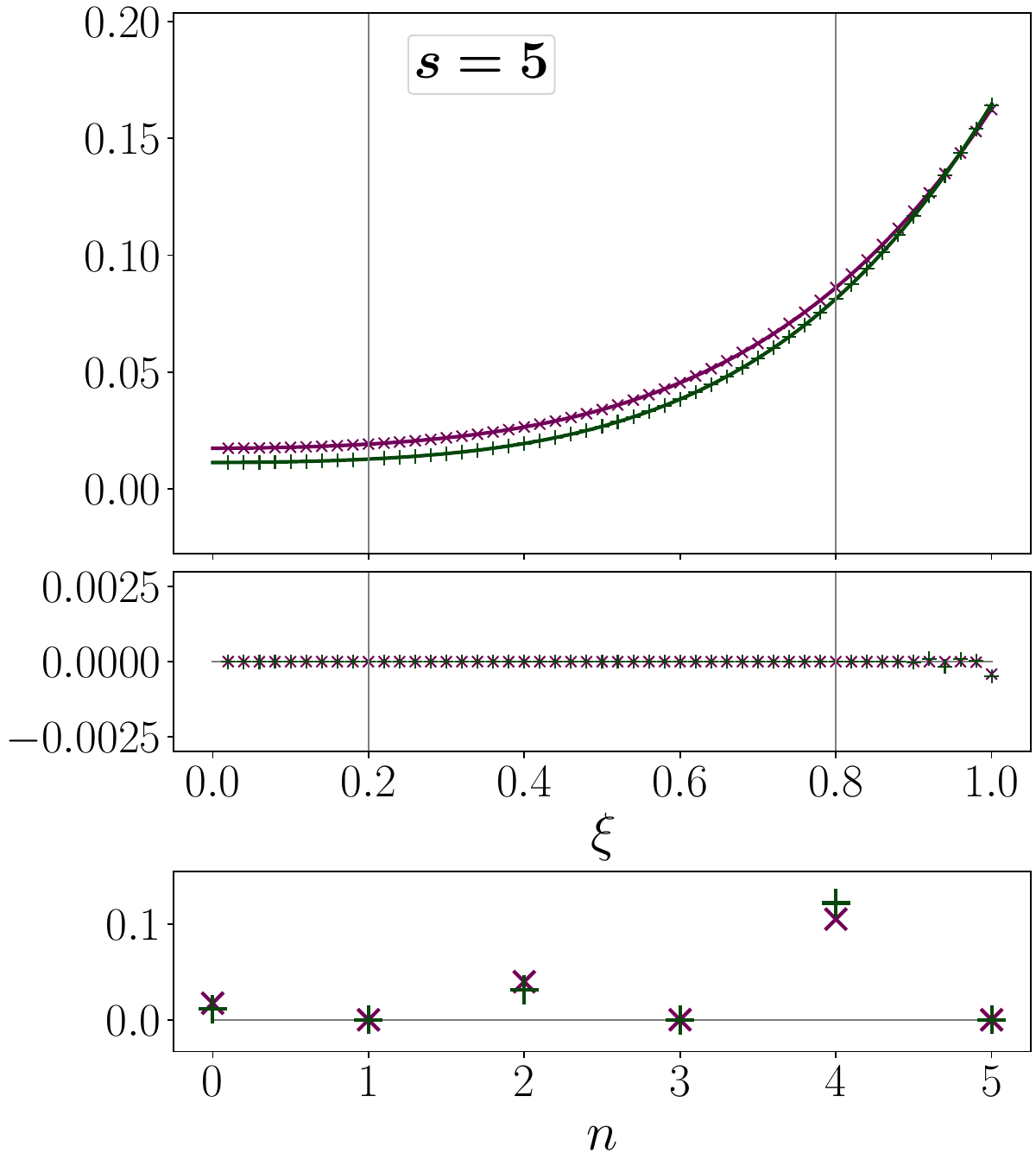}
  \caption{
    Polynomiality test for the $u$-quark distribution $H_u(x,\xi,Q^2)$. The evolution is performed between the charm and bottom quark masses, and polynomial fits are done over $\xi \in [0.2,0.8]$, which is delineated by gray vertical lines. The first three odd moments ($s=1,3,5$) are shown. The results of the polynomial fits to the numerical evaluations of the moments are shown at the initial scale (purple lines) and the final scale (green lines).
  }
  \label{fig:poly:umin}
\end{figure}

We perform polynomiality tests for both the initial-scale discretization of the GK model and for the evolved GPDs that the evolution code produces.
As an example, we show Fig.~\ref{fig:poly:umin} that that polynomiality holds for the up quark distribution $H^u(x,\xi,Q^2)$. 
For $\xi > 0.1$, polynomiality holds fairly well, but at smaller $\xi$ values the expected behavior breaks down.
In fact, polynomiality violations occur already at the initial scale.
This is a discretization effect, and since the violations are present at the initial scale, they cannot be attributed to evolution.
Additionally, the sum rule $A^u_{1,0} \equiv N_u = 2$, for the net number of $u$ quarks in proton, does not hold exactly, but is violated by $\sim 0.7\%$.
We find that a finer grid does not improve this violation, so the error is more likely due to the imprecision in the GK model parameters for the GPD.
Although the sum rule is slightly violated, the value of $A_{1,0}(0,Q^2)$ is not changed by evolution---as expected.

Similar tests can be performed for the singlet quark and gluon distributions, which we discuss in Appendix~\ref{sec:addbench}.
We do find that the preservation of polynomiality has limits, and that at $s \geq 7$ small but nonzero odd coefficients after evolution appear (see Fig.~\ref{fig:poly:S} in Appendix~\ref{sec:addbench}).
Since the odd coefficients should vanish identically, this is due to the accumulated numerical error in the discretized evolution.

We also perform a numerical test of the momentum sum rule (\ref{eqn:sum:mom}).
We find at the initial (charm quark mass) scale that:
\begin{align}
  \sum_q \langle x_q \rangle(m_c^2)
  =
  0.5525
  \,, \qquad
  \langle x_g \rangle(m_c^2)
  =
  0.4325
  \,, \qquad
  \Big( \sum_q \langle x_q \rangle + \langle x_g \rangle \Big)(m_c^2)
  =
  0.9850
  \,,
\end{align}
while at the evolved (bottom quark mass) scale :
\begin{align}
  \sum_q \langle x_q \rangle(m_b^2)
  =
  0.5160
  \,, \qquad
  \langle x_g \rangle(m_b^2)
  =
  0.4692
  \,, \qquad
  \Big( \sum_q \langle x_q \rangle + \langle x_g \rangle \Big)(m_b^2)
  =
  0.9852
  \,.
\end{align}
The momentum sum rule is already slightly violated (by around 1.5\%) at the initial scale, likely due to a mix of discretization error and imprecision in the GK model parameters.
The evolution code changes the answer for this moment by only 0.02\%, so introduces a much smaller error in the momentum sum rule.

\subsection{Comparison of evolution to PARTONS/APFEL++}

Next, we perform benchmarks for the evolution matrices themselves.
In contrast to the kernel benchmarks, we do not have access to a ``ground truth'' to compare to the evolved GPDs.
The next best solution is to compare the results of the methods to already accepted and validated GPD evolution codes.
An existing evolution code in $x$ space, also based on finite element methods, is available within the PARTONS framework~\cite{Berthou:2015oaw, Bertone:2023jeh}, which utilizes APFEL++~\cite{Bertone:2013vaa, Bertone:2017gds} to perform the evolution.
We use this code as a proxy for the ``ground truth,'' and and to compare against in the validation of our methods.
Two small modifications needed to be made to our code to make a fair comparison.
Firstly, our code uses the $\overline{\mathrm{MS}}$ bottom mass $m_b=4.18$~GeV, since the $\overline{\mathrm{MS}}$ factorization scheme is employed, whereas the PARTONS code uses $m_b=4.75$~GeV.
We thus needed to modify our code to use $m_b=4.75$~GeV when benchmarking against PARTONS.
Secondly, our code uses next-to-leading-log evolution of $\alpha_{\mbox{\tiny\rm{QCD}}}$, whereas PARTONS uses leading log, so we temporarily used leading log evolution as well in the comparison.

\begin{figure}
  \includegraphics[width=0.33\textwidth]{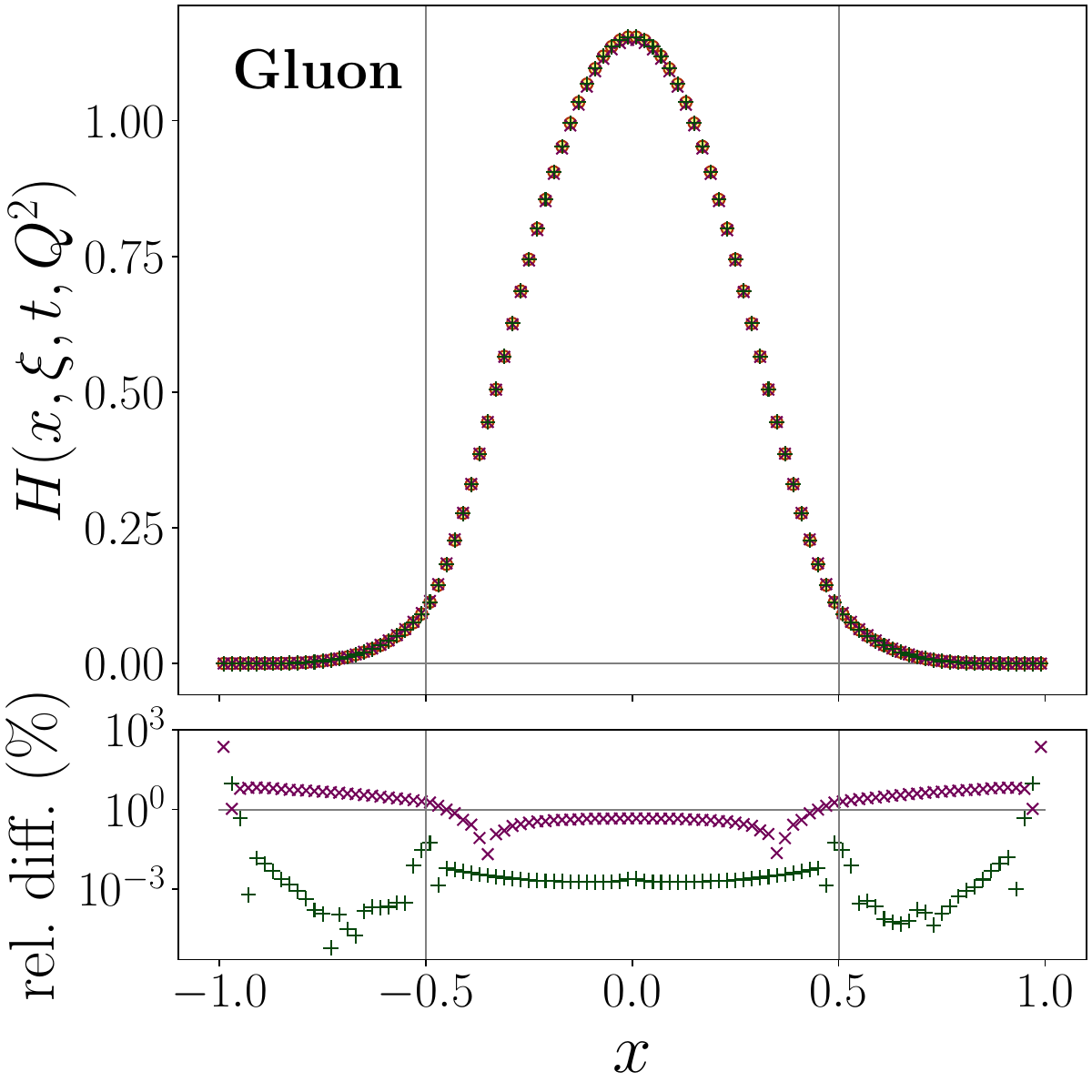}
  \includegraphics[width=0.33\textwidth]{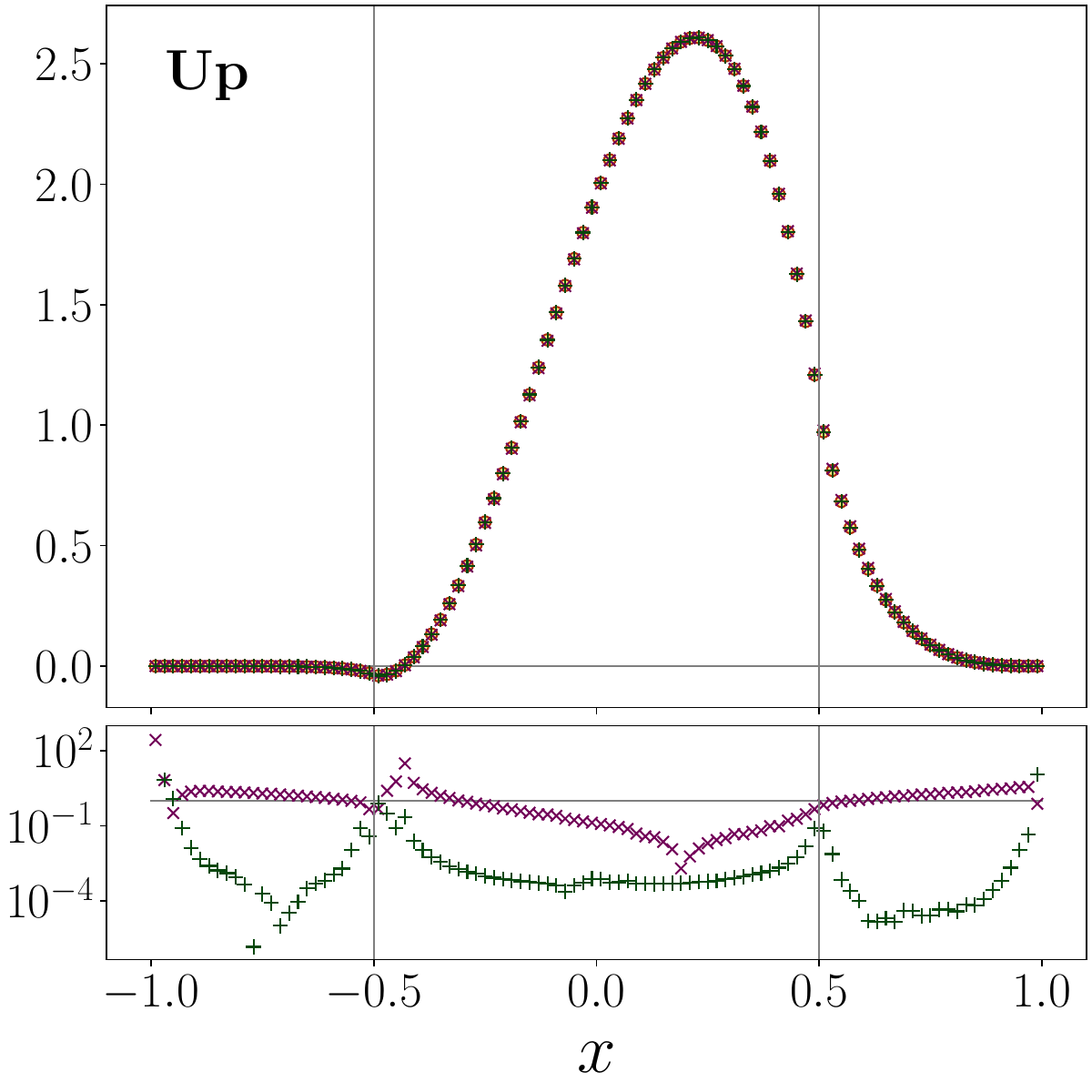}
  \includegraphics[width=0.33\textwidth]{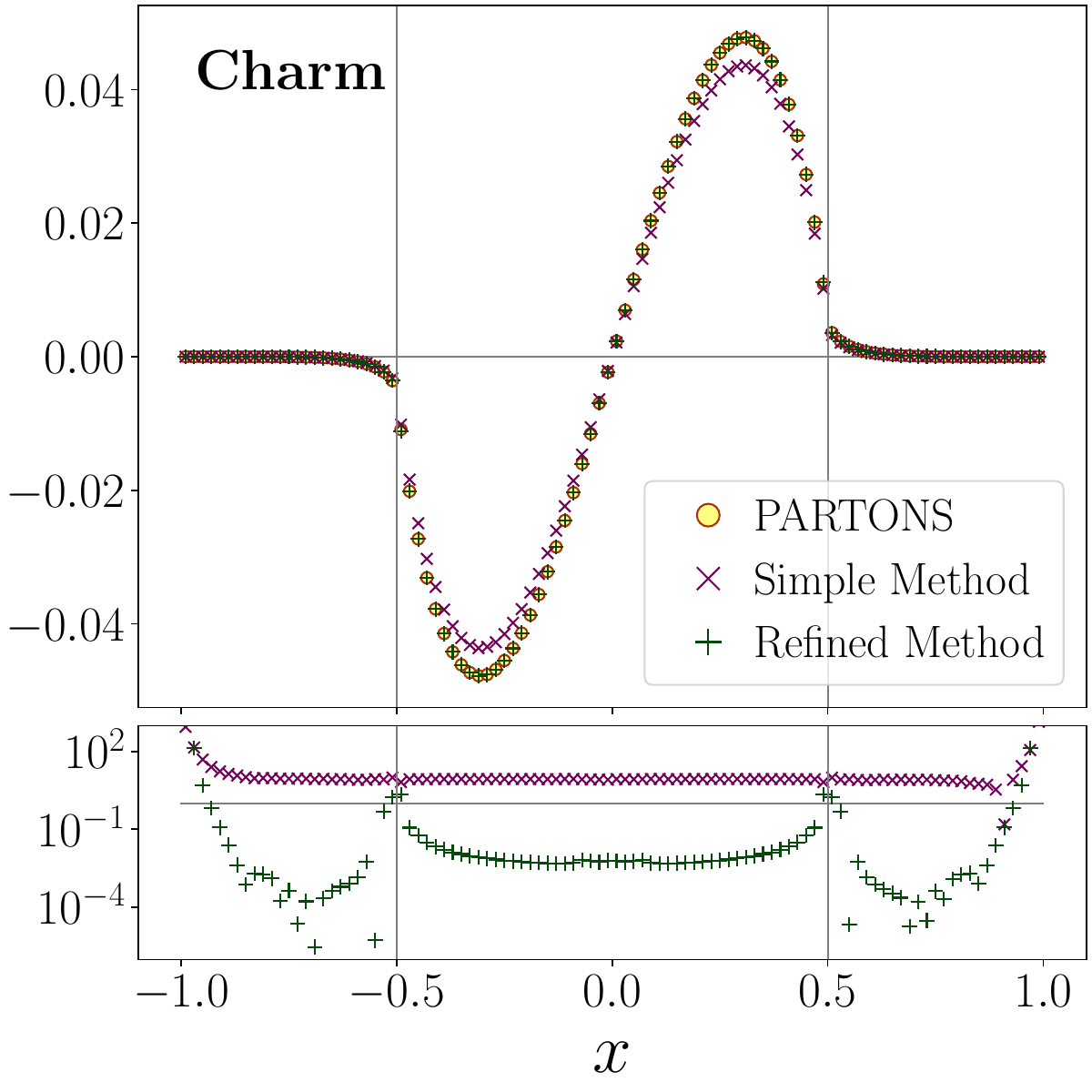}
  \caption{
    Comparison of evolved gluon (left panel), $u$-quark (middle panel) and $c$-quark (right panel) GPDs found via PARTONS/APFEL++ (yellow circles), the Simple Method (purple $\times$'s), and the Refined Method (green $+$'s), when evolution is performed from $m_c^2$ to 16~GeV$^2$. Since the PARTONS framework uses $m_b = 4.75$~GeV and leading log evolution of $\alpha_{\mbox{\tiny\rm{QCD}}}$, our code was modified to make this a fair comparison. This comparison uses $\xi=0.5$ and $t=0$.
  }
  \label{fig:bench:partons}
\end{figure}

The result of this benchmark is shown in Fig.~\ref{fig:bench:partons} for the gluon, up-quark and charm-quark GPDs.
The Simple Method and Refined Method results for the evolution were produced by using our own evolution matrices on the initial GPD scale provided, to ensure that the same initial scale GPD was evolved by all three codes.
The gluon and $u$-quark distributions are both phenomenologically important.
The charm-quark distribution is small, but is entirely generated here by evolution, and is thus a robust stress test of how much error accumulates in the evolution.
(The relative error in the other distributions can be masked by the evolution being a small correction to a consistent starting GPD.)
The Simple Method has reasonable agreement with the PARTONS result, with the relative difference hovering around 1\% (except in the charm quark, where it hovers around 10\%).
The Refined Method, however, has remarkably good agreement with the PARTONS code, except near the $x=\pm1$ and $x=\pm\xi$ boundaries.

At the $x=\pm1$ boundaries, the GPD itself drops extremely rapidly, and this region has little phenomenological impact.
It is easy for the relative error in a function that rapidly approaches zero to be large.
The $\sim 1\%$ discrepancy at $x=\pm\xi$ is more concerning, however; this is a region of extreme phenomenological importance in deeply-virtual Compton scattering.
We have already found (see Fig.~\ref{fig:bench:xi}, for instance) that the Refined Method has irreducible numerical error at exactly these points, and that the error was around 1\%.
We have also commented that this error occurs because we are approximating a nondifferentiable function with a smooth polynomial.
The 1\% discrepancy between the PARTONS result and the Refined Method is thus more likely numerical error in the Refined Method than in the PARTONS code.

Since the Refined Method is more accurate than the Simple Method at the level of the kernels, and since it agrees better with the already-validated PARTONS/APFEL++ code, we conclude that its result can be trusted more than the Simple Method's results.

\section{Summary and outlook}
\label{sec:end}

In this work we have described how finite element methods can be used to construct fast $x$-space computer code for renormalization group evolution of GPDs.
We considered in detail two specific methods: a Simple Method, which used Gauss-Legendre quadrature across the entire integration domain to estimate integrals, and modified cubic Hermite splines to interpolate to the Gaussian weight points; and a Refined Method which used Gauss-Kronrod quadrature on carefully selected subdomains, and piecewise Lagrange interpolation to interpolate to the weight points.
These separate methods follow from different ways of conceptualizing finite elements---the Simple Method from framing finite elements in terms of matrix operations, and the Refined Method from framing finite elements as linear operations on basis functions.

In both cases, the use of finite element methods allows GPD evolution to be approximated as a matrix differential equation.
The greatest computational bottleneck is in computing the kernel matrices that appear on the right-hand sides of the nonsinglet and singlet evolution equations (\ref{eqn:evo:NS}) and (\ref{eqn:evo:S}), respectively.
These matrices, however, are entirely independent of the particular GPD that is evolved.
Thus, the kernel matrices---and the evolution matrices that solve the matrix differential equation---only need to be computed once, and can be reused for a variety of initial-scale trial GPDs.
This makes the kernel method extremely apt for use in fits, as well as in machine learning where a large number of randomly-generated trial GPDs must be generated and evolved to compare with data.
In fact, since evolution can be framed as a tensor contraction operation, it is differentiable, allowing the gradient of some sort of loss function with respect to neural network parameters to be computed---which is of vital importance for the usability of these methods in machine learning.

The computer code may potentially be helpful as a tool for the quark-gluon tomography community, outside of just its machine learning applications.
Fast, easy-to-use evolution code is also useful to physicists to evolve model GPDs to empirically relevant $Q^2$ scales, and potentially for evolving lattice computations as well.
We have prepared a repository where the code from the Refined Method is available for download.
The package is called {\tt tiktaalik} and can be found on GitHub~\cite{Freese:2024tik}.

There are many opportunities to refine and extend the uses of the kernel method.
The most obvious extension is to include NLO corrections to the kernels, and an update of the Refined Method code to include NLO is underway.
Alternative grid spacings to improve accuracy at small $\xi$---as well as
at $x=\pm\xi$---are also being investigated.
A more outside-the-box extension could be to consider, for example, the evolution equation~(\ref{eqn:evo:NS}) as a constraint rather than as an equation to solve: violations of the equation could be added to a loss function which must be minimized.

\begin{acknowledgments}
  We gratefully acknowledge helpful discussions with
  Pi-Yueh Chuang,
  Christopher Cocuzza,
  Emil Constantinescu,
  Herv\'{e} Dutrieux,
  Peter Risse,
  Simone Rodini
  and Gabriel Santiago, 
  and thank Herv\'{e} Dutrieux for providing the initial scale GPD and its evolution by PARTONS/APFEL++.
  This work was supported by the DOE contract No.~DE-AC05-06OR23177, under which Jefferson Science Associates, LLC operates Jefferson Lab; the U.S. Department of Energy, Office of Science, Office of Nuclear Physics, contract no. DE-AC02-06CH11357; the Scientific Discovery through Advanced Computing (SciDAC) award {\it Femtoscale Imaging of Nuclei using Exascale Platforms}; and the Quark-Gluon Tomography (QGT) Topical Collaboration. The research described in this paper was conducted in part under the Laboratory Directed Research and Development Program at Thomas Jefferson National Accelerator Facility for the U.S. Department of Energy.
\end{acknowledgments}

\appendix
\section{Explicit formulas for kernels}
\label{sec:kernels}

In our computer codes we use the formulas for the LO evolution kernels as given by Belitsky, Freund and Mueller (BFM)~\cite{Belitsky:1999hf}.
Our notation differs slightly from theirs, so we restate the formulas in modified notation.
We use $x$ and $y$ to refer to (average) momentum fractions, and $\xi = (p^+ - p'^+)/(p^+ + p'^+)$ to signify skewness, together with the auxiliary variables:
\begin{subequations}
\begin{eqnarray}
  X = \frac{x + \xi}{2\xi}\,, \qquad
  Y = \frac{y + \xi}{2\xi}\,,
  \\
  \overline{X} = 1 - X\,, \qquad
  \overline{Y} = 1 - Y
  \,.
\end{eqnarray}
\end{subequations}
Note that BFM use the variables $x$, $y$, $\bar{x}$, and $\bar{y}$, where we use $X$, $Y$, $\overline{X}$, and $\overline{Y}$.
We also use the generalized step function:
\begin{align}
  \rho(X,Y)
  =
  \varTheta\Big(1 - \frac{X}{Y}\Big)
  \varTheta\Big(\frac{X}{Y}\Big)\,
  \mathrm{sgn}(Y)
  \,.
\end{align}
From BFM~\cite{Belitsky:1999hf}, the LO evolution kernels for helicity-dependent GPDs are as follows:
\begin{subequations}
\begin{align}
  K_{qq}^A(x,y,\xi)
  &=
  \left[
    K_{qq}^P(x,y,\xi)
    \right]_+,
  \\
  K_{qg}^A(x,y,\xi)
  &=
  \frac{T_F}{2\xi}
  \left(
  -
  \frac{X}{Y^2}\,
  \rho(X,Y)
  +
  \frac{\overline{X}}{\overline{Y}^2}\,
  \rho(\overline{X},\overline{Y})
  \right),
  \\
  K_{gq}^A(x,y,\xi)
  &=
  \frac{C_F}{2\xi}
  \left(
  \frac{X^2}{Y}\,
  \rho(X,Y)
  -
  \frac{\overline{X}^2}{\overline{Y}}\,
  \rho(\overline{X},\overline{Y})
  \right),
  \\
  K_{gg}^A(x,y,\xi)
  &=
  \frac{C_A}{2\xi}
  \left[
    K_{gg}^P(x,y,\xi)
    \right]_+
  -
  \left(
  \frac{\beta_0}{2}
  +
  \frac{7}{3} C_A
  \right)
  \delta(y-x)
  \,,
\end{align}
\end{subequations}
where the functions appearing inside the brackets $[...]_+$ are:
\begin{subequations}
\begin{align}
  K_{qq}^P(x,y,\xi)
  &=
  \frac{C_F}{2\xi}
  \left(
  \frac{X}{Y}\Big(1 + \frac{1}{Y-X}\Big)\,
  \rho(X,Y)
  +
  \frac{\overline{X}}{\overline{Y}}
  \Big(1 + \frac{1}{\overline{Y}-\overline{X}}\Big)\,
  \rho(\overline{X},\overline{Y})
  \right),
  \\
  K_{gg}^P(x,y,\xi)
  &=
  \frac{C_A}{2\xi}
  \left(
  \frac{X^2}{Y^2}\Big(2 + \frac{1}{Y-X}\Big)\,
  \rho(X,Y)
  +
  \frac{\overline{X}^2}{\overline{Y}^2}
  \Big(2 + \frac{1}{\overline{Y}-\overline{X}}\Big)\,
  \rho(\overline{X},\overline{Y})
  \right)
  \,,
\end{align}
\end{subequations}
with
\begin{align}
  \beta_0
  =
  -\frac{11}{3} C_A
  +
  \frac{4}{3} T_F n_{\text{fl}},
\end{align}
and $n_{\text{fl}}$ is the number of active flavors.
The factor $-\frac{7}{3} C_A$ appearing with the $\delta$ function is the extra piece related to BFM's modification of the plus prescription to ensure the correct forward limit for the gluon splitting function.

The formulas for the helicity-independent evolution kernels, again from BFM~\cite{Belitsky:1999hf}, are given by:
\begin{subequations}
\begin{align}
  K_{qq}^V(x,y,\xi)
  &=
  K_{qq}^A(x,y,\xi),
  \\
  K_{qg}^V(x,y,\xi)
  &=
  K_{qg}^A(x,y,\xi)
  +
  \frac{C_F}{2\xi}
  \left(
  -
  2\frac{X}{Y^2}\Big(2\overline{X}Y - X\Big)\,
  \rho(X,Y)
  +
  2\frac{\overline{X}}{\overline{Y}^2}
  \Big(2X\overline{Y} - \overline{X}\Big)
  \rho(\overline{X},\overline{Y})
  \right),
  \\
  K_{gq}^V(x,y,\xi)
  &=
  K_{gq}^A(x,y,\xi)
  +
  \frac{C_F}{2\xi}
  \left(
  2\frac{X^2}{Y}\Big(2\overline{X}Y - \overline{Y}\Big)\,
  \rho(X,Y)
  -
  2\frac{\overline{X}^2}{\overline{Y}}\Big(2X\overline{Y} - Y\Big)\,
  \rho(\overline{X},\overline{Y})
  \right),
  \\
  K_{gg}^V(x,y,\xi)
  &=
  K_{gg}^A(x,y,\xi)
  +
  \frac{C_A}{2\xi}
  \left(
  2\frac{X^2}{Y^2}\Big(2\overline{X}Y + Y - X\Big)\,
  \rho(X,Y)
  +
  2\frac{\overline{X}^2}{\overline{Y}^2}
  \Big(2X\overline{Y} + \overline{Y} - \overline{X}\Big)\,
  \rho(\overline{X},\overline{Y})
  \right)
  \,.
\end{align}
\end{subequations}
At LO, the following simple relationships hold for obtaining the singlet kernels:
\begin{subequations}
\begin{align}
  K_{SS}^X(x,y,\xi)
  &=
  K_{qq}^X(x,y,\xi),
  \\
  K_{Sg}^X(x,y,\xi)
  &=
  2 n_{\text{fl}}
  K_{qg}^X(x,y,\xi),
  \\
  K_{gS}^X(x,y,\xi)
  &=
  K_{gS}^X(x,y,\xi).
\end{align}
\end{subequations}

In contrast to Vinnikov~\cite{Vinnikov:2006xw}, we do not find that it is necessary to massage these formulas (for example, by partial fraction decomposition) into in order to obtain stable numerical integrals involving these kernels.
In fact, we find that even the most numerically challenging integral---given by
\begin{align*}
  \int_{-1}^1 \d y \,
  \big( K^P(x,y,\xi) - K^P(y,x,\xi) \big),
\end{align*}
as prescribed by the plus prescription (\ref{eqn:plus})---can be performed numerically by breaking the $y$ integration domain into the pieces 
    $[-1, -|x|]$, $[-|x|, |x|]$ and $[|x|, 1]$,
and using adaptive Gaussian integration on each of these pieces.
The numerical accuracy and stability of this integral will prove vital to a future update of our code that includes the NLO kernels.
However, at LO these integrals can be done analytically, and for prudence we use the analytic results in our codes.
We find:
\begin{subequations}
\begin{align}
  \label{eqn:qq:cst}
  \int_{-1}^1 \d y \,
  \Big( K_{qq}^P(x,y,\xi) - K_{qq}^P(y,x,\xi) \Big)
  &=
  C_F 
  \left\{
    \begin{array}{lcl}
      \frac{3}{2}
      +
      \log\frac{(1-x)^2}{x^2-\xi^2}
      +
      \frac{x-\xi}{2\xi}\,
      \log\frac{1+\xi}{x+\xi}
      -
      \frac{x+\xi}{2\xi}\,
      \log\frac{1-\xi}{x-\xi}
      &:&
      \xi < x
      \\
      \frac{3}{2}
      +
      \log\frac{1-x^2}{\xi^2-x^2}
      +
      \frac{1-\xi-x}{2\xi}
      \log\frac{\xi+x}{1+\xi}
      +
      \frac{1-\xi+x}{2\xi}
      \log\frac{\xi-x}{1+\xi}
      &:&
      -\xi < x < \xi
      \\
      \frac{3}{2}
      +
      \log\frac{(1+x)^2}{x^2-\xi^2}
      -
      \frac{x+\xi}{2\xi}
      \log\frac{1+\xi}{-x+\xi}
      +
      \frac{x-\xi}{2\xi}
      \log\frac{1-\xi}{-x-\xi}
      &:&
      x < -\xi
    \end{array}
    \right.
  \\
  \label{eqn:gg:cst}
  \int_{-1}^1 \d y \,
  \Big( K_{gg}^P(x,y,\xi) - K_{gg}^P(y,x,\xi) \Big)
  &=
  C_A 
  \left\{
    \begin{array}{lcl}
      \frac{7}{3}
      -
      \frac{2x(1-x)}{1-\xi^2}
      +
      \log\frac{(1-x)^2}{1-\xi^2}
      &:&
      \xi < x
      \\
      \frac{7}{3}
      -
      \frac{2x^2}{\xi(1+\xi)}
      +
      \log\frac{1-x^2}{(1+\xi)^2}
      &:&
      -\xi < x < \xi
      \\
      \frac{7}{3}
      +
      \frac{2x(1+x)}{1-\xi^2}
      +
      \log\frac{(1+x)^2}{1-\xi^2}
      &:&
      x < -\xi
    \end{array}
    \right.
    \,.
\end{align}
\end{subequations}

\section{Interpolation scheme for Simple Method}
\label{sec:cubic_spline}

In Sec.~\ref{sec:simple}, in order to translate from the grid on which the GPD was defined to the grid required for fixed Gaussian quadrature, we need to interpolate from one grid to the other. 
The particular interpolation scheme we employ is a modification of piecewise cubic Hermite polynomials, which we describe in detail here.

The piecewise cubic Hermite polynomials use information about the function at two grid points, in addition to the derivative of the function, to interpolate between the two points. 
More exactly, if we have two grid points, $x_L$ and $x_R$, their function values $y_L=y(x_L)$ and $y_R=y(x_R)$, and their derivatives, $S_L = {\partial y(x_L)}/{\partial x}$ and $S_R={\partial y(x_R)}/{\partial x}$. 
The value of the function $y(x)$ evaluated between those two points is approximately~\cite{van2000introduction}
\begin{eqnarray}
\label{eq cubic interpolation}
    y(x) \approx A + B(x-x_L) + C(x-x_L)^2 + D(x-x_L)^2(x-x_R),
\end{eqnarray}
with
\begin{align}
    A &= y_L, \nn\\
    B &= S_L, \nn\\
    C &= \frac{y'-S_L}{x_R-x_L}, \\
    D &= \frac{S_L + S_R - 2y'}{(x_R-x_L)^2}, \nn\\
    y'&=\ \frac{y_R-y_L}{x_R-x_L}. \nn
\end{align}
The problem with using this interpolation scheme directly is that we do not have direct access to the exact derivatives, $S_L$ and $S_R$. 
To know the exact value of the derivative would require knowing the exact function, $y(x)$; however, our goal in this section is to devise an interpolation method that works independently of any particular function $y(x)$.

Our modification is to approximate these derivatives in a manner that does not require information about which specific function $y(x)$ is being interpolated.
We approximate these derivatives using a finite difference:
\begin{subequations}
\label{eq cubic derivative choice}
\begin{eqnarray}
    S_L &\approx& \frac{y_R-y_{L-1}}{x_R-x_{L-1}}, \\
    S_R &\approx& \frac{y_{R+1}-y_{L}}{x_{R+1}-x_{L}},
\end{eqnarray}
\end{subequations}
where $x_{L-1}~(y_{L-1})$ is the (function evaluated at the) grid point to the left of $x_L$, and $x_{R+1}~(y_{R+1})$ is the (function evaluated at the) grid point to the right of $x_R$.
With this substitution, $y(x)$ can be expressed purely as a linear combination of the function evaluated at the four grid points $x_{L-1}, x_{L}, x_{R}$ and $x_{R+1}$. The coefficients for this interpolation scheme are given by:
\begin{subequations}
\begin{eqnarray}
    a(x) &=& -\frac{(x-x_L)}{(x_R-x_{L-1})} + \frac{(x-x_L)^2}{(x_R-x_L)(x_R-x_{L-1})} - \frac{(x-x_L)^2(x-x_R)}{(x_R-x_L)^2(x_R-x_{L-1})},
    \\
    b(x) &=& 1 - \frac{(x-x_L)^2}{(x_R-x_L)^2} - \frac{(x-x_L)^2(x-x_R)}{(x_R-x_L)^2(x_{R+1}-x_L)} + \frac{2(x-x_L)^2(x-x_R)}{(x_R-x_L)^3},
    \\
    c(x) &=& \frac{(x-x_L)}{(x_R-x_{L-1})} + \frac{(x-x_L)^2}{(x_R-x_L)^2} - \frac{(x-x_L)^2}{(x_R-x_L)(x_R-x_{L-1})} + \frac{(x-x_L)^2(x-x_R)}{(x_R-x_L)^2(x_R-x_{L-1})} - \frac{2(x-x_L)^2(x-x_R)}{(x_R-x_L)^3}, 
    \nn\\
         & &
    \\
    d(x) &=& \frac{(x-x_L)^2(x-x_R)}{(x_R-x_L)^2(x_{R+1}-x_L)}.
\end{eqnarray}
\end{subequations}
The interpolation of point $x_L<x<x_R$ is then given by:
\begin{eqnarray}
    y(x) \approx a(x)\, y_{L-1} + b(x)\, y_L + c(x)\, y_R + d(x)\, y_{R+1}.
\end{eqnarray}
This interpolation scheme works well for most grid points but fails when it reaches the end of the grid. 
Our approximations for $S_L$ and $S_R$ are ill-defined when $x_{L-1}$ and $x_{R+1}$ do not exist. 
To remedy this, we make one final modification to this interpolation scheme.
The fix is to simply use a brute-force cubic interpolation when interpolating near the first or last points on the grid.
That is, a unique cubic polynomial can be defined if the function is known at four points.
When $x_0<x<x_1$, the coefficients of the interpolation are given by:
\begin{eqnarray}
    a_0 &=& x_1x_2x_3/\big((x_1-x_0)(x_2-x_0)(x_3-x_0)\big),\nonumber\\
    a_1 &=& x_0x_2x_3/\big((x_0-x_1)(x_1-x_2)(x_1-x_3)\big),\nonumber\\
    a_2 &=& x_0x_1x_3/((x_0-x_2)(x_2-x_1)(x_2-x_3)),\nonumber\\
    a_3 &=& x_0x_1x_2/((x_0-x_3)(x_3-x_1)(x_3-x_2)),\nonumber\\
    b_0 &=& (x_1x_2+x_1x_3+x_2x_3)/((x_0-x_1)(x_0-x_2)(x_0-x_3)),\nonumber\\
    b_1 &=& (x_0x_2+x_0x_3+x_2x_3)/((x_1-x_0)(x_1-x_2)(x_1-x_3)),\nonumber\\
    b_2 &=& (x_0x_1+x_0x_3+x_1x_3)/((x_2-x_0)(x_2-x_1)(x_2-x_3)),\nonumber\\
    b_3 &=& (x_0x_1+x_0x_2+x_1x_2)/((x_3-x_0)(x_3-x_1)(x_3-x_2)),\nonumber\\
    c_0 &=& (x_1+x_2+x_3)/((x_1-x_0)(x_2-x_0)(x_3-x_0)),\nonumber\\
    c_1 &=& (x_0+x_2+x_3)/((x_0-x_1)(x_2-x_1)(x_3-x_1)),\nonumber\\
    c_2 &=& (x_0+x_1+x_3)/((x_0-x_2)(x_1-x_2)(x_3-x_2)),\nonumber\\
    c_3 &=& (x_0+x_1+x_2)/((x_0-x_3)(x_1-x_3)(x_2-x_3)),\nonumber\\
    d_0 &=& 1/((x_0-x_1)(x_0-x_2)(x_0-x_3)),\nonumber\\
    d_1 &=& 1/((x_1-x_0)(x_1-x_2)(x_1-x_3)),\nonumber\\
    d_2 &=& 1/((x_2-x_0)(x_2-x_1)(x_2-x_3)),\nonumber\\
    d_3 &=& 1/((x_3-x_0)(x_3-x_1)(x_3-x_2)),\nonumber\\
    p_0(x) &=& a_0 + b_0 x + c_0 x^2 + d_0 x^3,\nonumber\\
    p_1(x) &=& a_1 + b_1 x + c_1 x^2 + d_1 x^3,\nonumber\\
    p_2(x) &=& a_2 + b_2 x + c_2 x^2 + d_2 x^3,\nonumber\\
    p_3(x) &=& a_3 + b_3 x + c_3 x^2 + d_3 x^3,\nonumber
\end{eqnarray}
with the interpolated point given by:
\begin{eqnarray}
    y(x)\approx p_0(x) y_0 + p_1(x) y_1 + p_2(x) y_2 + p_3(x)y_3.
\end{eqnarray}
A cubic interpolation for values of $x$ between the last two grid points is solved for in a similar manner.

\begin{figure}
    \centering
    \includegraphics[width=0.9\linewidth]{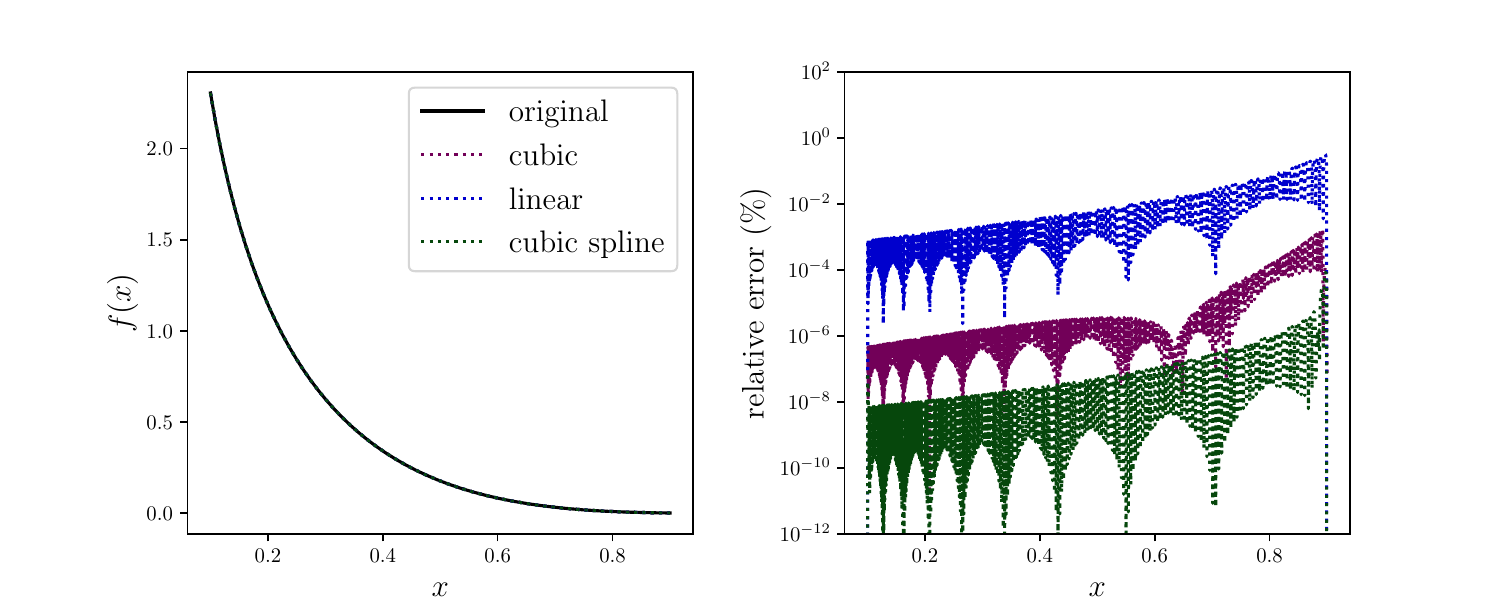}
    \caption{
    (Left) A plot of the function $f(x) = x^{-0.5}(1-x)^3$ (black) along with three interpolations: cubic with modified Hermite polynomials (purple), linear interpolation (blue), and cubic spline interpolation (green). 
    (Right) Relative error of the various interpolation schemes by taking the ratio of the interpolation to the analytic function evaluated on the denser grid.}
    \label{fig:interpolation_comparison}
\end{figure}

We benchmark this modified piecewise Hermite cubic polynomial against two other methods of interpolation, comparing to the simpler linear interpolation as well as the state-of-the-art cubic spline interpolation. 
Using a test function $f(x) = x^{-0.5}(1-x)^3$, with the starting grid a set of 300 logarithmically spaced points from $x=0.1$ to $x=0.9$, we interpolate onto a grid that is ten times as dense, but still logarithmically spaced. 
The result is shown in \ref{fig:interpolation_comparison}.
Our interpolation performs much better than linear interpolation. 
It does not perform as well as the state-of-the-art cubic spline interpolation, although this is satisfactory for our purposes.
One might propose using cubic spline interpolation instead, however, it is currently not known how to recast cubic spline interpolation as a matrix transformation.

\section{Additional plots of accuracy benchmarks}
\label{sec:addbench}

There are a number of other interesting and worthwhile accuracy benchmarks to consider, in addition to those given in Sec.~\ref{sec:benchmarks}, which we summarize in this Appendix.

\subsection{Number of Gaussian weights points}

\begin{figure}
  \includegraphics[width=0.5\textwidth]{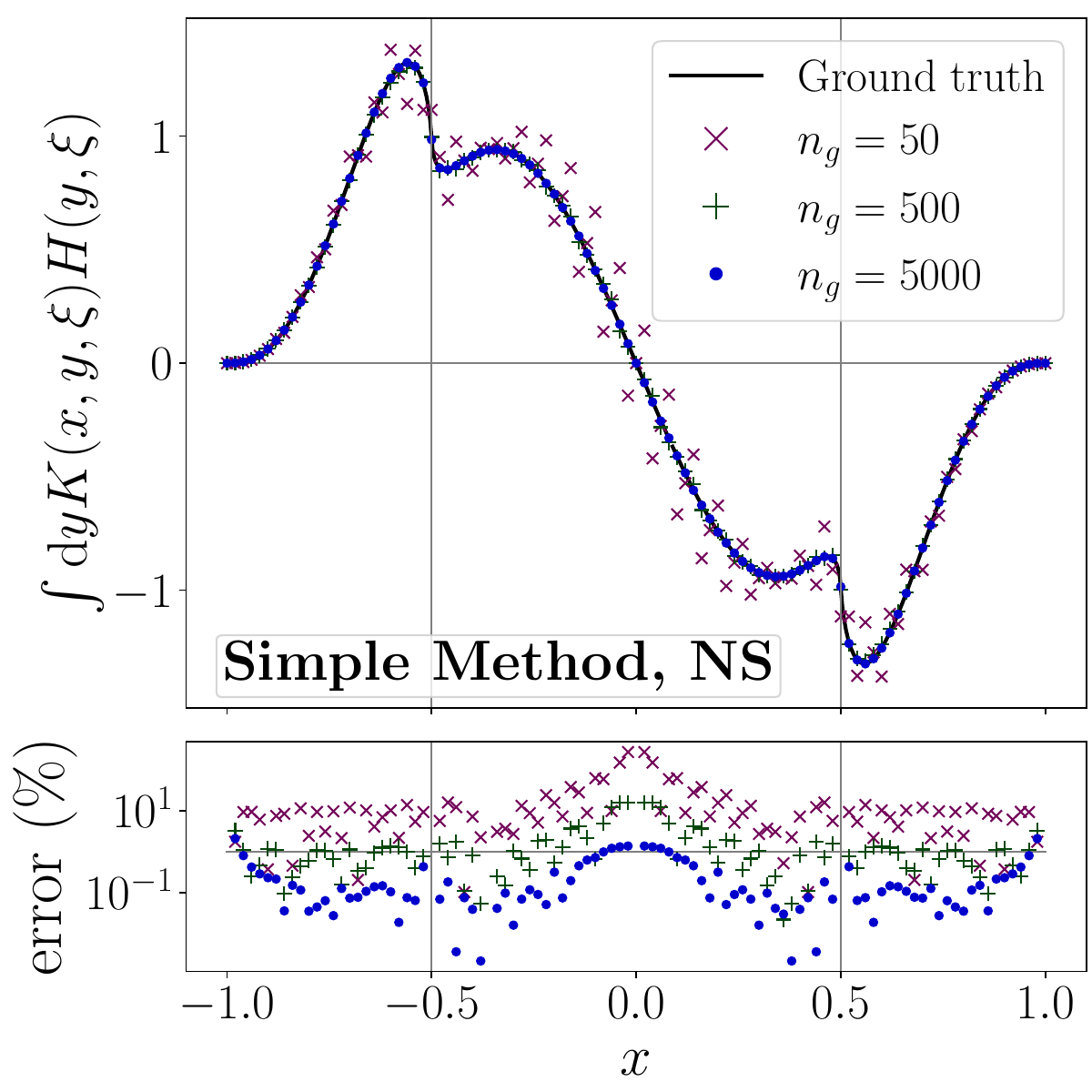}
  \caption{
    Accuracy benchmark for the right-hand side of the evolution equation (\ref{eqn:evolution}) at LO in the nonsinglet sector. The simple matrix method of Sec.~\ref{sec:simple} is compared to a ``ground truth,'' determined with the GK model~\cite{Kroll:2012sm} using adaptive quadrature. For the Simple Method, we use $N_x = 101$, and compare the result for varying numbers of Gaussian weight points $n_g$.
  }
  \label{fig:bench:gauss}
\end{figure}

We first study how the number of Gaussian weight points $n_g$ affects the numerical accuracy of the Simple Method.
Figure~\ref{fig:bench:gauss} shows the right-hand side of the evolution equation (\ref{eqn:evolution}) in the nonsinglet sector at LO, using the nonsinglet GPD $H_{T_3}(x,\xi=0.5,t=0,Q^2)$ in the GK model~\cite{Kroll:2012sm} as an example.
The results are striking: an extremely large number of weight points are necessary to obtain $< 1\%$ accuracy over most of the $x$ range.
Even at $n_g=5000$, there are relative errors on the order of a percent for $x \sim 0$ and $x \sim \pm 1$, although the ``ground truth'' itself is small in these regions, so large relative error is to be expected.

\subsection{Number of $x$ points}

\begin{figure}
  \includegraphics[width=0.49\textwidth]{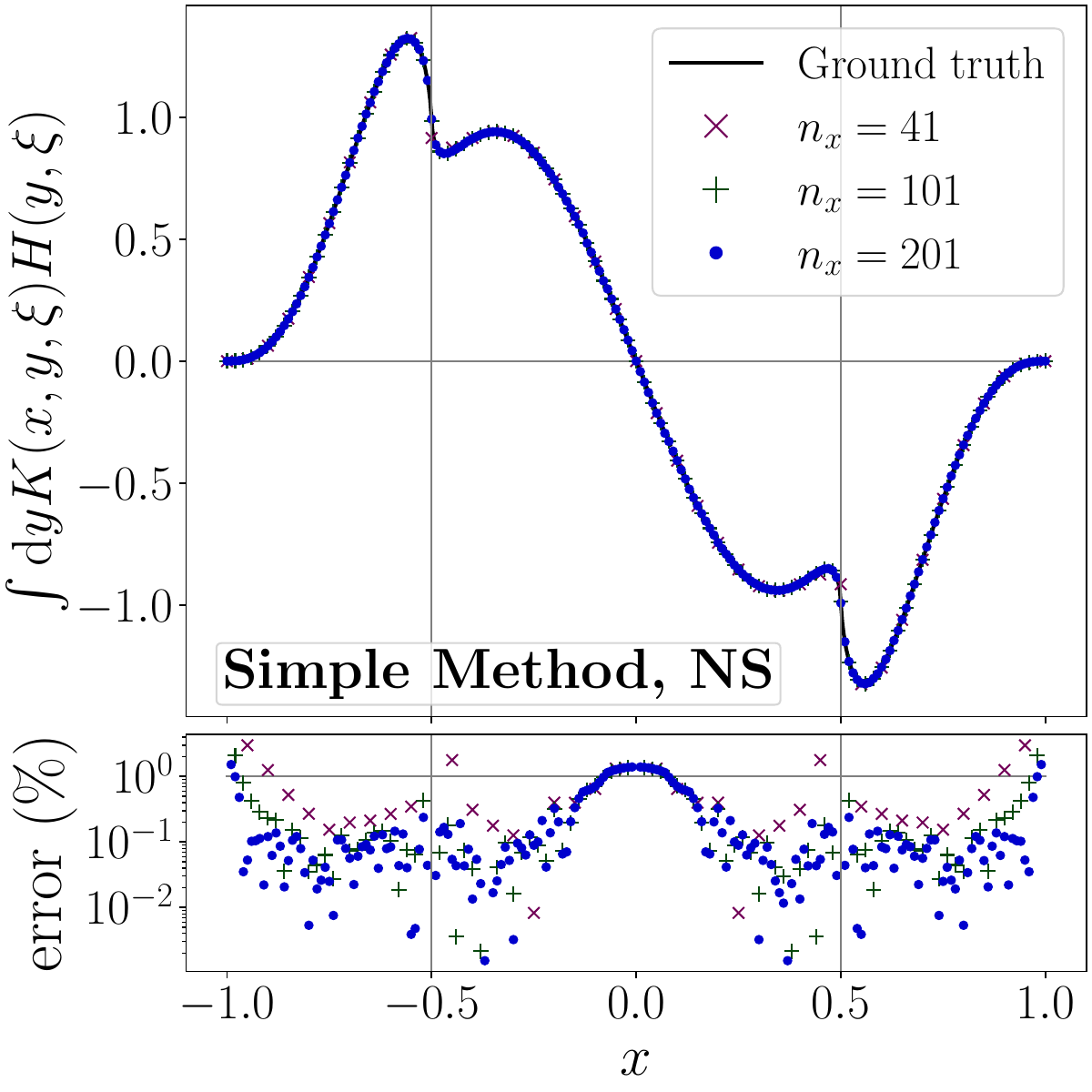}
  \includegraphics[width=0.49\textwidth]{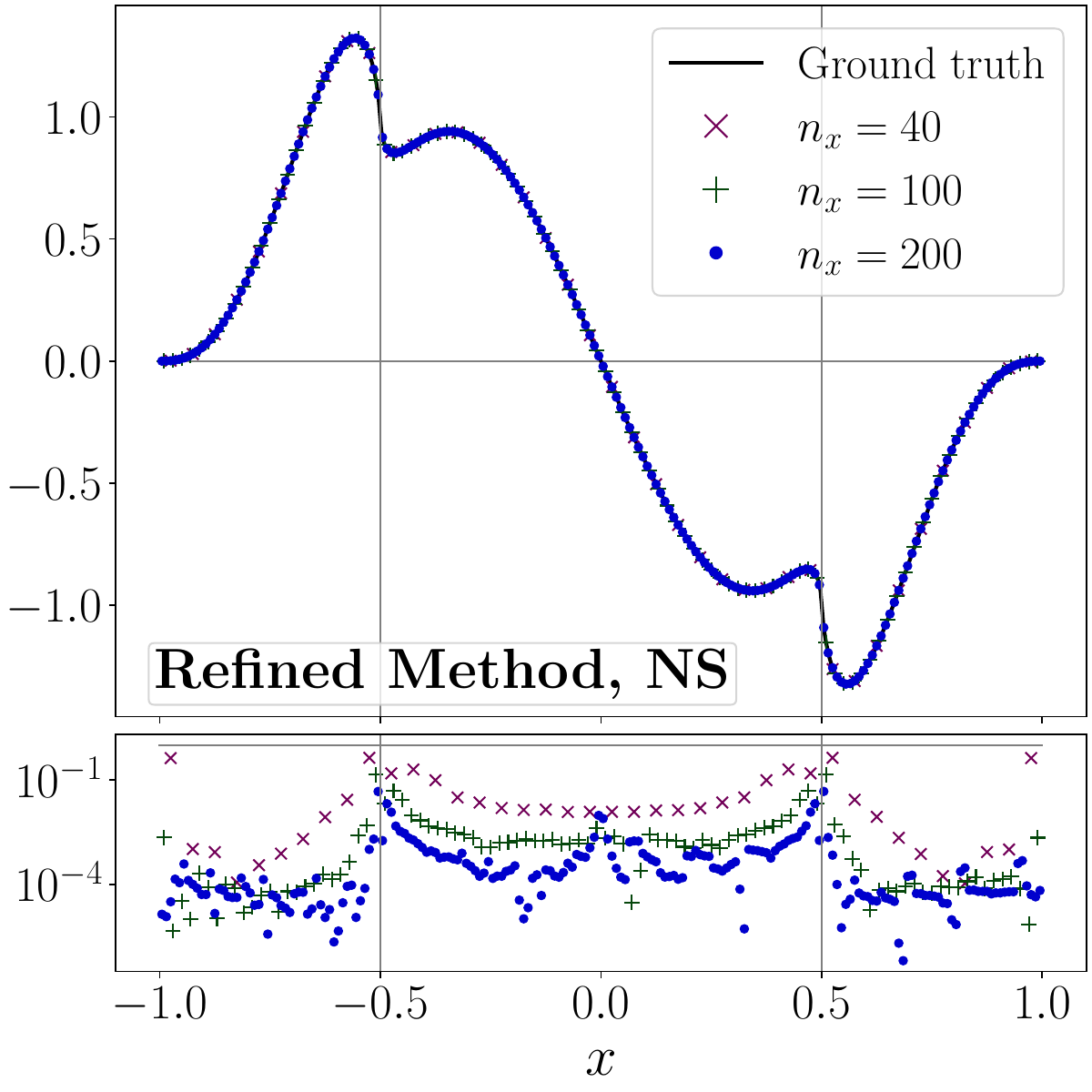}
  \caption{
    Accuracy benchmark for the right-hand side of the evolution equation
    (\ref{eqn:evolution}) at LO in the nonsinglet sector. This figure compares both the Simple Method (left panel) and the Refined Method (right panel) to a ``ground truth,'' the latter determined with the GK model~\cite{Kroll:2012sm} using adaptive quadrature. For the Simple Method, we use $n_g = 5000$ Gaussian weight points. The kernel method results are considered at varying values of $N_x$.
  }
  \label{fig:bench:nx}
\end{figure}

We next study how the accuracy of both methods depends on the density of the $x$ grid. This is shown in the nonsinglet sector at $\xi=0.5$ in Fig.~\ref{fig:bench:nx}. The discretization of $\xi$ in both methods is the same, but the Simple Method aligns the $x$ and $\xi$ grids, while the Refined Method staggers them---hence the $n_x$ considered in both methods is off by 1. The Simple Method is not able to achieve sub-percent accuracy over its entire $x$ range. By contrast, the Refined Method achieves sub-percent accuracy over the entire $x$ range---including $x \sim 0$ and $x \sim \pm 1$---with a coarser pixelation in $x$ space. In fact, the Refined Method at $n_x=40$ surpasses the Simple Method at $n_x=201$ in accuracy.

\subsection{Singlet kernels}

We next consider the singlet kernels, and produce companion plots for the nonsinglet benchmark in Fig.~\ref{fig:bench:xi}.
At LO, the $qq$ splitting function in the singlet sector is the same as in the nonsinglet sector; thus, only the $qg$, $gq$ and $gg$ kernels need to be benchmarked.

\begin{figure}
  \includegraphics[width=0.33\textwidth]{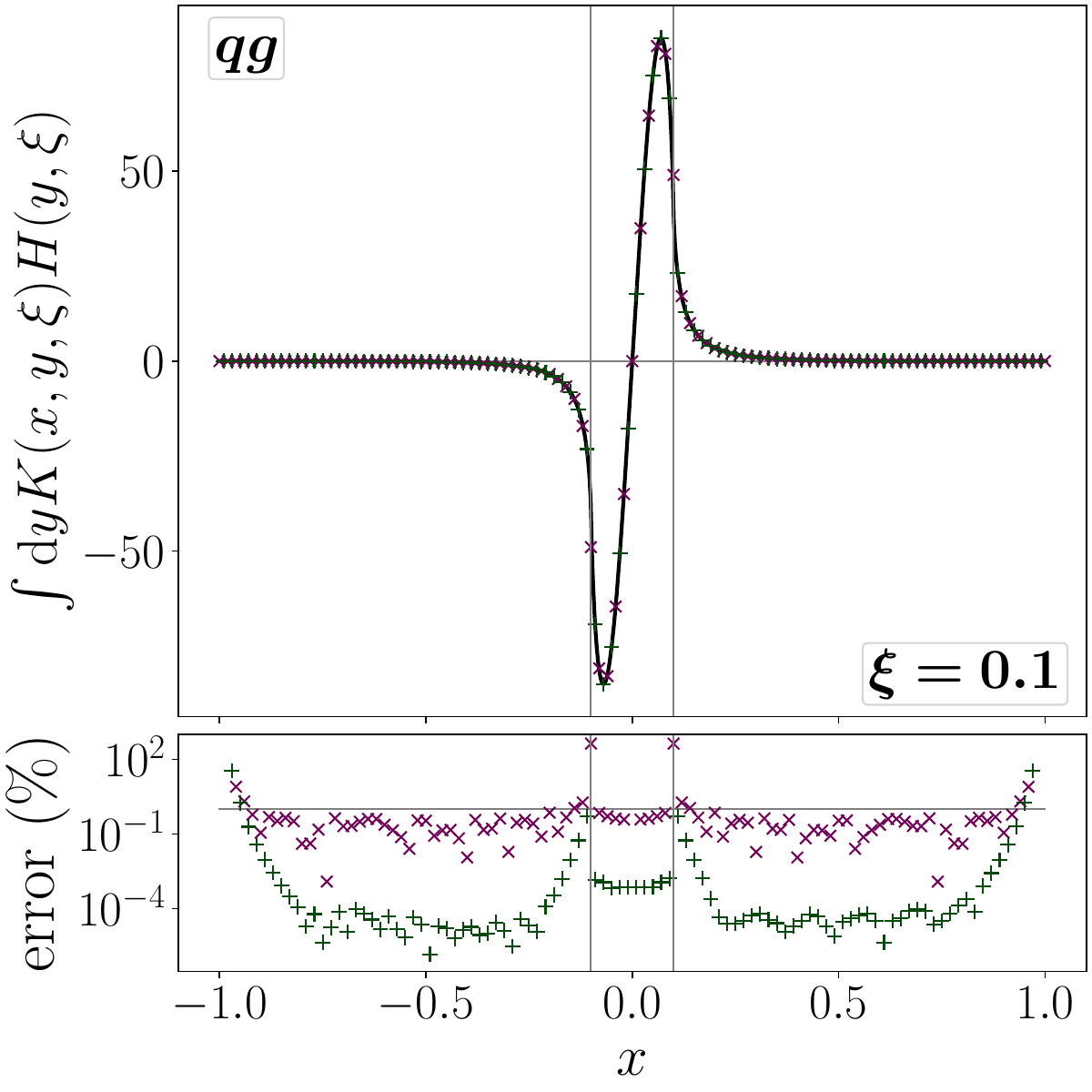}
  \includegraphics[width=0.33\textwidth]{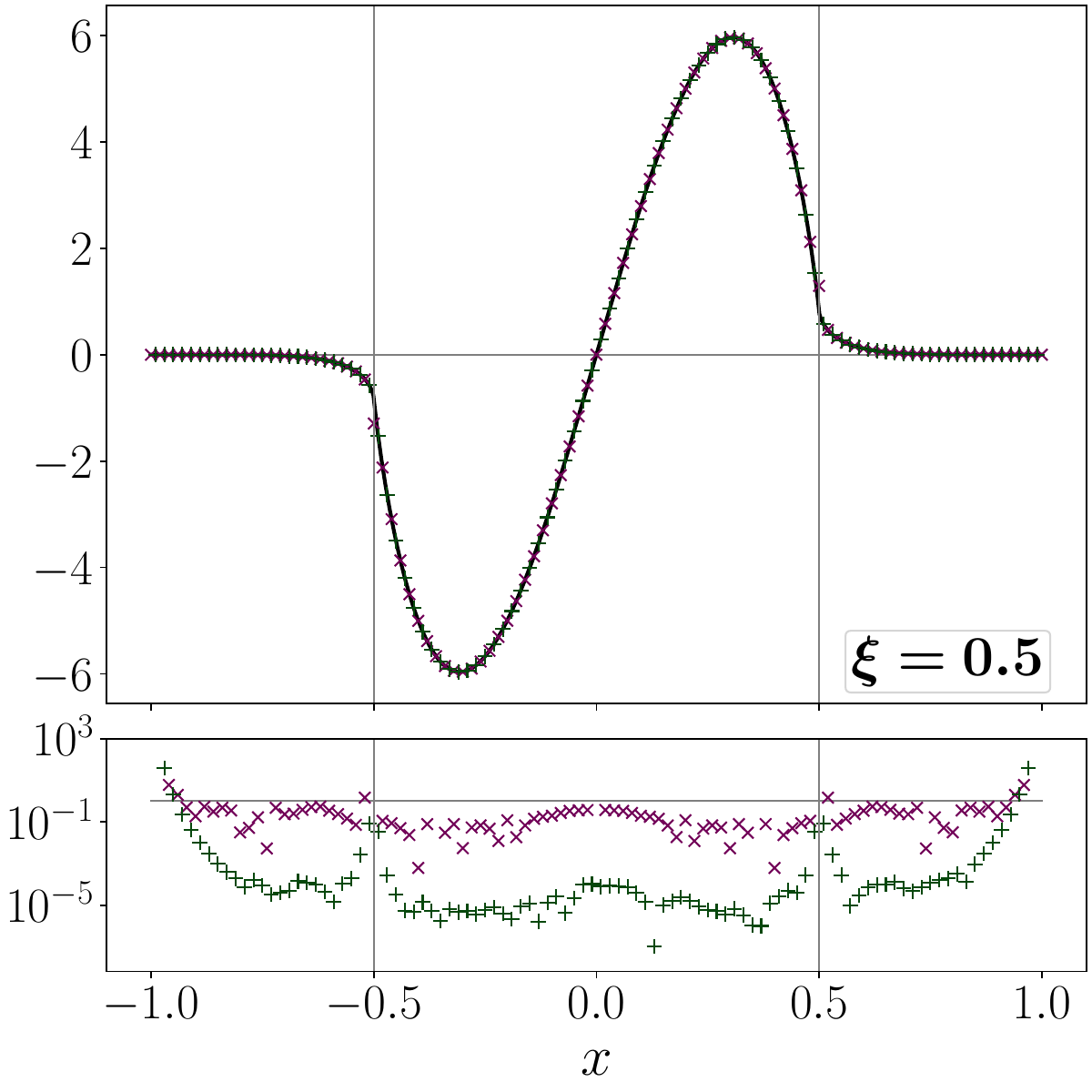}
  \includegraphics[width=0.33\textwidth]{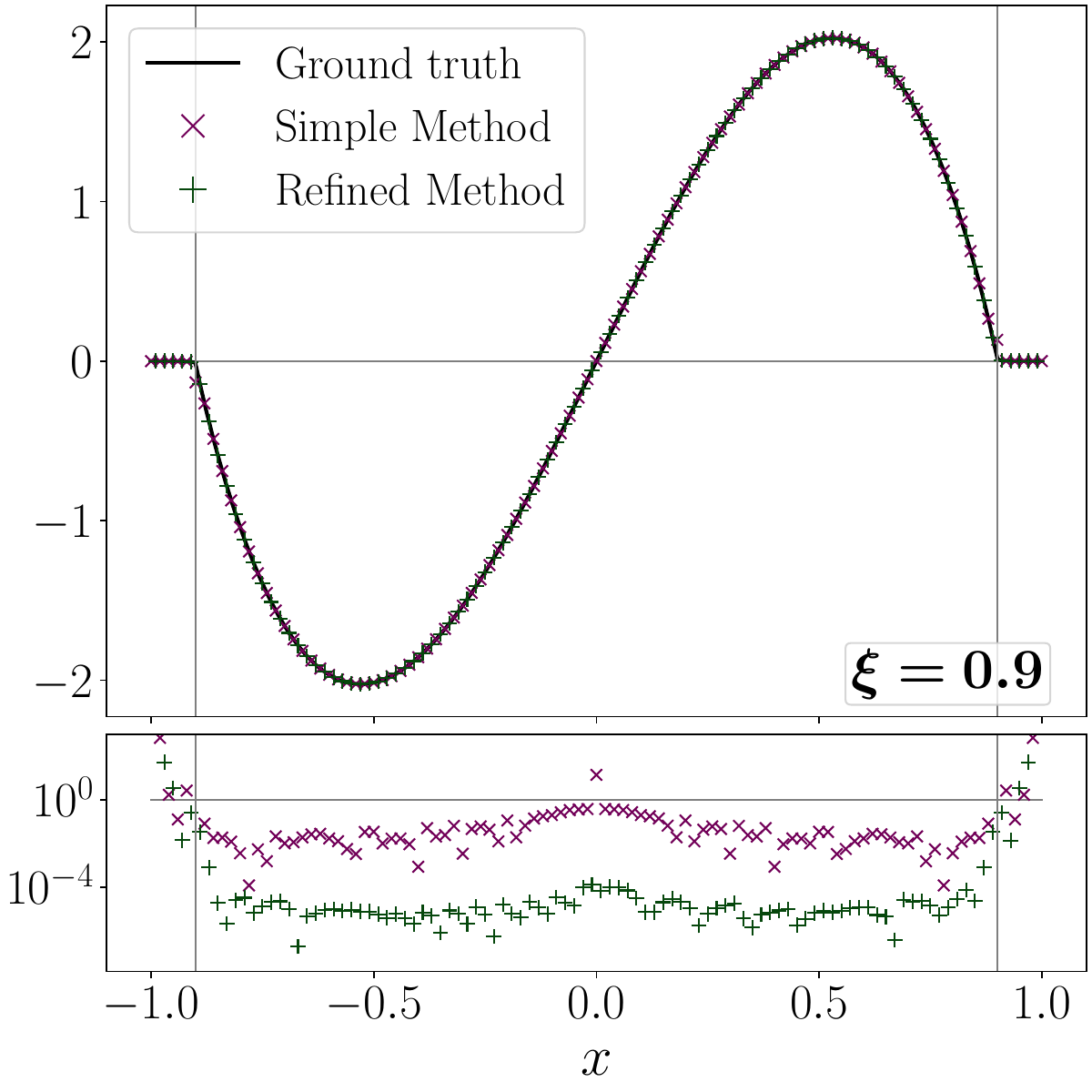}
  \caption{
    Accuracy benchmark for the $qg$ (quark from gluon) splitting function in the singlet evolution equation (\ref{eqn:evo:S}) at LO. This figure compares both the Simple Method (purple $\times$'s) and the Refined Method (green $+$'s) to a ``ground truth,'' the latter determined with the GK model \cite{Kroll:2012sm} using adaptive quadrature. For the Simple Method, we use $n_g = 5000$ Gaussian weight points and $N_x = 101$; for the Refined Method, we use $N_x = 100$. The left, middle and right panels show the results for $\xi=0.1$, $\xi=0.5$, and $\xi=0.9$, respectively.
  }
  \label{fig:bench:qg}
\end{figure}

First, in Fig.~\ref{fig:bench:qg} we show results for the right-hand side of the \emph{singlet} evolution equation (\ref{eqn:evo:S})---in particular, for the $qg$ (quark from gluon) contribution thereto, using the GK model \cite{Kroll:2012sm} as an example.
The ``ground truth'' is determined by directly integrating the GK model gluon GPD with the respective LO kernel.
These integrals must be performed numerically, but through adaptive Gaussian quadrature via quadpack, the results can be treated as arbitrarily precise relative to the errors incurred through the kernel method.
As in the nonsinglet sector, the Refined Method is significantly more accurate than the Simple Method.
Both methods produce about 1\% accuracy, except near special points, such as a zero crossing at $x=0$, or the endpoints $x=\pm\xi$ or $x=\pm 1$.

\begin{figure}
  \includegraphics[width=0.33\textwidth]{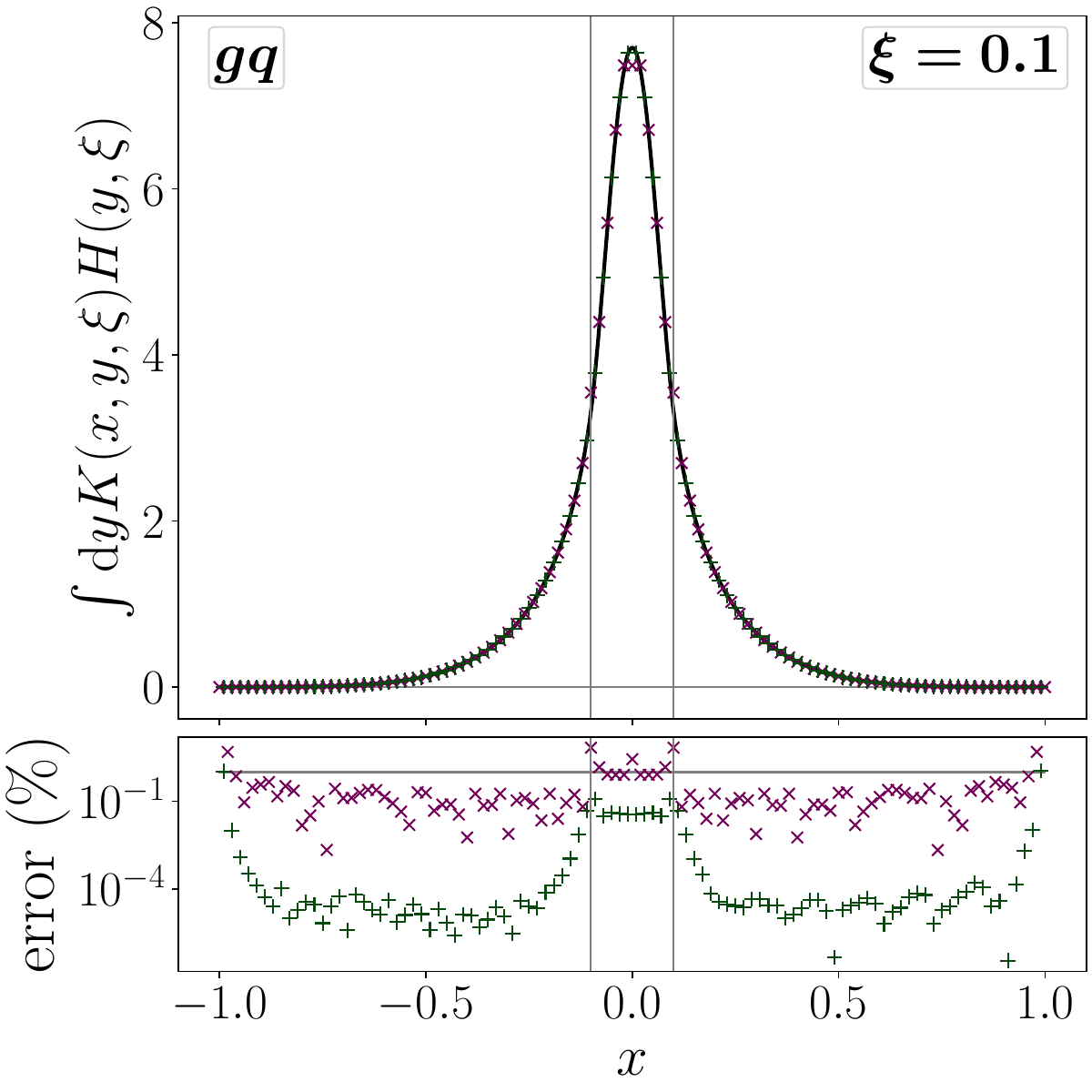}
  \includegraphics[width=0.33\textwidth]{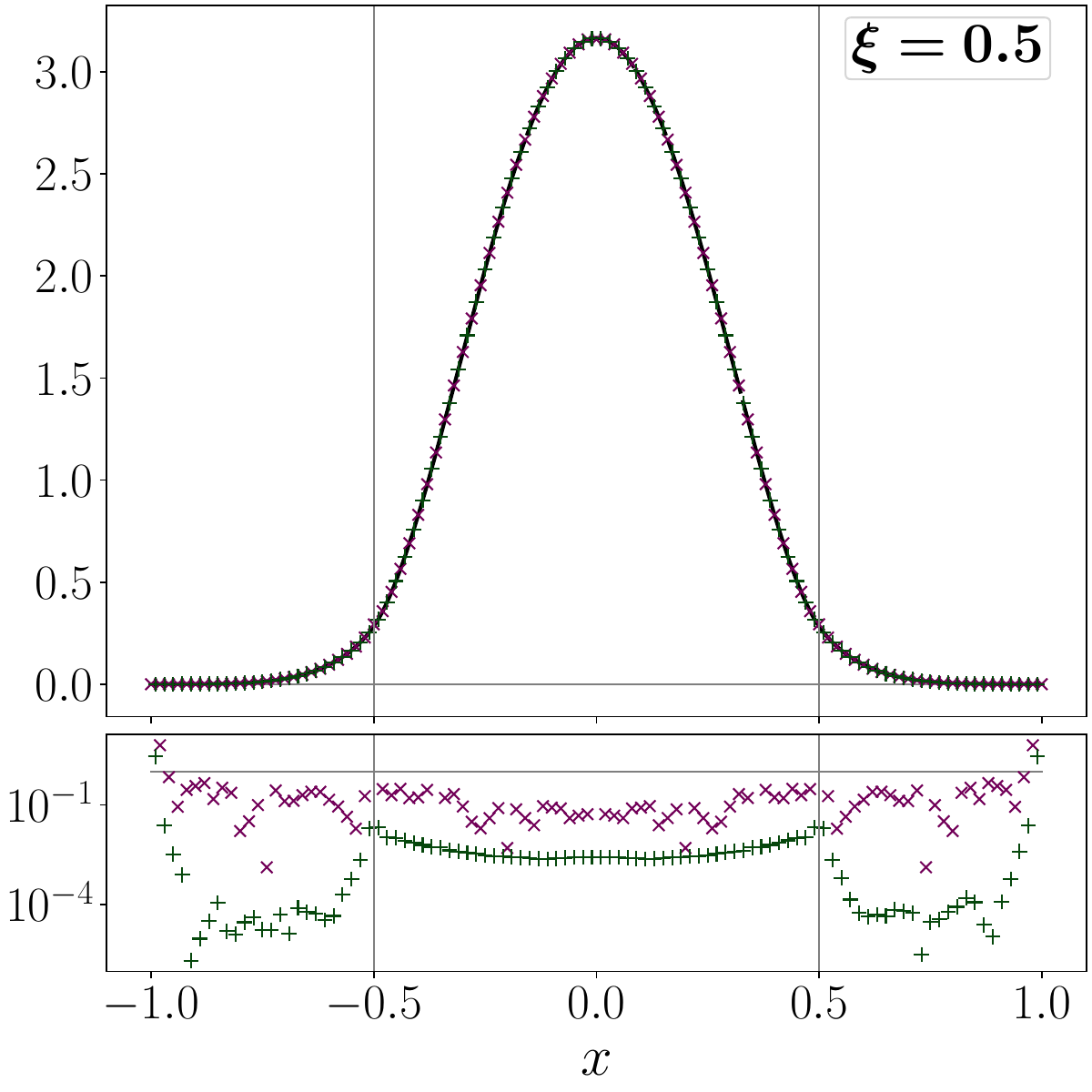}
  \includegraphics[width=0.33\textwidth]{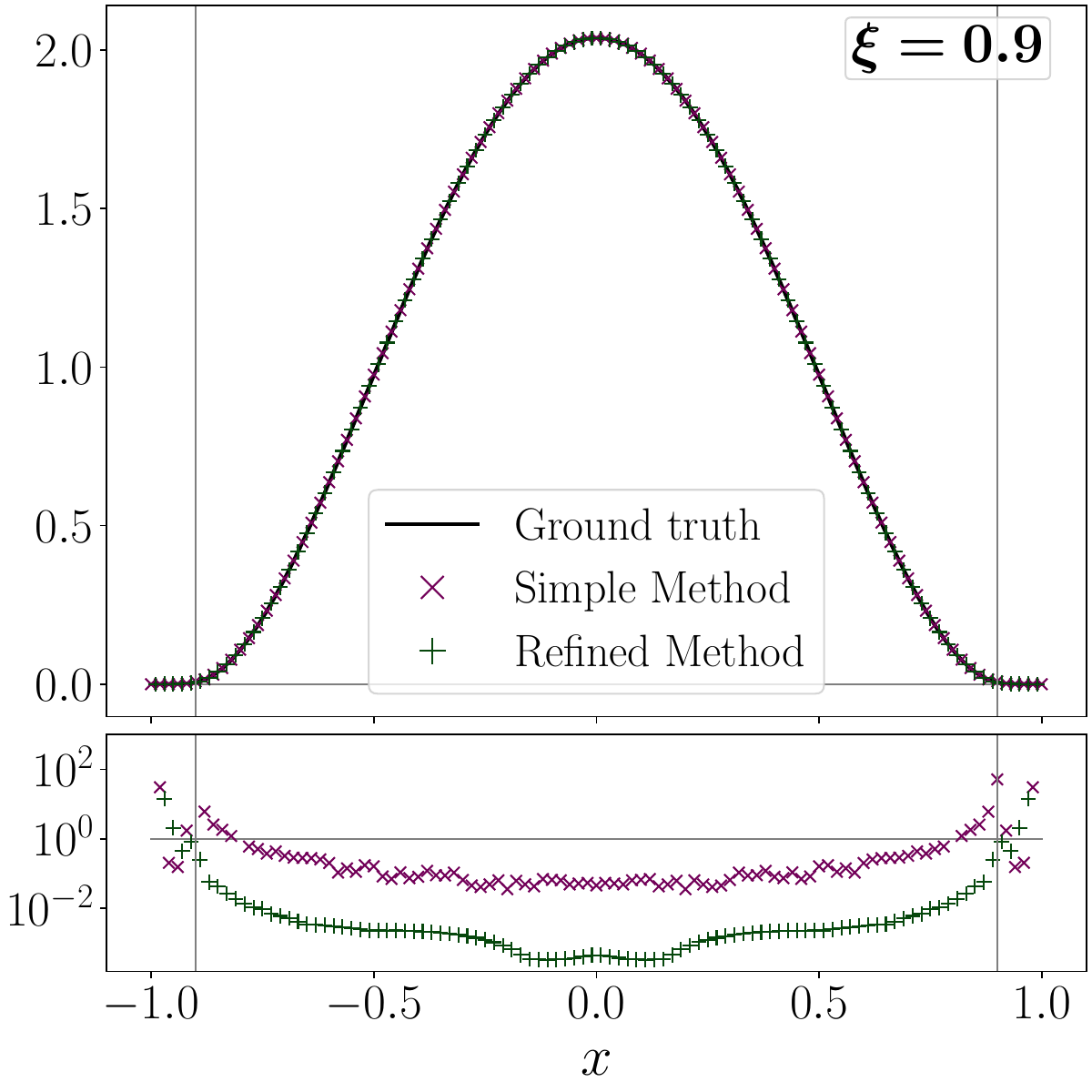}
  \caption{
    Accuracy benchmark for the $gq$ (gluon from quark) splitting function in the singlet evolution equation (\ref{eqn:evo:S}) at LO. This figure compares both the Simple Method (purple $\times$'s) and the Refined Method (green $+$'s) to a ``ground truth,'' the latter determined with the GK model~\cite{Kroll:2012sm} using adaptive quadrature. For the Simple Method, we use $n_g = 5000$ Gaussian weight points and $N_x = 101$; for the Refined Method, we use $N_x = 100$. The left, middle and right panels show $\xi=0.1$, $\xi=0.5$ and $\xi=0.9$ respectively.
  }
  \label{fig:bench:gq}
\end{figure}

Next, in Fig.~\ref{fig:bench:gq}, we show results using the $gq$ (gluon from quark) kernel.
The Refined Method again surpasses the Simple Method, the latter of which shows poorly accuracy in the ERBL region at $\xi=0.1$, suggesting the Refined Method is more suitable for phenomenology in this domain.
Both methods struggle with accuracy at $x \sim \pm 1$, where the ``ground truth'' rapidly falls to $0$.

\begin{figure}
  \includegraphics[width=0.33\textwidth]{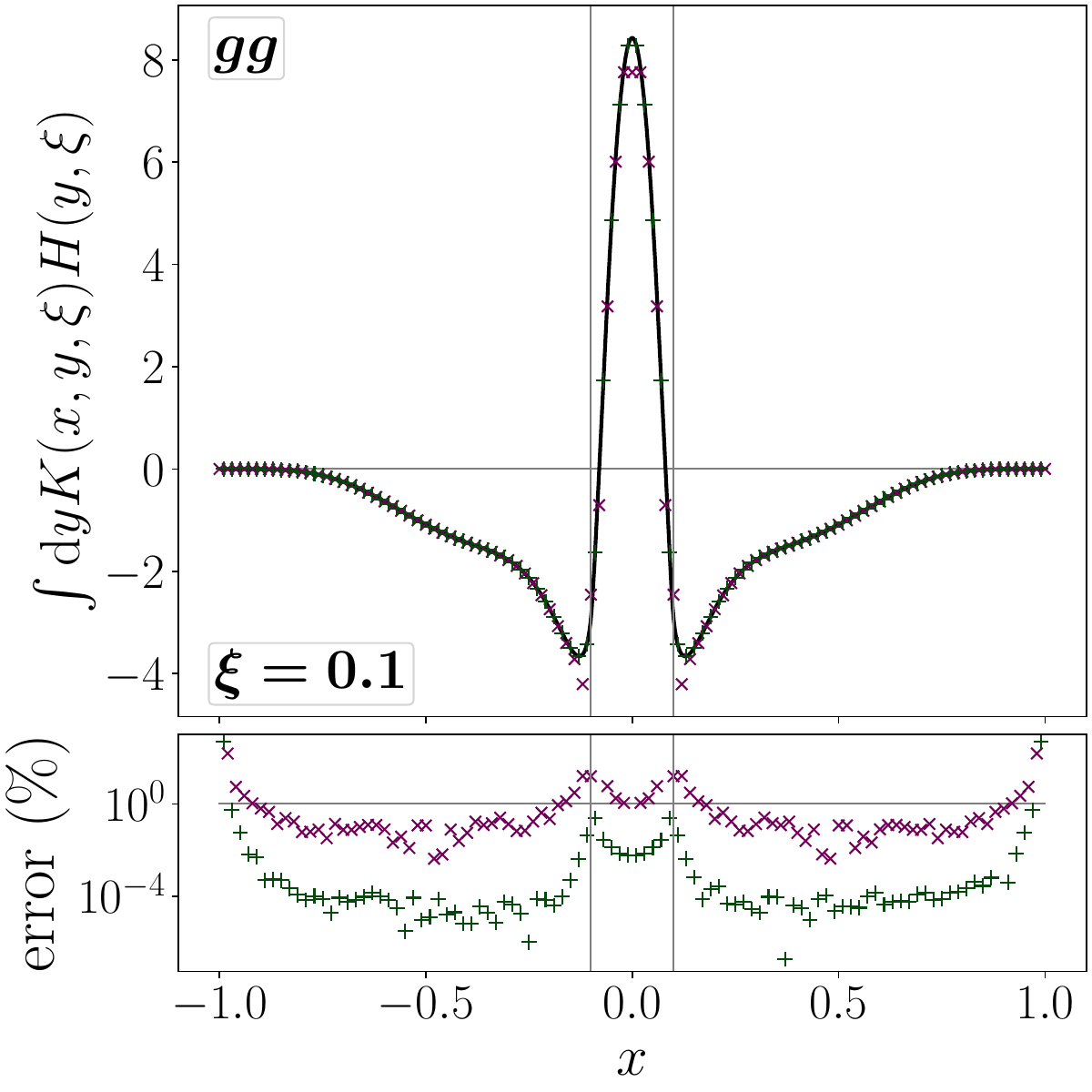}
  \includegraphics[width=0.33\textwidth]{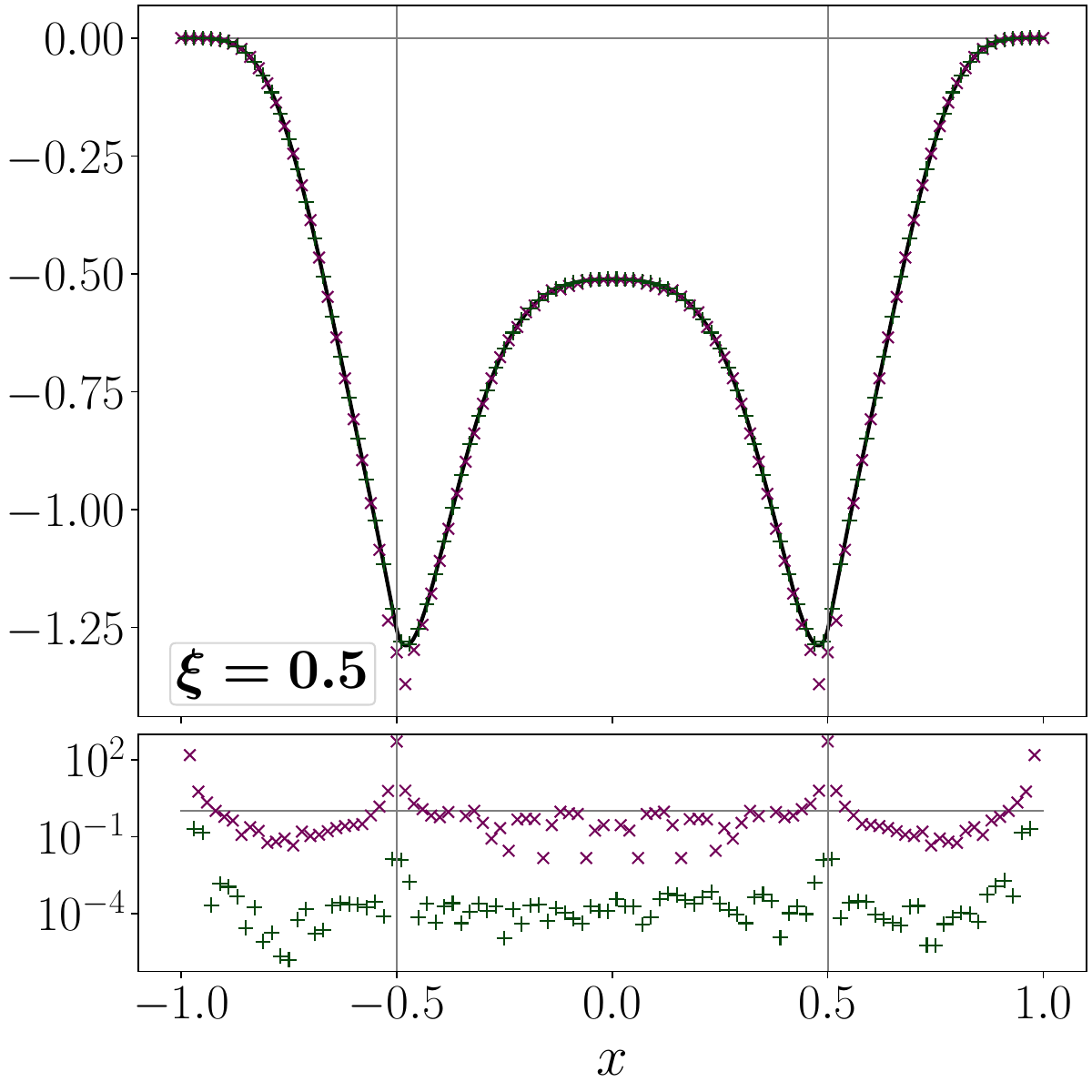}
  \includegraphics[width=0.33\textwidth]{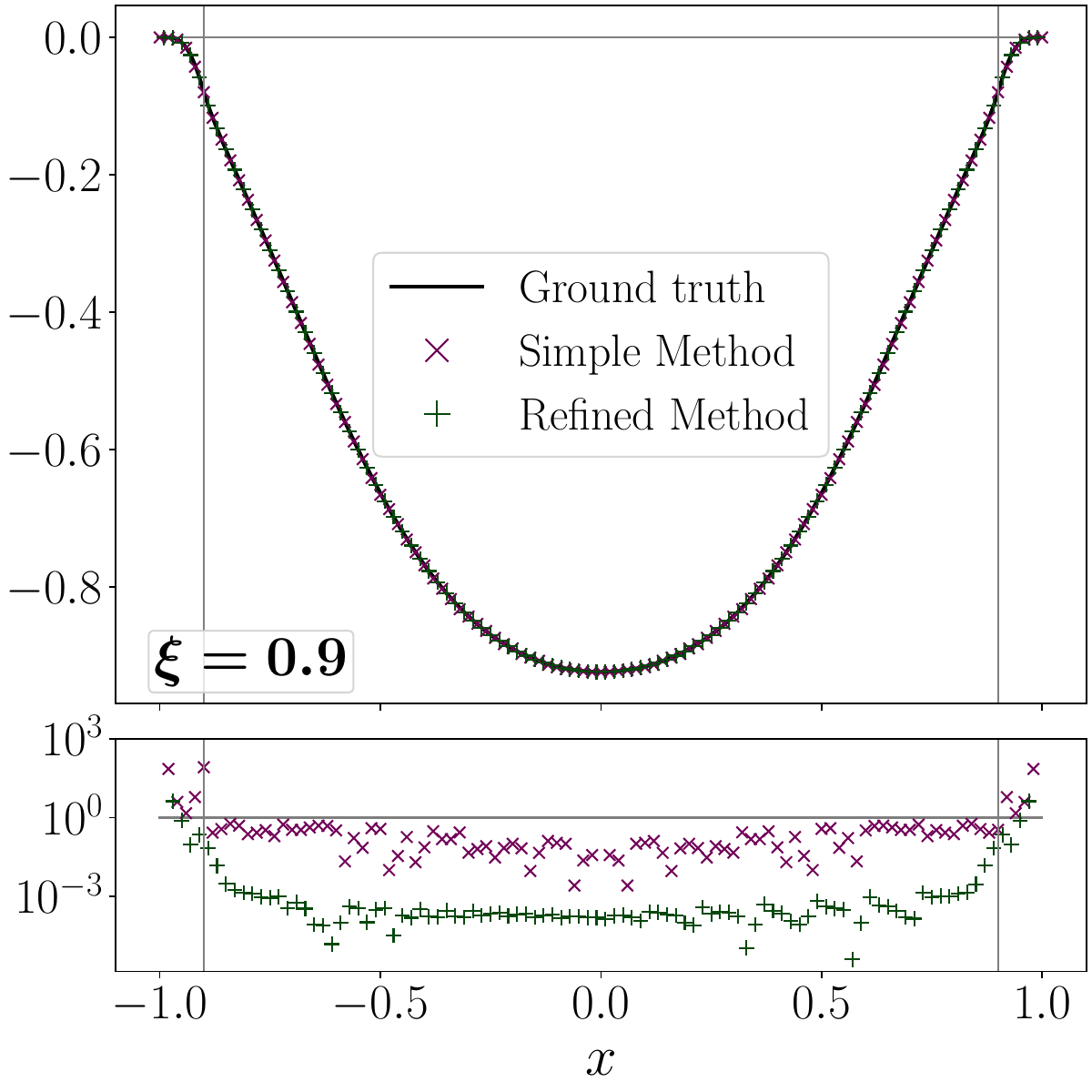}
  \caption{
    Accuracy benchmark for the $gg$ (gluon from gluon) splitting function in
    the singlet evolution equation (\ref{eqn:evo:S}) at LO.
    This figure compares both the Simple Method (purple $\times$'s)
    and the Refined Method (green $+$'s) to a ``ground truth,''
    the latter determined with the GK model~\cite{Kroll:2012sm}
    using adaptive quadrature.
    For the Simple Method, we use $n_g = 5000$ Gaussian weight points
    and $N_x = 101$; for the Refined Method, we use $N_x = 100$.
    The left, middle and right panels show $\xi=0.1$, $\xi=0.5$
    and $\xi=0.9$ respectively.
  }
  \label{fig:bench:gg}
\end{figure}

Lastly, in Fig.~\ref{fig:bench:gg} we show an accuracy benchmark for the $gg$  splitting function.
Much like in our $gq$ benchmark, the Refined Method is able to deliver sufficient accuracy (away from the $x=\pm1$ endpoints), and outperforms the Simple Method by orders of magnitude.
The Simple Method has especially large errors around the vicinity of the DGLAP-ERBL boundary, and greater than 1\% error in the ERBL region at $\xi=0.1$.
The Refined Method thus appears more suitable for phenomenological application.

\subsection{Breakdown at small skewness}

\begin{figure}
  \includegraphics[width=0.46\textwidth]{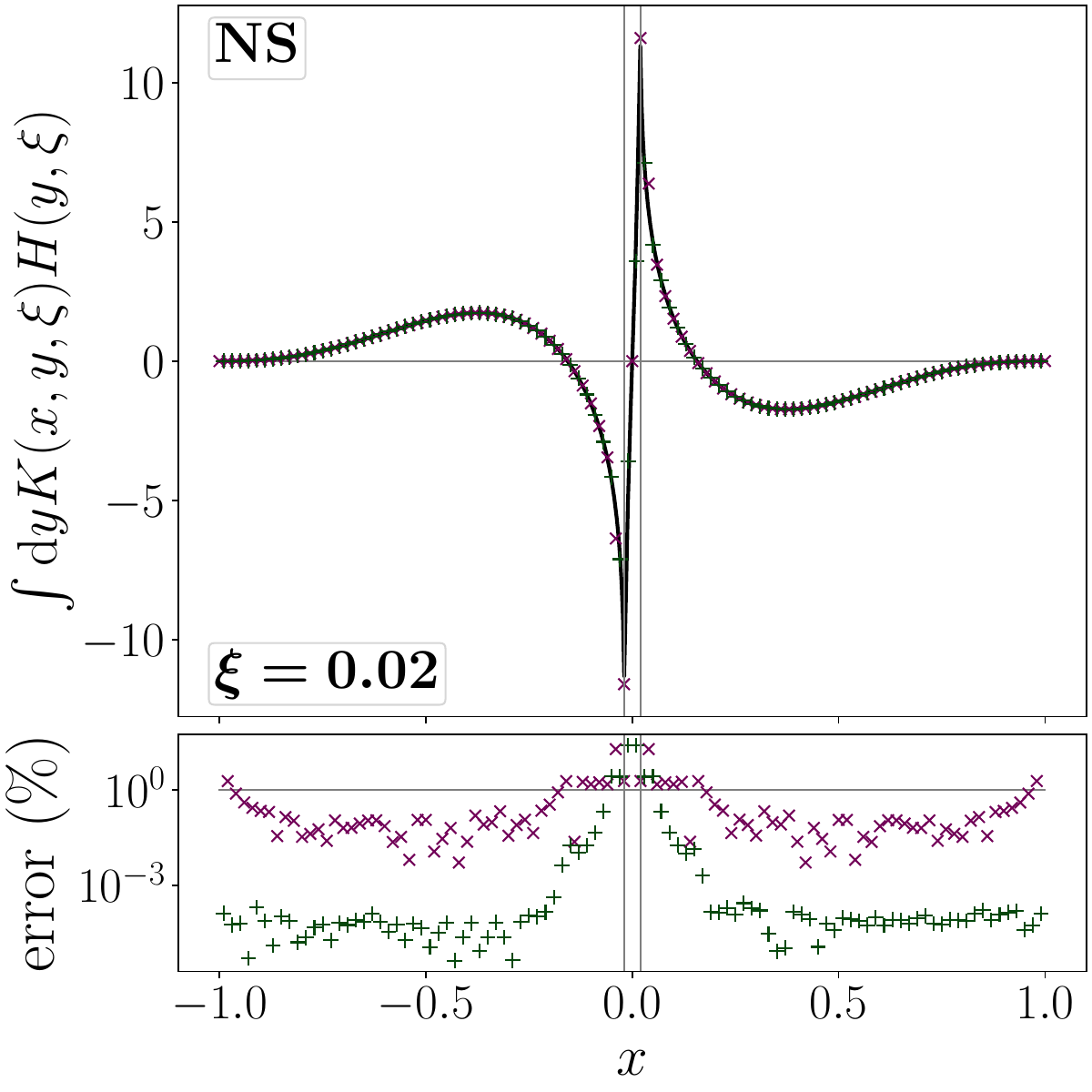}
  \includegraphics[width=0.46\textwidth]{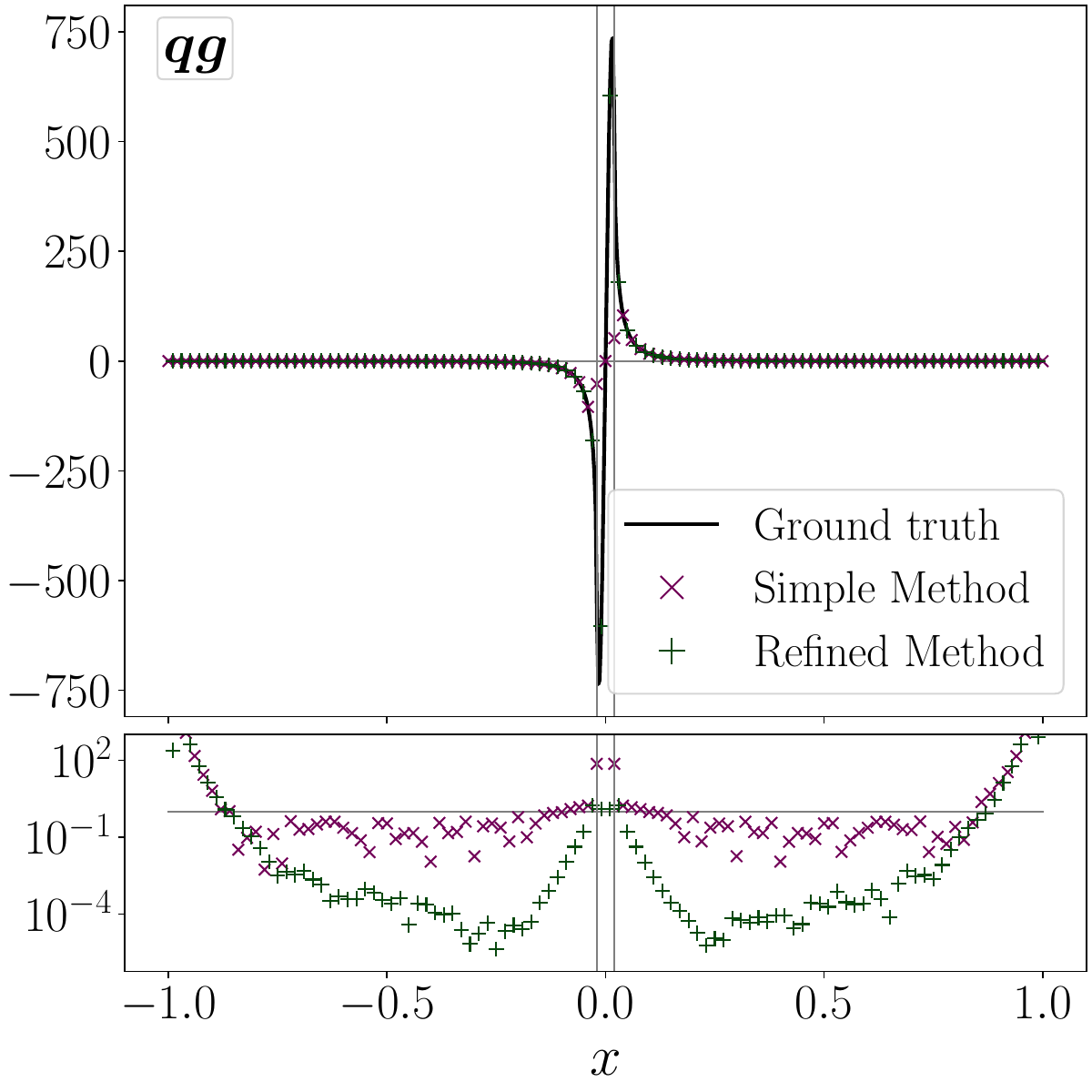}
  \includegraphics[width=0.46\textwidth]{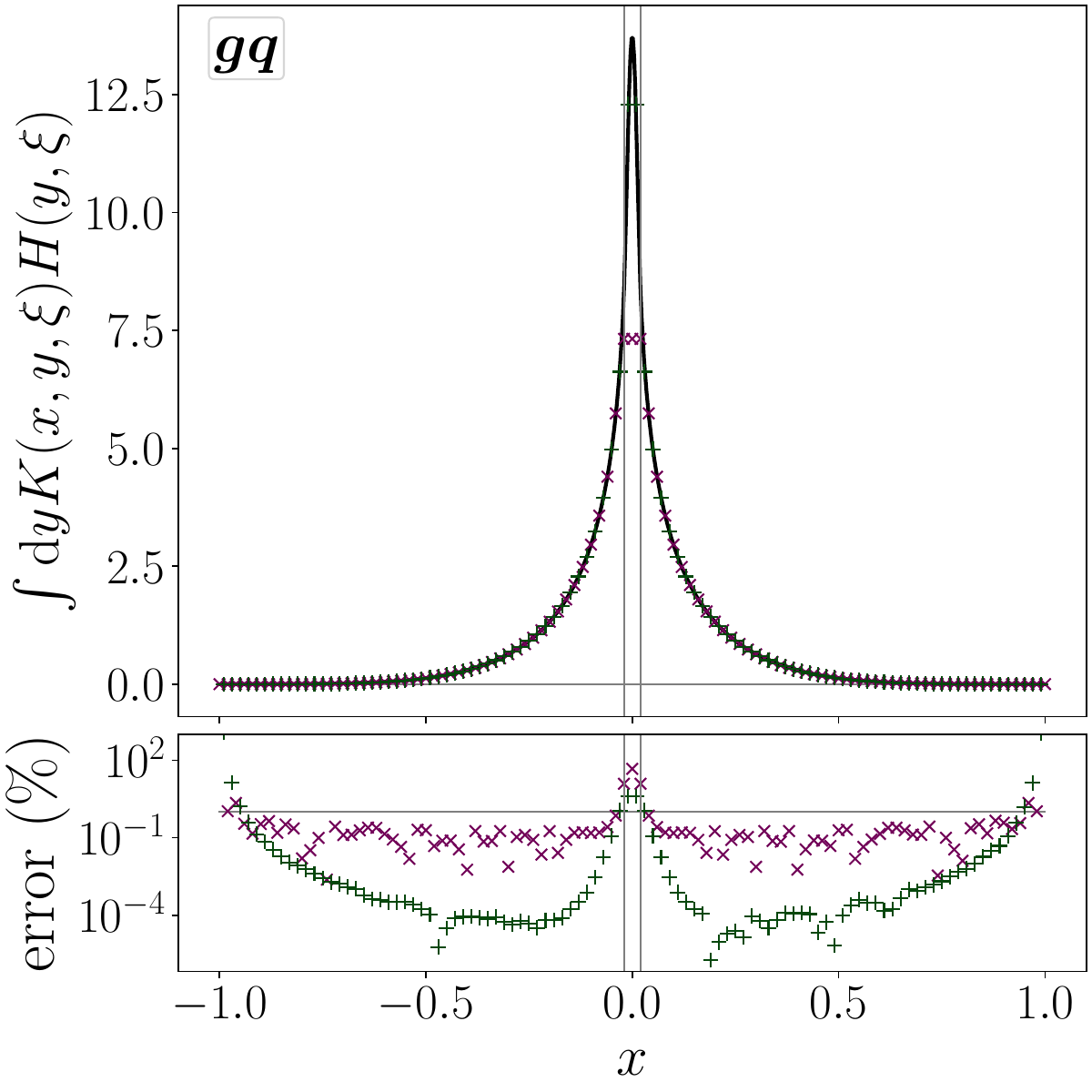}
  \includegraphics[width=0.46\textwidth]{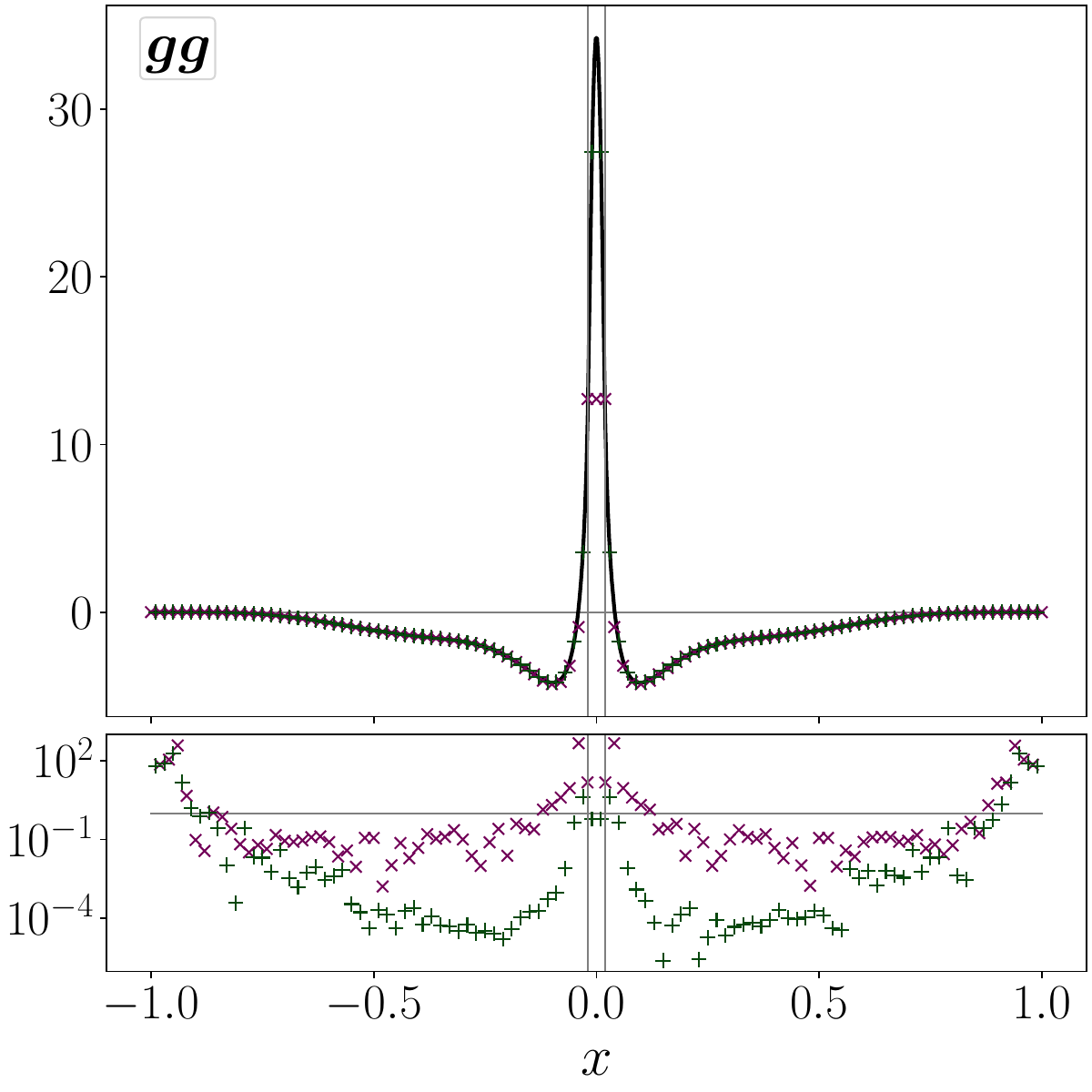}
  \caption{
    Accuracy benchmarks of the nonsinglet (top-left), quark-gluon (top-right), 
    gluon-quark (bottom-left), and gluon-gluon singlet (bottom-right) kernels
    at $\xi=0.02$, representing small skewness.
  }
  \label{fig:smol}
\end{figure}

In Fig.~\ref{fig:smol}, we provide an accuracy benchmark for our evolution codes at $\xi=0.02$, a stress test of the Methods.
For $|x| \gg 0$, at least, the Methods show the same level of accuracy that they do for larger skewness values: around $\sim1\%$ error in the Simple Method, and much better accuracy in the Refined Method, except sometimes at the $x\sim\pm1$ endpoints where the GPD itself rapidly falls.
However, at small $|x|$, the error in both Methods grows significantly.
It is important to disclose that the codes break down in this area, and advise potential users of the Refined Method code that its results cannot be trusted when both $x$ and $\xi$ are very small.

The cause of these inaccuracies is the steep increase in the GPD at small $|x|$.
As $\xi \rightarrow 0$, the behavior of GPDs approaches that of collinear parton distributions---for which it is well-known that evolution is more accurate with a logarithmically-spaced grid.
Thus, improvements at very small $\xi$ could be attained through future updates to the code to utilize a logarithmically spaced grid (similar to that used by APFEL~\cite{Bertone:2013vaa}).

\subsection{Additional polynomiality tests}

\begin{figure}
  \includegraphics[width=0.33\textwidth]{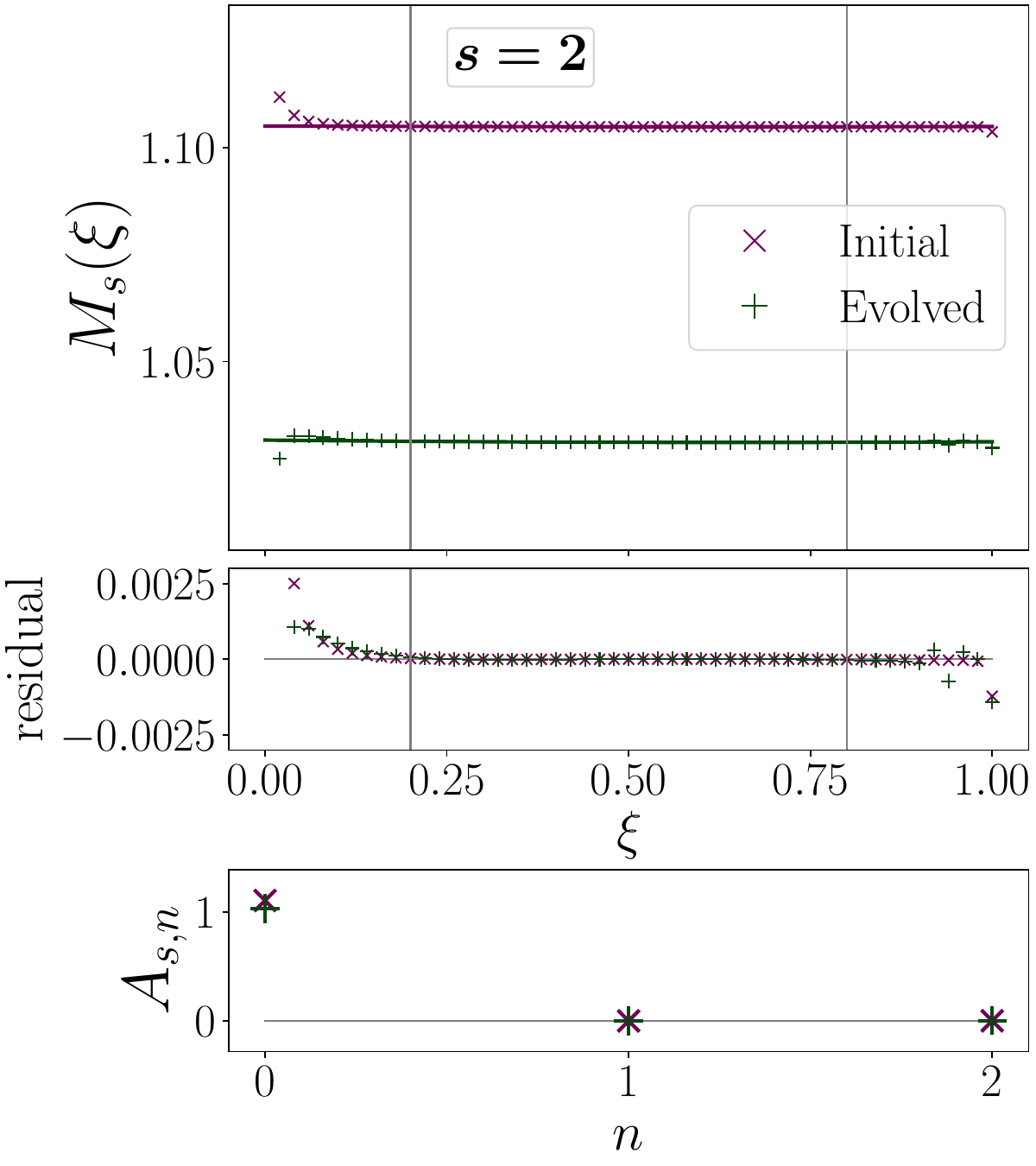}
  \includegraphics[width=0.33\textwidth]{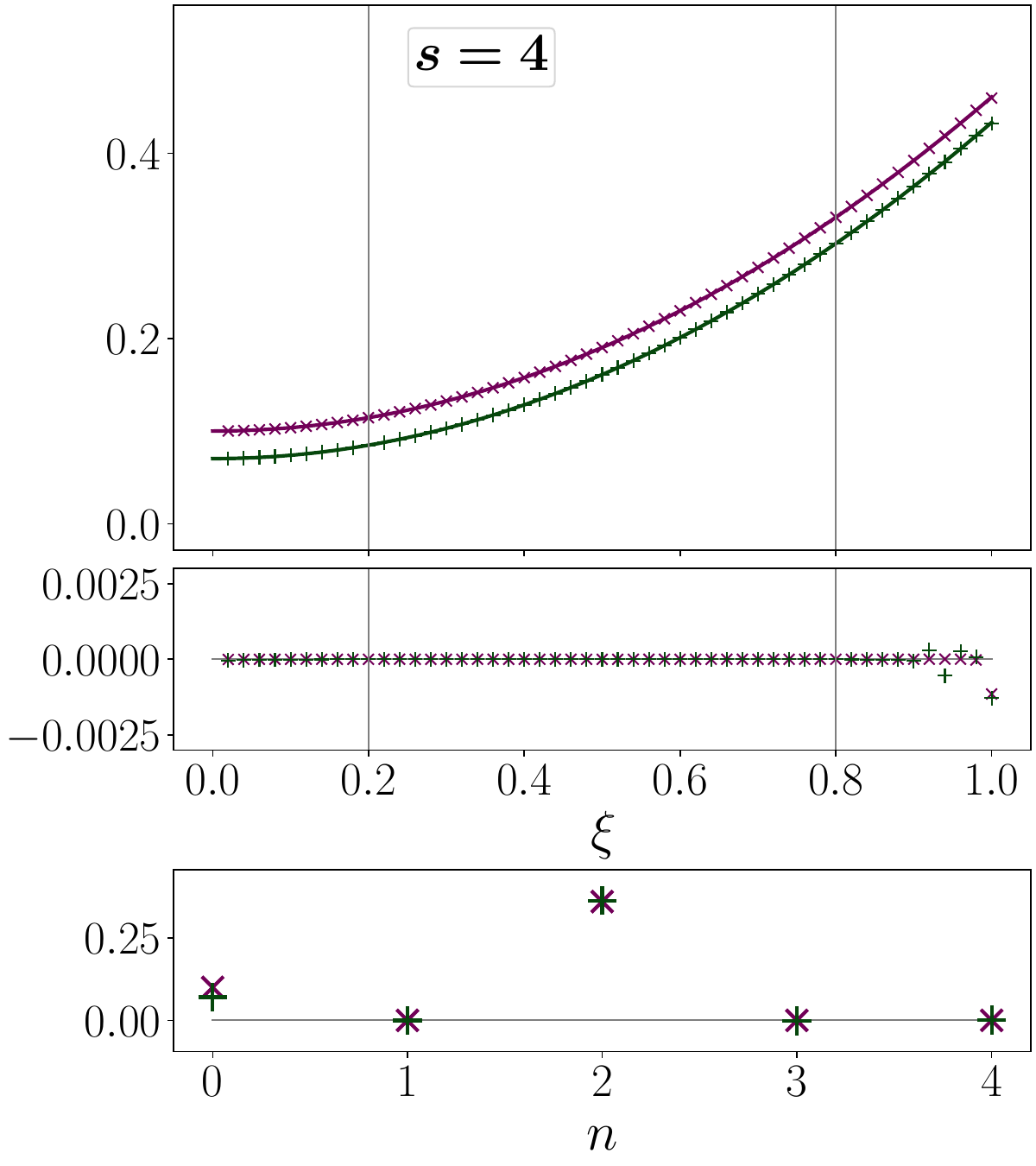}
  \includegraphics[width=0.33\textwidth]{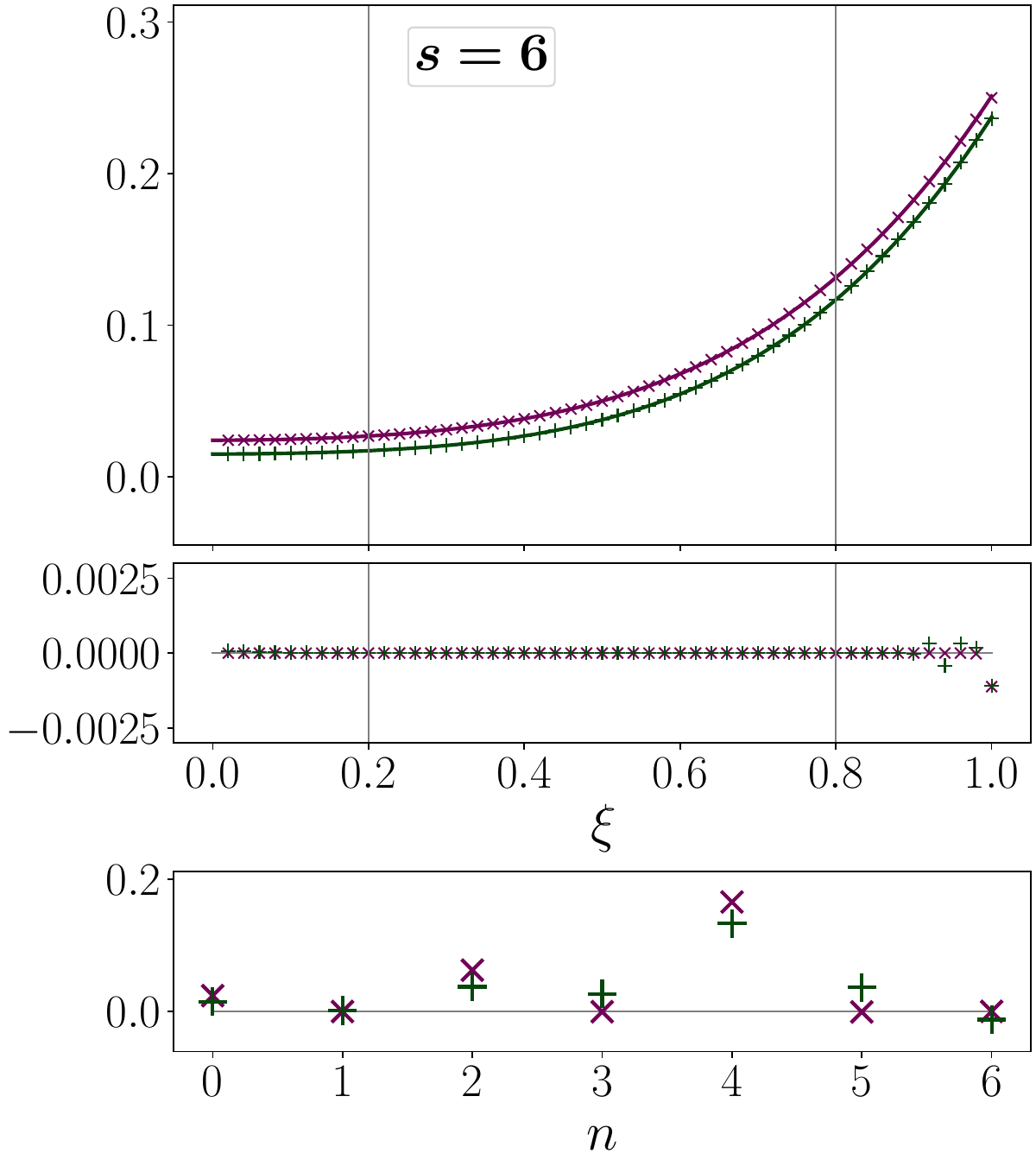}
  \caption{
    As in Fig.~\ref{fig:poly:umin}, but for the singlet quark distribution.
    Since odd moments of the singlet distribution vanish,
    the first three even moments ($s=2,4,6$) are shown.
  }
  \label{fig:poly:S}
\end{figure}

In Fig.~\ref{fig:poly:S} we show a polynomiality test for the singlet quark GPD.
Since odd moments of the singlet quark GPD are zero, we show the first 3 even moments.
For $s=2,4$, polynomiality is obeyed to very good precision, except at small $\xi < 0.1$.
For $s=6$, the expected polynomial behavior breaks down after evolution: all odd coefficients should be zero due to time reversal symmetry, but we find small, nonzero odd coefficients after evolution.
This is an indication of accumulated numerical error, likely due to discretization.

\begin{figure}
  \includegraphics[width=0.33\textwidth]{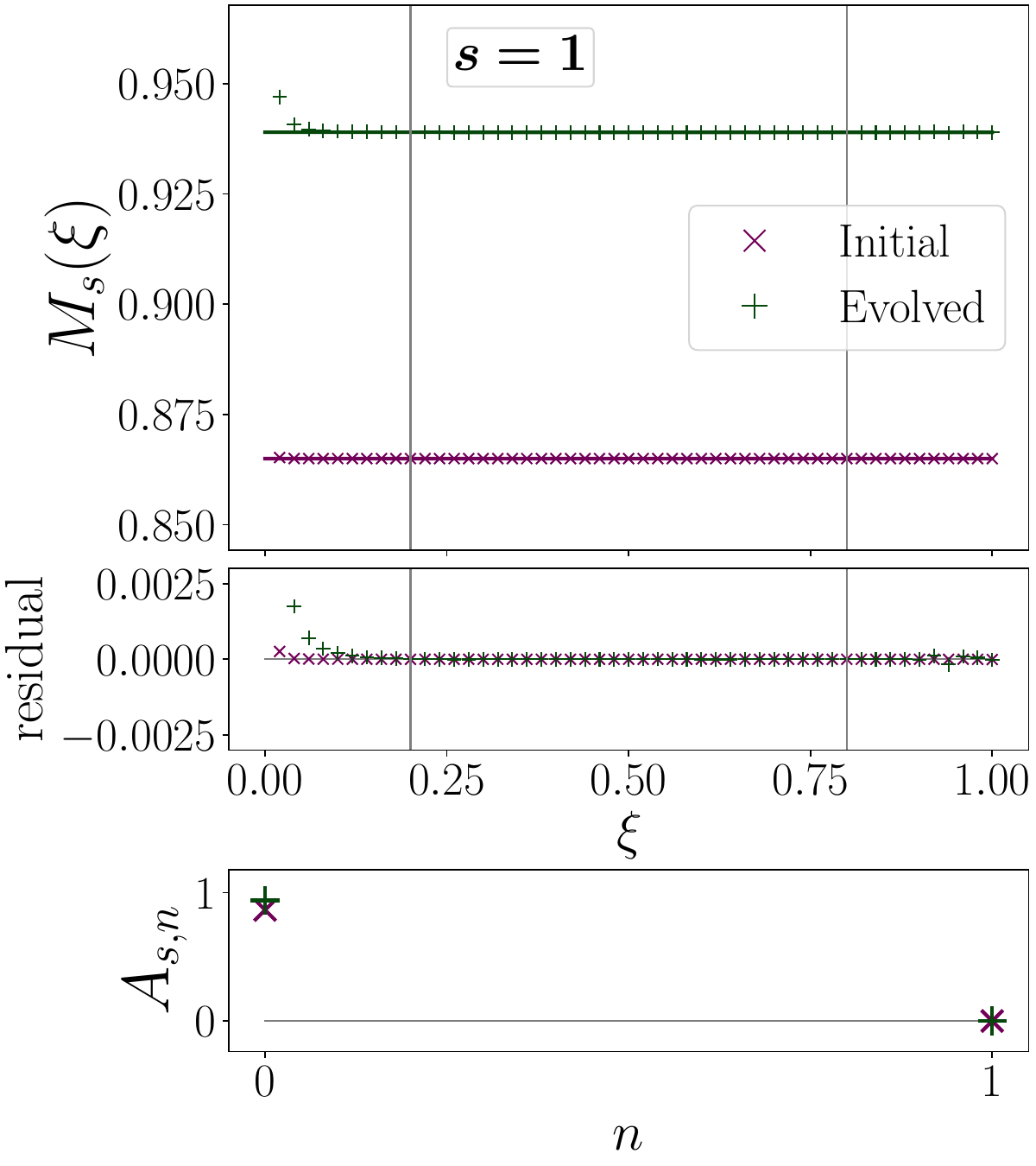}
  \includegraphics[width=0.33\textwidth]{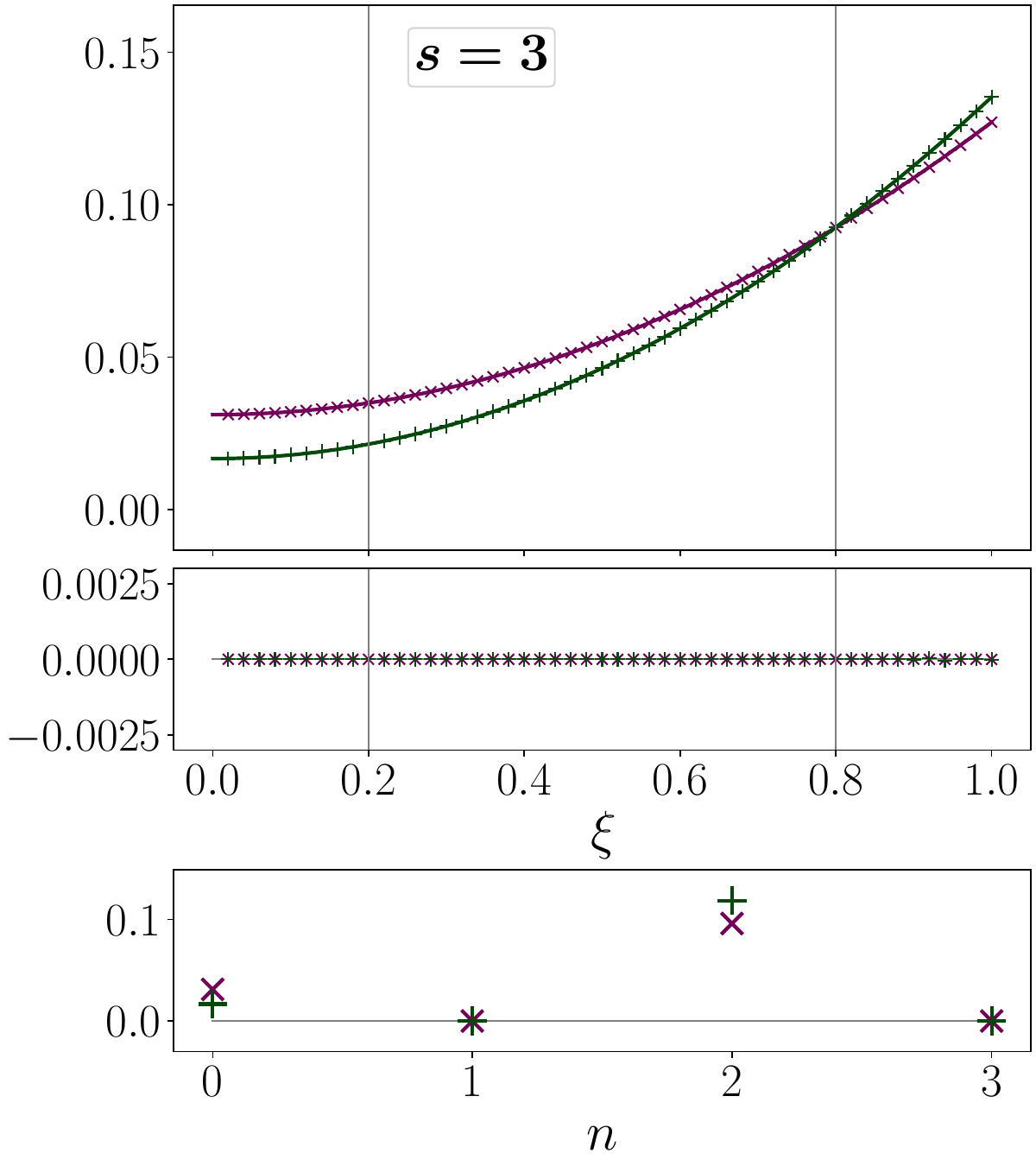}
  \includegraphics[width=0.33\textwidth]{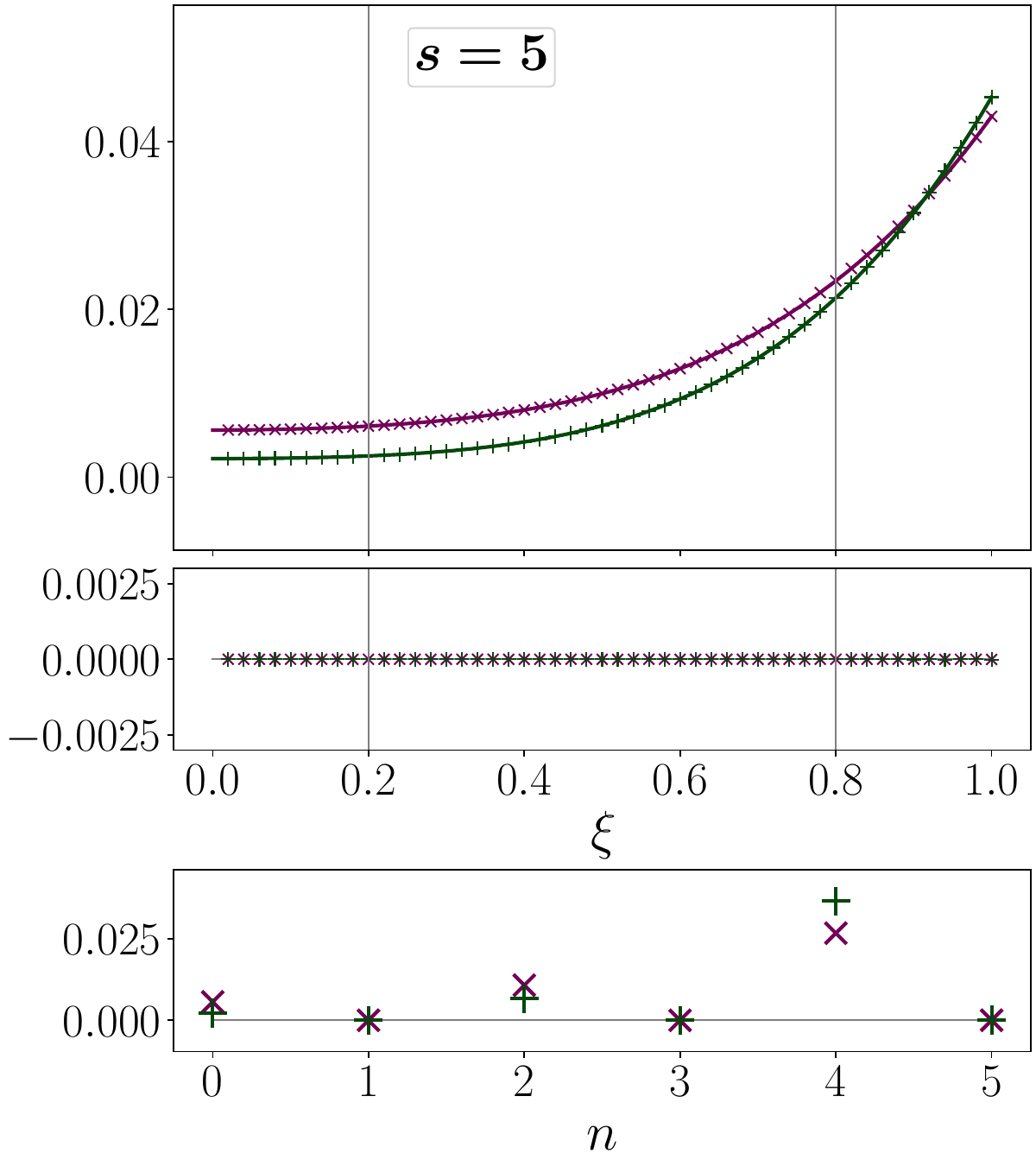}
  \caption{
    As in Fig.~\ref{fig:poly:umin}, but for the gluon distribution.
    Since even moments of the gluon distribution vanish,
    the first three odd moments ($s=1,3,5$) are shown.
  }
  \label{fig:poly:G}
\end{figure}

Finally, in Fig.~\ref{fig:poly:G} we show a polynomiality test for the gluon GPD. Since even moments of the gluon GPD are zero, we show the first three odd moments. 
We find that polynomiality is obeyed to very good precision, except at small $\xi < 0.1$. 
One must consider $s \geq 7$ for gluon GPDs to find numerical breakdown of the expected behavior (where odd coefficients become nonzero after evolution).

\clearpage
\bibliography{references.bib}

\end{document}